\documentclass[prl,reprint,showpacs,preprintnumbers,amsmath,amssymb,footnotebib]{revtex4-1}
\usepackage{epsfig}
\usepackage{graphicx}
\usepackage{textcomp}
\usepackage{tabularx}
 
\begin{document}
\title{Fragility of reaction-diffusion models to competing advective processes}

\author{Oleg Kogan}
\affiliation{Laboratory of Atomic and Solid State Physics, Cornell University, Ithaca, NY, 14853}
\email{obk5@cornell.edu}
\author{Kevin O'Keeffe}
\affiliation{Center for Applied Mathematics, Cornell University, Ithaca, NY, 14853}
\author{Christopher R. Myers}
\affiliation{Laboratory of Atomic and Solid State Physics, and Institute of Biotechnology, Cornell University, Ithaca, MY 14853}

\begin{abstract}
We study the coupling of a Fisher-Kolmogorov-Petrovsky-Piskunov (FKPP) equation to a separate, advection-only transport process.  We find that the front dynamics can be described by an FKPP-like equation only at sufficiently fast diffusion or large coupling strength.  For such parameter regimes, we find a mapping to an effective FKPP equation.  We also find that FKPP equation is fragile with respect to the coupling to an advection-only mechanism, discover conditions when the front width diverges, and when front speed is insensitive to the coupling.  At zero diffusion in this mean-field description, the downwind front speed goes to a finite value as the coupling goes to zero.
\end{abstract}

\maketitle

The Fisher-Kolmogorov-Petrovsky-Piskunov (FKPP) equation, originally introduced to describe the population dynamics of the spread of advantageous genes \cite{FKPP_Pop_Gen} has found applications in a very wide range of contexts that include ecology \cite{Math_Bio}, epidemiology \cite{Math_Bio}, population biology \cite{DNelson review},  chemical kinetics \cite{Chem_Kin}, extreme-value statistics \cite{Extreme_Vals}, disordered systems \cite{Disordered_Systems}, and even high energy physics \cite{High_En}.  It describes reaction-diffusion processes involving saturation-limited growth and diffusion, and admits front-like solutions - known as Fisher waves - that invade a linearly-unstable state.  

\begin{figure}[ht]
\includegraphics[width=2.3in]{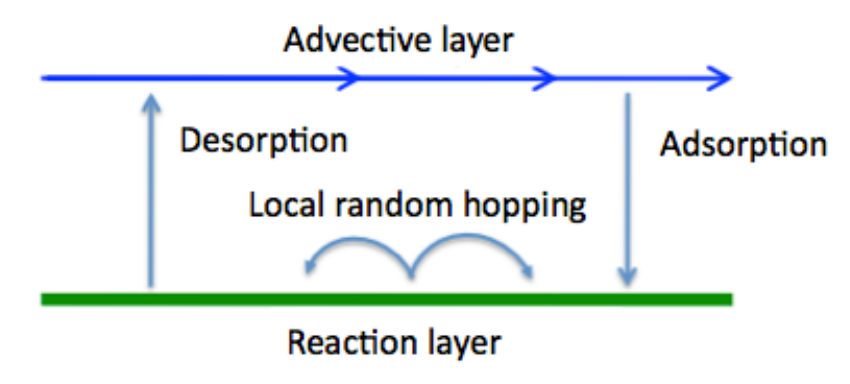}
\caption{(Color online) Reaction-diffusion process coupled to advection.}
\label{fig:schematic}
\end{figure}

Often, reaction-diffusion processes are coupled to an additional advective process.   In some models, an advective term is added to the FKPP equation \cite{DNelson review, Thip}, such that reactions, advection and diffusion occur simultaneously. 
In other instances, however, advection takes place in a separate competing transport channel, often associated with
a flow over a substrate on which reaction-diffusion processes take place.  Examples include heterogeneous catalysis on surfaces under flow \cite{Catalysis, Gervais},  population ecology in streams \cite{Stream_ecology, Pachepsky}, microbial population dynamics in the digestive tract \cite{Freter, Cremer}, and the long-range aerial spread of fungal plant pathogens - the original motivation for this work \cite{Aerobiology, Plot-scale, LR_transport}.
%
In all of these examples, there is a reaction-diffusion region, such as a catalytic substrate or a biological growth layer, and an advection region alongside it - with adsorption and desorption taking material on and off the growth layer - as depicted schematically in Fig.~\ref{fig:schematic}.  In this Letter we study this specific scenario whereby advection competes with separate reaction-diffusion processes.

As we demonstrate, presence of a competing transport mechanism causes the results of the FKPP model to be fragile -- giving a finite change in results due to an infinitesimal coupling. We thus identify a perturbation that causes a reaction-diffusion model to fail. In contrast, at finite coupling it is sometimes possible to map the coupled process to an effective FKPP equation with an advective term and suitably adjusted parameters.
%

\emph{Theoretical framework.}
Let $\rho(\mathbf{x},t)$ and $\sigma(\mathbf{x},t)$  denote the number 
density 
in the advective and reactive layers respectively.  The advective layer has an imposed velocity field $\mathbf{v}$.  Then, ignoring finite-number fluctuations, the spatio-temporal dynamics of $\rho(\mathbf{x},t)$ and $\sigma(\mathbf{x},t)$ will evolve from initial conditions $\rho(\mathbf{x},0)>0$ and $\sigma(\mathbf{x},0) >0$ according to
\begin{eqnarray}
\label{eq:modelCB1}
\frac{d\rho}{dt} &=& -\nabla \cdot \left(\mathbf{v} \rho \right) + \alpha \sigma - \beta \rho, \\
\label{eq:modelCB2}
\frac{d\sigma}{dt} &=& \delta f(\sigma) - \alpha \sigma + \beta \rho + D\nabla^2 \sigma.
\end{eqnarray}
In contrast to the FKPP model with an advective term, here any one particle either reacts and diffuses or advects within a small time interval $\Delta t$.   

Here $\alpha>0$ and $\beta>0$ are rates 
of mass transfer between the two layers, and $\delta$ is a characteristic reaction rate - all with dimensions $\left[time^{-1}\right]$.  In this paper we focus on one reactant, so Eqs.~(\ref{eq:modelCB1})-(\ref{eq:modelCB2}) are specific to growth-type reactions, and the reaction layer will be called the growth layer (GL).  The function $f(\sigma)$ is a dimensionless growth rate. $D$ is a diffusion constant on the GL.  

We focus 
on one dimension with a constant advective velocity $v_0$.  We assume $f(\sigma)$ is a concave and smooth function 
that admits one unstable state at $\sigma = 0$, where $df/d\sigma = 1$ and $f = 0$, and a stable state at $\sigma = \sigma_{max}$.  In some cases we will use a logistic model as a concrete example: $f(\sigma) = \sigma(1 - \sigma/\sigma_{max})$.  A natural length scale is $v_0/\delta$ - the distance traveled by the advective layer (AL) per characteristic growth time.   Rescaling $x$ by $v_0/\delta$, $t$ by $1/\delta$, and $\sigma$, $\rho$ by $\sigma_{max}$, we are left with three parameters: $a \equiv \alpha/\delta$, $b \equiv \beta/\delta$ and $\mathcal{D} = \left(\delta D/v_0^2\right)$;  we will continue to use the letters $\rho$, $\sigma$, $x$, $t$ and $f$.  All dimensionless speeds will be in units of $v_0$.
It will prove useful to first study the $\mathcal{D}=0$ case, and then consider the effect of diffusion.  

\emph{Zero diffusion.}
Consider a patch of GL around position $x_0$.  It produces new mass, and loses mass to the AL at rate $a$.  Once there, the mass is swept along at speed $1$ by the advection.  
All the while, mass is continuously shed onto the parts of the GL at $x>x_0$ with rate $b$.  The returned mass resumes growth at these new locations of the GL, while at the same time continuing re-desorbtion back onto the AL, and so on.  

Without the deposition of new mass from the AL, the dynamics on the GL 
unfolds independently at each $x$ following an initial condition (IC).  The advective layer effectively couples different locations of the growth layer.  The dynamics of the GL at each $x$ is driven by the AL, which itself is a result of accumulation of the upstream GL density.  The state $\sigma=0$, $\rho=0$ is linearly unstable to perturbations over a low wavevector range.  The nonlinearity limits the growth.  Thus, an IC that decays to $0$ as $x \rightarrow \pm  \infty$ leads to traveling fronts.  

\begin{figure}[ht]
\includegraphics[width=3.4in]{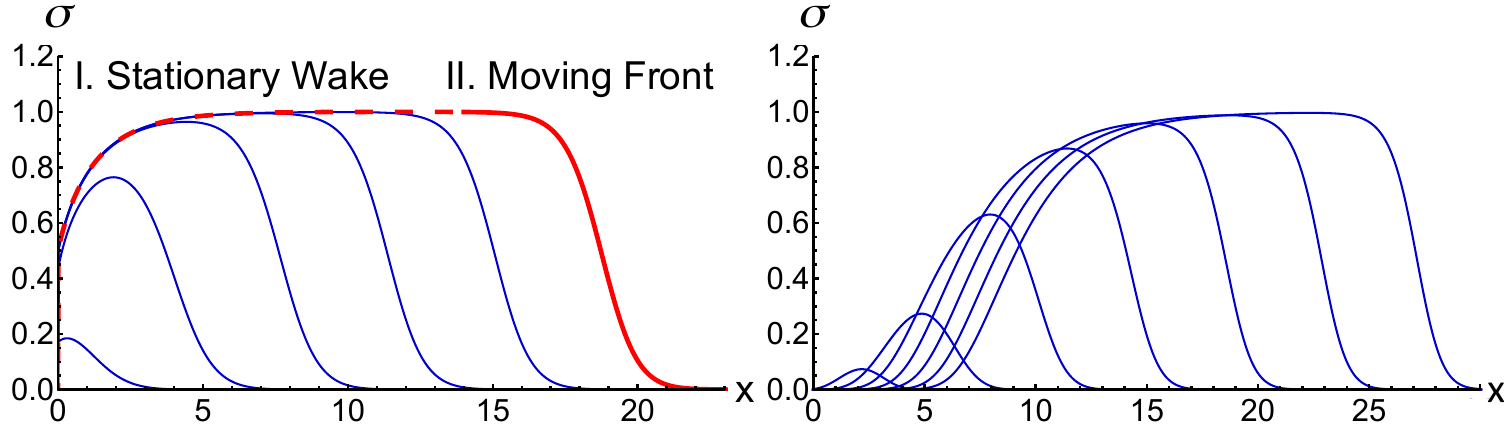} 
\caption{(Color online) Evolution of the GL profile from a $\delta$-function IC with a logistic growth model and no diffusion.  Left:  $a=0.5$, $b=1$ at $t=5$, $10$, $15$, $20$, $25$ and $30$.  Right, $a=2$, $b=1$, same $t$s.  In both cases, the IC launches uniformly translating fronts (UTF).  Early transients are not shown.}
\label{fig:example}
\end{figure}

Typical $\sigma(x)$ profiles are depicted in Fig.~\ref{fig:example};  $\rho(x)$ is qualitatively similar.  Here the advective velocity is directed rightward.  We will only consider ICs with a finite support; its left edge set at $x=0$.  Fig.~\ref{fig:example} depicts profiles at various times, evolving from a $\delta$-function IC at $x=0$, but qualitatively similar picture holds for all ICs with a finite support.  Depending on parameters, the profiles have one or two moving fronts.  

If the desorption rate is slower than the growth rate ($a<1$), there is an asymptotically stationary part of the profile left behind a single front propagating in the advective direction.  
It is depicted as a thick dashed part of the profile in the left panel of Fig.~\ref{fig:example} (section ``I''), while the moving front is depicted as the thick solid curve (section ``II'').   When $a \sigma > f(\sigma)$ for any $0<\sigma<1$ - regardless of the convexity of $f(\sigma)$ - there are two moving fronts - one leading (``downwind'') and one trailing (``upwind''), with a  plateau in between - Fig.~\ref{fig:example}, right panel. 

We now study the long-time behavior of moving fronts.  
We define the front speed $s$ as the speed of $x(t)$ that 
satisfies $\rho(x,t) = c_0$.  
In many systems, $s$ is determined by the growth of the leading edge of the profile where the linear approximation is valid \cite{von Saarloos review}.  Such fronts are called ``pulled''.   We proceed with this assumption - it will be validated by the comparison with the numerical solutions of Eqs.~(\ref{eq:modelCB1})-(\ref{eq:modelCB2}).  So, we let $f(\sigma) \rightarrow \sigma$ in Eqs.~(\ref{eq:modelCB1})-(\ref{eq:modelCB2}), solve the resulting equation, and compute $s$.  Due to the linearity of the equation, the resulting speed is independent of the value of $c_0$.  The speed defined by $\sigma(x,t) = c_0$ is identical, since $\rho$ and $\sigma$ are both governed by the same dispersion relation.

For the physically important IC $\sigma_0(x) = M \delta(x)$ and $\rho_0(x) = 0$ (profiles evolving from this IC are denoted by $*$), we can obtain a long-time asymptotically exact solution to linearized equations:
\begin{eqnarray}
\label{eq:LinearSol}
\rho^*(x,t) &=& \left\{ \begin{array}{l} aMe^{-\kappa \left(x-wt\right)}I_0\left(2\sqrt{ab}\sqrt{x(t-x)} \right), 0<x<t \\ 0, \mbox{   otherwise}\end{array}\right.  \nonumber \\
\kappa &=& 1-a+b, w = (1-a)/(1-a+b)
\end{eqnarray}
To obtain this, we found the dispersion relation of the linearized equations, computed the Fourier integral on a contour in the complex plane around a branch cut, and approximated the result by the modified Bessel function $I_0$ for $t > (ab)^{-1/2}$.  
Details, and $\theta^*(x,t)$ are in Supplementary Materials (SM) \cite{SM}, Section SII.  Using 
$I_0(z) \sim \frac{e^{z}}{\sqrt{2\pi z}}$ for large $z$, the front speeds are given by 
\begin{eqnarray}
\label{eq:spm}
s^*_{\pm} = \left(1+ \frac{b}{(1 \pm \sqrt{a})^2}\right)^{-1}
\end{eqnarray}
for $a \geq 1$.  
The $+/-$ represent the downwind/upwind profiles respectively.   For $a \leq 1$, $s^*_- = 0$ (upwind front does not move with zero diffusion and $a \leq 1$), but Eq.~(\ref{eq:spm}) applies for $s^*_+$.   Note: when $a=0$, Eq.~(\ref{eq:spm}) does not apply, since $\rho^* = 0$ and  $\sigma^* = 0$; when only $b=0$, only $\sigma^* = 0$ for $x>0$ (SM \cite{SM}, Section SII).  
Eq.~(\ref{eq:spm}) can also be obtained by the saddle-point method, which only requires that the Fourier Transform of the IC does not contain poles (SM \cite{SM}, Section SIII). Therefore, our result applies to any IC with a finite support.  

\begin{figure}[ht]
\includegraphics[width=3.3in]{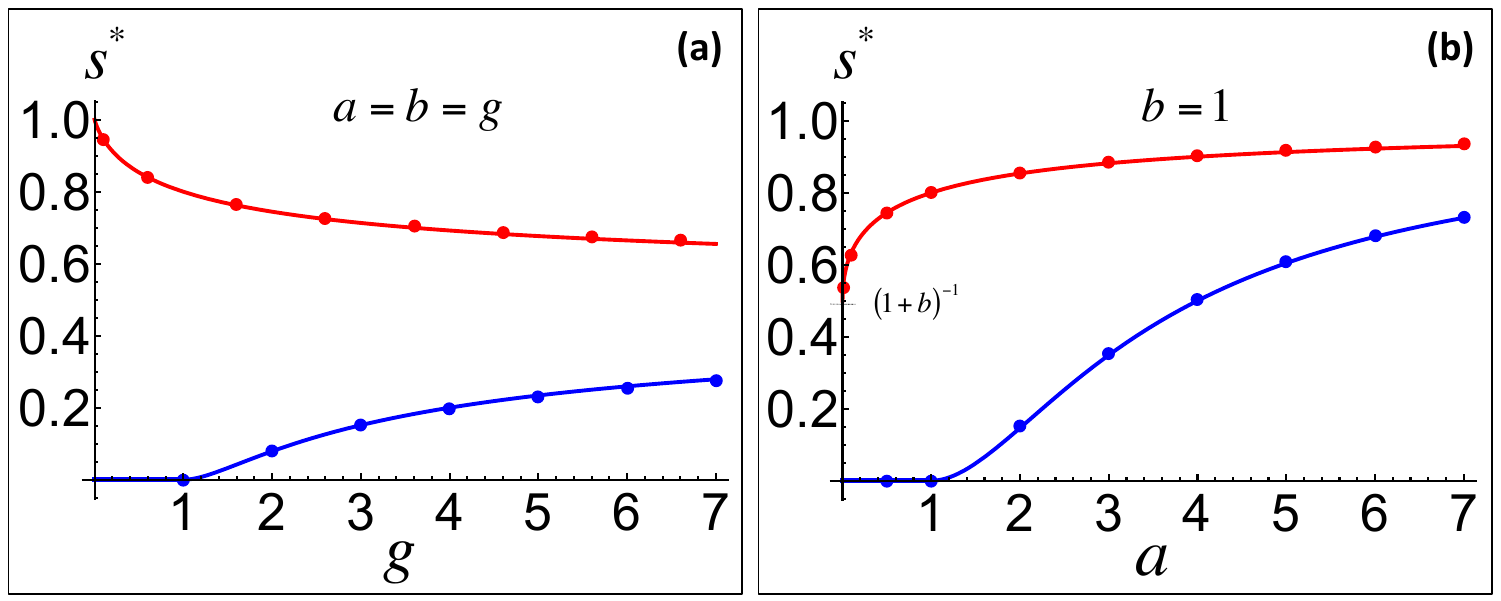}
\caption{(Color online) (a): front speeds vs.~the interlayer coupling $a=b=g$ for a $\delta$-function IC in the GL.  Upper curve (red) - downwind front speed $s^*_+$, lower curve (blue) - upwind front speed $s^*_-$.  
For $g <1$ there is a stationary profile behind the downwind front.    Solid dots are from the numerical solution of Eqs.~(\ref{eq:modelCB1})-(\ref{eq:modelCB2}).
(b): front speeds vs.~$a$ at $b=1$.  Now $s^*_+ \rightarrow (1+b)^{-1}$ as $a \rightarrow 0$.  The first two dots are at $a=0.01$ and $a=0.1$.} 
\label{fig:1}
\end{figure}

The finite speed of the downwind front as couplings approach (but $\neq$) zero, see Fig.~\ref{fig:1}, is the key prediction of the mean-field theory when $\mathcal{D}=0$.  The match with numerical calculations \cite{Footnote5prime} supports the validity of the pulled front assumption.  

The characteristic front width is $1/|\lambda^*_{\pm}|$, where
\begin{equation}
\label{eq:width}
\lambda^*_{\pm} = \frac{1+a+b \pm 2\sqrt{a}}{1 \pm \sqrt{a}},
\end{equation}
is the negative of the spatial growth rate of tails of the solution in Eq.~(\ref{eq:LinearSol}) at $c_0 \ll 1$.  Defined this way, $\lambda^*_+ >0$, while $\lambda^*_- < 0$ for $a>1$;  upwind front is not moving for $a<1$, and $\lambda^*_-$ becomes meaningless.  The value of $c_0$ may affect the time to attain $s^*_{\pm}$ and $\lambda^*_{\pm}$, but not their values.  

The prediction that $s^*_+ \rightarrow \mathrm{const} >0$ as $a\rightarrow 0$ at fixed $b$ is most surprising.  A parcel of mass that enters the AL - for however brief a period of time - will travel with speed $1$ downstream, and because this is a continuum theory, there will always be mass present in the AL.  So, the seeding process advances with speed $1$.  The speed of the front is defined at a constant density contour, so in general it is $<1$.  Note that $a \rightarrow 0$ causes $\rho\rightarrow 0$.

At sufficiently small $a$, the particle density in the AL becomes so small that the continuum theory breaks down.  Our predictions will not apply when the number of particles within a region of AL of the width of the feature size, i.e. front width, becomes $\sim O(1)$.  In physical units, the maximum number of particles in the AL within $\Delta x$ is $(a/b)\sigma_{\mathrm{max}}\Delta x$.  Therefore, mean-field theory predictions will hold as long as $\frac{a}{b} \gg \frac{\delta \lambda(a,b)}{\sigma_{\mathrm{max}} v_0}$, where $\sigma_{\mathrm{max}}$ is the carrying capacity on the GL per unit length.  
Otherwise, a stochastic treatment is needed.  This threshold can be extremely small due to large $\sigma_{max}$.  For example, in applications to fungal pathogen transport by wind, we estimate $\frac{\delta \lambda(a,b)}{\sigma_{\mathrm{max}} v_0} \sim O(10^{-14})$.  

When $a=b \equiv g \rightarrow \infty$, the time spent by a typical particle on the GL is $\ll$ growth time, so the speed is determined by the fraction of time spent in the AL.  
Thus, $s^* \rightarrow 1/2$ as $g \rightarrow \infty$.  When $a \neq b$, $s^*_{\pm} \rightarrow (1+b/a)^{-1}$ as $a \rightarrow \infty$.  
(see discussion of the zero growth case at the end of Section SII of SM \cite{SM}).

No time-invariance of a front shape was assumed, only that it is ``pulled''.  We see, however, that the decay rate in Eq.~(\ref{eq:width}) is indeed a constant.  If one seeks a uniformly-translating front (UTF) solution for $\rho$ and $\sigma$ that depend on $x-st$, there is a continuous family of solutions, each characterized by a decay rate $\lambda$ for a given $s$ (details in SM \cite{SM}, Section SI).  The solutions $(\lambda^*_{\pm}, s^*_{\pm})$ obtained above is one point in this family.  This suggests 
that the front evolves to a UTF form.  
The resulting UTF shapes match numerically-obtained profile shapes.  

\emph{Diffusion - a competing transport mechanism.}
Although we could not solve the linearized equations when $\mathcal{D} \neq 0$, progress can be made with a UTF ansatz.  Letting $\sigma(x,t) =\tilde{\sigma}(x-st)$ and $\rho(x,t) = \frac{a}{b} \tilde{\rho}(x-st)$ valid in the vicinity of the front, linearizing the resulting equation around $(\tilde{\rho}=0,\tilde{\sigma} = 0)$, and substituting an eigen-solution $\tilde{\rho} = A e^{-\lambda z}$, $\tilde{\sigma} = B e^{-\lambda z}$ ($\lambda>0$ describes the downwind front, and $\lambda<0$, describes the upwind front), we obtain the following equation relating the decay rate of the leading edge with speed $s$:
\begin{equation}
\label{eq:slambdanonzeroD}
s\lambda = (1-a) + \frac{ab}{b + (s-1)\lambda} + \mathcal{D}\lambda^2
\end{equation}
The front ``vicinity'' can be defined by $|x-x_{\mathrm{front}}| \lesssim 1/\lambda$, and $x_{\mathrm{front}}$ is a characteristic point on the front, such as the inflection point.  

\begin{figure}[ht]
\includegraphics[width=2.9in]{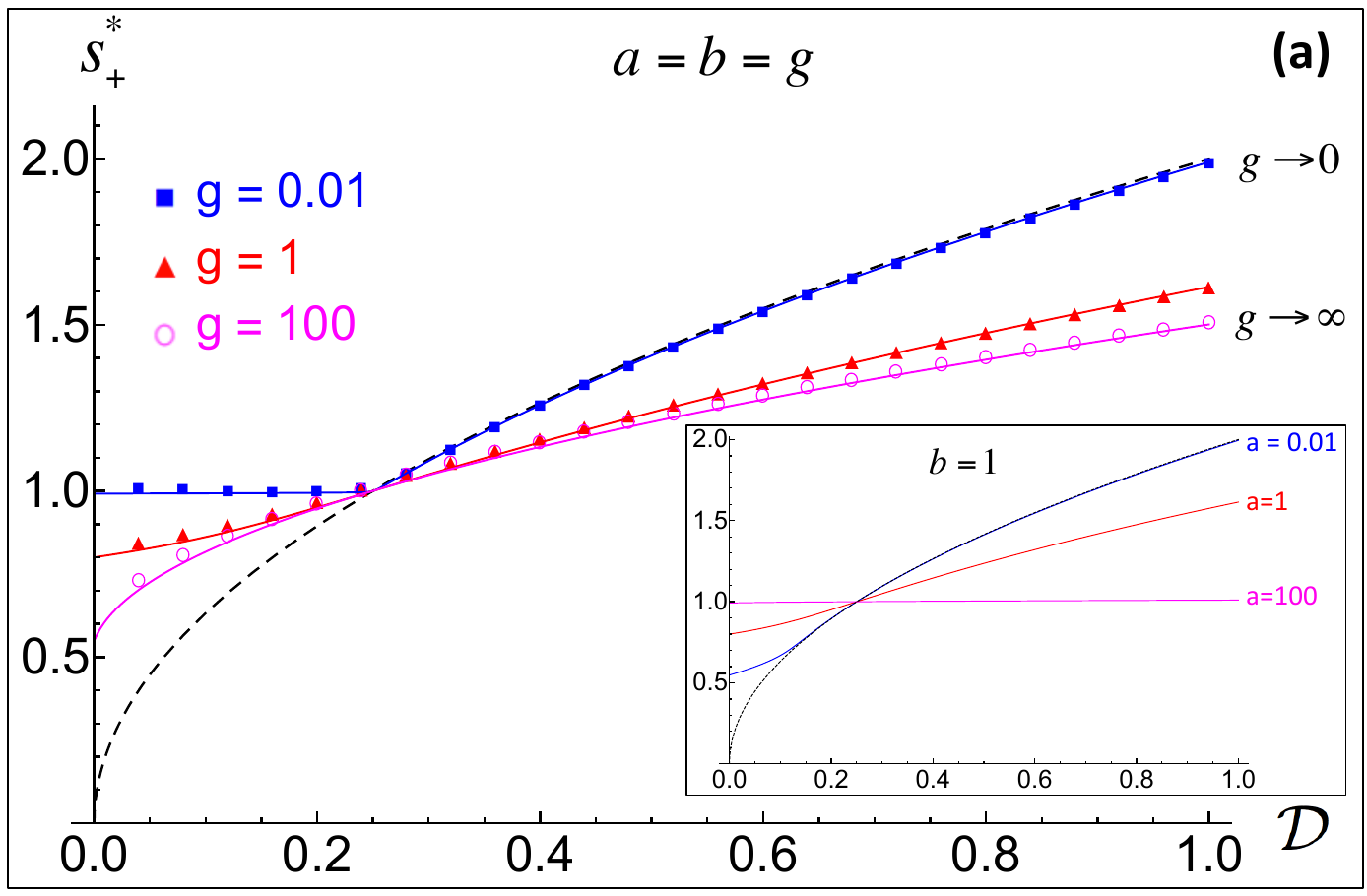}
\includegraphics[width=2.9in]{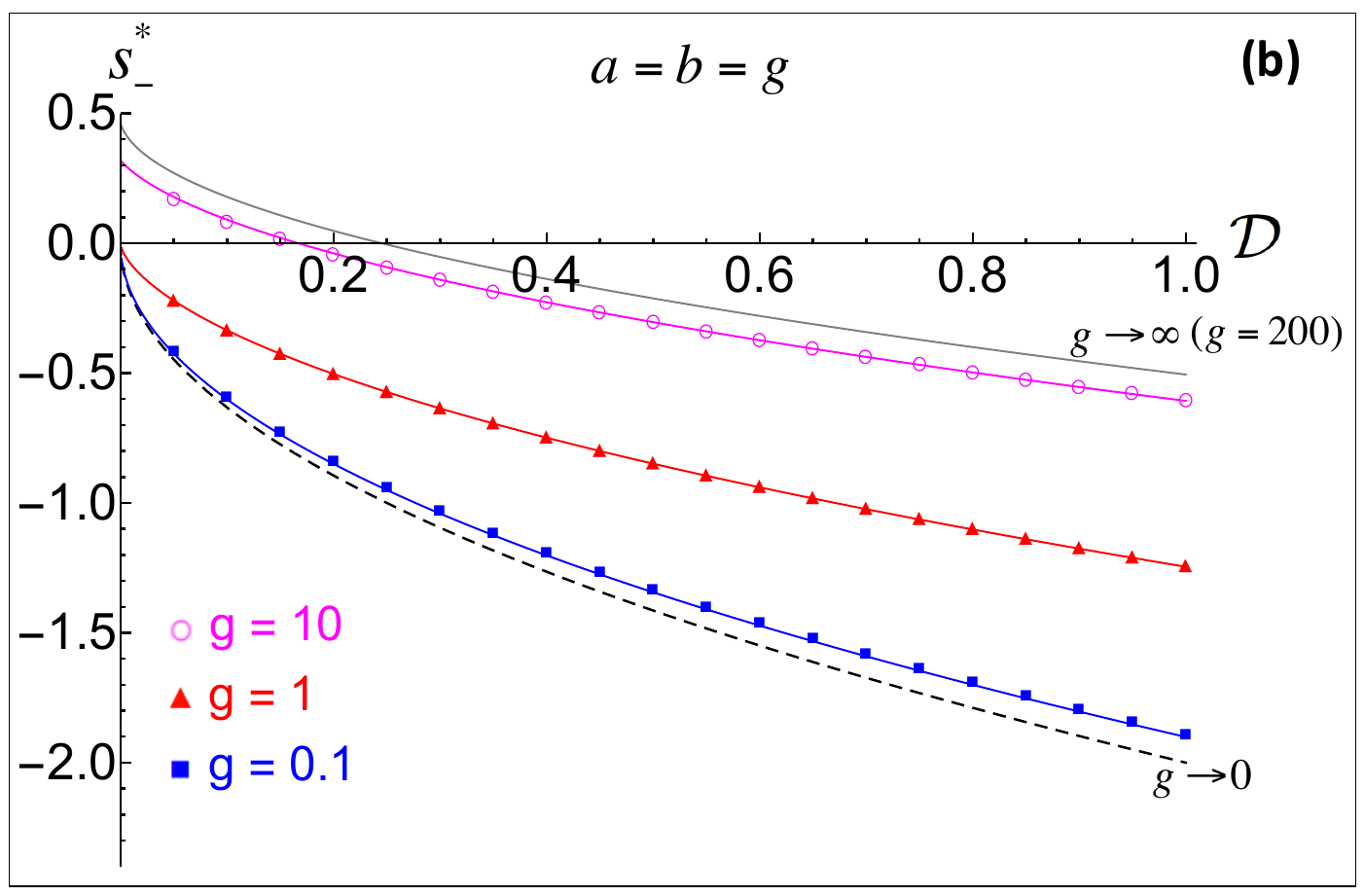}
\caption{(Color online) The dimensionless speed of the downwind (a), and upwind (b) fronts vs.~$\mathcal{D}$ for several coupling values.  The symbols were obtained by numerical solutions of Eqs.~(\ref{eq:modelCB1})-(\ref{eq:modelCB2}), while continuous curves are theory - see main text.  Dashed curve is $\pm 2\sqrt{\mathcal{D}}$ - FKPP speed in units of $v_0$.}
\label{fig:3}
\end{figure}

The resulting $s(\lambda)$ has multiple branches.  
The theory of pulled fronts \cite{von Saarloos review} predicts that for ICs that decay faster than $e^{-\lambda^*_+x}$, with $\lambda^*_+>0$ the minimum point of largest branch of $s(\lambda)$ (``steeply decaying'' ICs), the selected decay rate of the downwind front will evolve to be $\lambda^*_+$, and its speed will be $s^*_+ = s(\lambda^*_+)$ \cite{Footnote6} (this was indeed so in the $\mathcal{D} = 0$ case).  
The resulting $s^*_+ (a,b,\mathcal{D})$ is displayed in Fig.~\ref{fig:3}a.  Equivalently, the maximum of the lowest branch for $\lambda <0$ describes the selected state $(\lambda^*_-, s^*_-)$ of the upwind front resulting from steeply growing ICs, Fig.~\ref{fig:3}b.  We now study the properties of each front (details in SM \cite{SM}, Section SI) resulting from steep ICs, including a $\delta$-function IC.  We maintain $a, b \neq 0$.

Adding diffusion in the GL does not change $s^*_+$ discontinuously.  This may be viewed as a consequence of the finite speed for any non-zero $a$ and $b$ when $\mathcal{D}=0$.  The front speed on the GL decoupled from the AL, i.e.~of the FKPP model, is the well-known Fisher speed \cite{DNelson review}, given in the physical units by $s_{F} = 2\sqrt{D\delta} = 2v_0\sqrt{\mathcal{D}}$, which  $\rightarrow 0$ as $\mathcal{D} \rightarrow 0$, while $s^*_+$ from Eq.~(\ref{eq:spm}) stays finite \cite{Footnote7}.  
Moreover, when the Fisher speed (of the GL) $\ll$ the advective speed (i.e. when $\mathcal{D} \ll 1/4$), $s^*_+$ does not scale like $\sqrt{\mathcal{D}}$; it is given approximately by Eq.~(\ref{eq:spm}) plus a correction $b\mathcal{D}/(1+\sqrt{a})$.  So the advective mechanism dominates at small $\mathcal{D}$ and finite $b$.  

On the other hand, at large $\mathcal{D}$ 
there is an asymptotic 
\begin{equation}
\label{eq:downwind_asymptotic} s^*_+ \sim  v_{\mathrm{\mathrm{eff}}}  + 2\sqrt{\mathcal{D}_{\mathrm{\mathrm{eff}}}},
\end{equation}
which we recognize to be $s^*_+$ of the FKPP model $\dot{\sigma} = -v_{\mathrm{eff}} \sigma' + \mathcal{D}_{\mathrm{eff}} \sigma'' + f(\sigma)$.
To get $v_{\mathrm{eff}}$ and $\mathcal{D}_{\mathrm{eff}}$, we notice that FKPP $\lambda^*_{\pm} = \pm \frac{1}{\sqrt{\mathcal{D}}}$, so we let $\lambda^*_{\pm} \sim \frac{c(a,b)}{\sqrt{\mathcal{D}}}$ and solve for $c(a,b)$ that makes this ansatz correct as $\mathcal{D} \rightarrow \infty$.  

The resulting, $\mathcal{D}_{\mathrm{eff}} \rightarrow \mathcal{D}$ and $v_{\mathrm{eff}} \rightarrow 0$ 
when $a \rightarrow 0$ at $b=\mathrm{const}$ or $b \rightarrow \infty$ at $a=\mathrm{const}$, i.e. particles are forced to stay on the GL.  On the other hand, $\mathcal{D}_{\mathrm{eff}} \rightarrow 0$ and  $v_{\mathrm{eff}} \rightarrow 1$ as $a \rightarrow \infty$ at $b=\mathrm{const}$ or $b \rightarrow 0$ at $a=\mathrm{const}>1$, i.e. particles are forced to stay on the AL.  
At $a$ and $b \ll 1$, the crossover into this asymptotic regime is sharp, and takes place at $\mathcal{D}=1/4$.  For the crossover to penetrate to a given $\mathcal{D} \ll {1/4}$, $b$ has to scale as $\mathcal{D}^{-1/2}$; this also applies to the upwind front (SM, end of Section I).   
For $a/b=\mathrm{const}$, $s^*_{\pm} \rightarrow \frac{a/b \pm 2\sqrt{\mathcal{D}}}{1+a/b}$ $\forall \mathcal{D}$ as $b \rightarrow \infty$. 

At $a=1$, $c \sim b^{1/4}$ for small $b$.  So, $\lim_{b \rightarrow 0} c = 0$, $\lambda^*_+ \sim 1/\mathcal{D}$, and  Eq.~(\ref{eq:downwind_asymptotic}) becomes invalid; the crossover $\mathcal{D}$ into the $s^*_+ \sim \mathcal{D}^{1/2}$ behavior diverges as $b^{-1/2}$.  Sufficiently far from $(a=1,b=0)$, $c \approx 1$, which gives
\begin{eqnarray}
\label{eq:Deff} \mathcal{D}_{\mathrm{eff}} &\approx& \frac{\left[2-a-b+\sqrt{(a+b)^2 + 4(1+b-a)}\right]^2}{16}\mathcal{D}, \\
\label{eq:veff} v_{\mathrm{eff}} &\approx& \frac{1}{2} + \frac{a-b-2}{2\sqrt{(a+b)^2 + 4(1+b-a)}}.
\end{eqnarray}

A prominent feature of Fig.~\ref{fig:3}(a) is the intersection at $\mathcal{D} = 1/4$, where $s^*_+$ is coupling-independent.  This can happen if $a\rho = b\sigma$ (see Eqs.~(\ref{eq:modelCB1})-(\ref{eq:modelCB2})), which 
is only possible at $\mathcal{D} = 1/4$, when Fisher speed $=$ advective speed.  To see that $a\rho$ does equal to $b\sigma$ at $\mathcal{D}=1/4$, we can show that for ICs that evolve to a UTF, $s^*_+ =1$ for any $a$ or $b$ only when $\mathcal{D}=1/4$;  $a\rho = b\sigma$ follows.  

For the upwind front, the two transport mechanisms are opposing.  Advection alone can not propagate the front for $a<1$, so diffusion is essential for front movement.   For $a$ slightly $<1$ and $\mathcal{D} \ll \frac{(1-a)^3}{27b^2}$, $s^*_- \sim -2\sqrt{1-a}\sqrt{\mathcal{D}}$.  At $a=1$, $s^*_- \sim -\frac{3b^{1/3}\mathcal{D}^{2/3}}{2^{2/3}}$ before crossing over into $\sim -\sqrt{\mathcal{D}}$ behavior.  This crossover $\mathcal{D}$ diverges as $b^{-1/2}$ when $b \rightarrow 0$. 
For $a>1$, there is a critical value $\mathcal{D}_{\mathrm{\mathrm{stall}}}$, such that this front propagates in the advective direction for $\mathcal{D} < \mathcal{D}_{\mathrm{\mathrm{stall}}}$;  $\mathcal{D}_{\mathrm{\mathrm{stall}}}$ increases with $a$ and decreases with $b$.  For small $a-1>0$, $\mathcal{D}_{\mathrm{\mathrm{stall}}} = (a-1)^3/(8b^2)$.   Adding diffusion in the GL also does not change $s^*_-$ discontinuously - for $a>1$, it is given by Eq.~(\ref{eq:spm}) plus a correction 
$b\mathcal{D}/(1-\sqrt{a})$. 
At large $\mathcal{D}$, $\lambda^*_- \sim -\frac{c(a,b)}{\sqrt{\mathcal{D}}}$ and
\begin{equation}
\label{eq:upwind_asymptotic} s^*_- \sim v_{\mathrm{eff}}  - 2\sqrt{\mathcal{D}_{\mathrm{eff}}},
\end{equation}
except at $(a=1, b\rightarrow 0)$, with identical $c(a,b)$, $\mathcal{D}_{\mathrm{eff}}$ and $v_{\mathrm{eff}}$ as for the downwind front.  
This is $s^*_-$ of the model $\dot{\sigma} = -v_{\mathrm{eff}} \sigma' + \mathcal{D}_{\mathrm{eff}} \sigma'' + f(\sigma)$.  Using Eqs.~(\ref{eq:Deff})-(\ref{eq:upwind_asymptotic}) we see that at large $a$ and $b$, $\mathcal{D}_{\mathrm{\mathrm{stall}}} \sim \left(\frac{a}{2b}\right)^2$, so when $a/b=\mathrm{const}$, $\mathcal{D}_{\mathrm{stall}}$ has a limiting value.  
The upwind front width diverges at $a=1$ as $\sim b^{-1/3}$ when $b \rightarrow 0$ at finite $\mathcal{D}$.  

The competition between advection and diffusion results in rich phenomenology.  The mapping to an effective FKPP model with advection is generally possible only at large desorption rate or fast diffusion.  Yet, the AL renormalizes even a very large $\mathcal{D}$, so the AL can never be ignored!  For the downwind front with $\mathcal{D} < 1/4$, 
infinitesimal coupling of the FKPP model to an advection equation causes a finite change in front properties (see Fig.~\ref{fig:3}a), indicating fragility of the mean-field FKPP model.  

The existence of a critical point and fragility found here suggest that similar effects may take place in other phenomena involving a combination of advection, diffusion, and reactions taking place in multiple layers \cite{Catalysis}-\cite{LR_transport}, \cite{Fedotov}.  It remains to explore the interplay of these critical points and fluctuations when the departure from the mean-field model is considered. 



\begin{acknowledgments}
We thank David Schneider, whose insights and feedback have greatly influenced and improved this work; M. C. Cross for valuable feedback, S. H. Strogatz, David R. Nelson, Thiparat Chotibut, N. Mahowald, W. van Saarloos, Stephen Ellner, and Alan Hastings for useful discussions. This work was supported in part by the Science \& Technology Directorate, Department of Homeland Security, interagency Agreement No.~HSHQDC-10-X-00138, and a seed grant from the Atkinson Center for a Sustainable Future at Cornell University.
\end{acknowledgments}

{}

\newpage
\onecolumngrid
\vspace{6in}

\begin{center}
\Large{Supplementary Materials for ``Fragility of reaction-diffusion models to competing advective processes''}
\end{center}
\normalsize

\setcounter{equation}{0}
\setcounter{figure}{0}
\setcounter{table}{0}
\setcounter{page}{1}
\makeatletter
\renewcommand{\theequation}{S\arabic{equation}}
\renewcommand{\thefigure}{S\arabic{figure}}
\renewcommand{\thetable}{S\Roman{table}}
\renewcommand{\bibnumfmt}[1]{[S#1]}
\renewcommand{\citenumfont}[1]{S#1}
\onecolumngrid

\section{SI.  Uniformly Translating Fronts (UTF)}
\label{sec:UTFsec}
Substituting the UTF ansatz $\sigma(x,t) =\tilde{\sigma}(x-st)$ and $\rho(x,t) = \frac{a}{b} \tilde{\rho}(x-st)$ into Eqs.~(1)-(2) of the main text, yields
\begin{eqnarray}
\label{eq: TW_system1} (1-s)\frac{d\tilde{\rho}}{dz} &=& -b \tilde{\rho} + b \tilde{\sigma}, \\
\label{eq: TW_system2} -s \frac{d \tilde{\sigma}}{dz} &=& f(\tilde{\sigma}) - a \tilde{\sigma} + a \tilde{\rho} + \mathcal{D} \frac{d^2 \tilde{\sigma}}{d z^2},
\end{eqnarray} 
where $z=x-st$.  This is a three-dimensional dynamical system in coordinates $\tilde{\sigma}$, $\tilde{\rho}$, and $\tilde{u} = \frac{d \tilde{\sigma}}{dz} $, and has fixed points at $(\tilde{\sigma} = 0, \tilde{\rho} = 0, \tilde{u} = 0)$ and $(\tilde{\sigma}  = 1, \tilde{\rho} =1, \tilde{u} = 0)$, which are respectively stable and unstable (note: $t$ decreases with increasing $z$).  Only the heteroclinic solution connecting the two goes from $z = -\infty$ to $z=+\infty$, so $\rho(x)$ or $\sigma(x)$ will have a sigmoidal shape.   Thus, if an IC evolves to a UTF, it will be a front-like solution.  The existence of stationary solutions, such as section I in Fig.~2 of the main text, implies that a UTF cannot exist for all $x$ and $t$, so a UTF describes the vicinity of a moving front, defined more precisely below.

Solutions of nonlinear Eqs.~(\ref{eq: TW_system1})-(\ref{eq: TW_system2}) are parametrized by $s$, which determines the phase portrait in the $(\tilde{\rho}, \tilde{\sigma})$ space.  It is customary to characterize solutions by the eigenvalues around the state $(0,0)$, which describes the tail of a UTF.  Instead of expressing the eigenvalues as functions of $s$, we follow the standard convention \cite{von Saarloos review_SM} and express $s$ as a function of -eigenvalue $\equiv \lambda$.   It is easiest to do this by linearizing Eqs.~(\ref{eq: TW_system1})-(\ref{eq: TW_system2}) around $(\tilde{\sigma} = 0,\tilde{\rho} = 0)$, and substituting an eigen-solution $\tilde{\rho} = A e^{-\lambda z}$, $\tilde{\sigma} = B e^{-\lambda z}$.  
The result is 
\begin{equation}
\label{eq:s_of_lambda_equation}
s\lambda = (1-a) + \frac{ab}{b + (s-1)\lambda} + \mathcal{D}\lambda^2
\end{equation}

\noindent
There are two solutions, which we label $s_{1}(\lambda)$ and $s_{2}(\lambda)$.  We devote the rest of this section to the study of equation  Eq.~(\ref{eq:s_of_lambda_equation}).

\subsection{Zero diffusion.}
\label{sec:UTFsecD=0}
\begin{figure}[ht]
\includegraphics[width=3.5in]{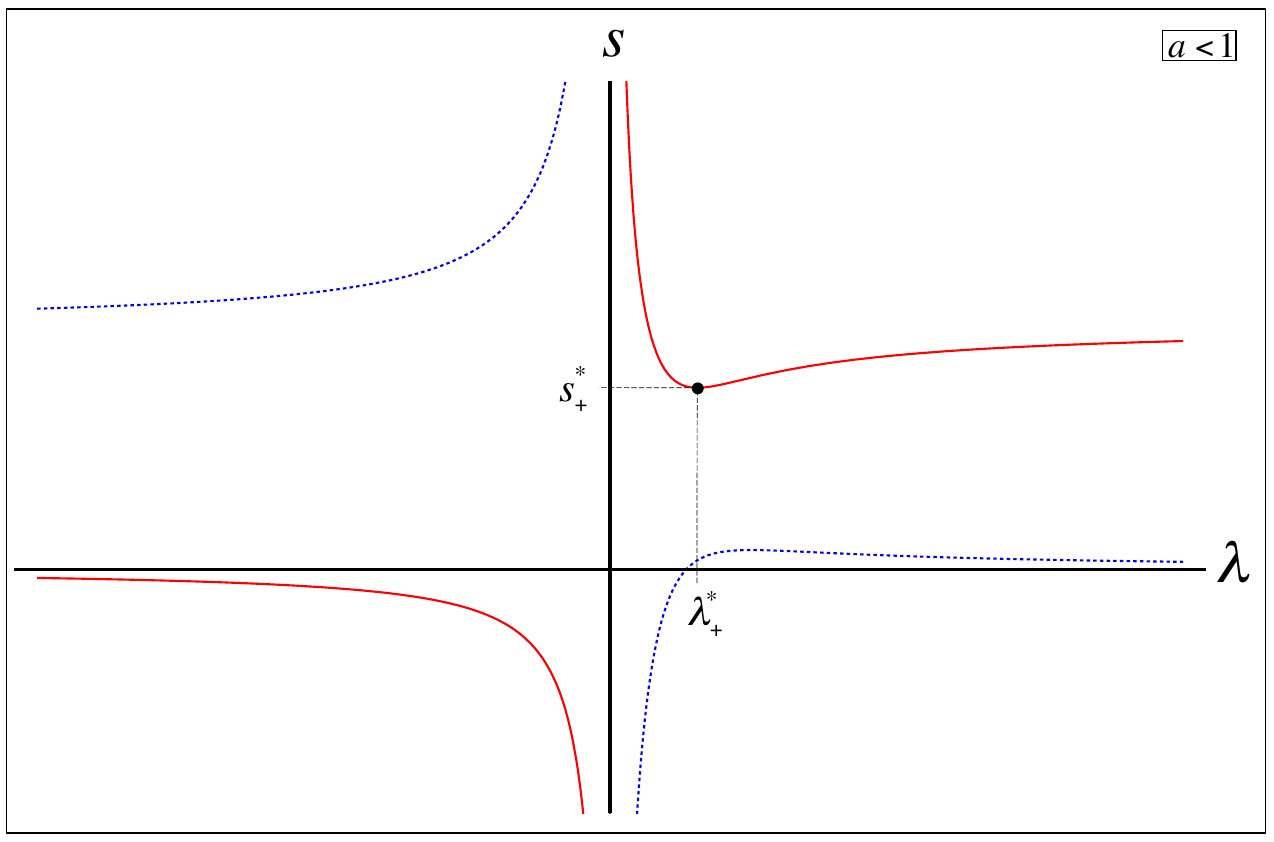}
\includegraphics[width=3.5in]{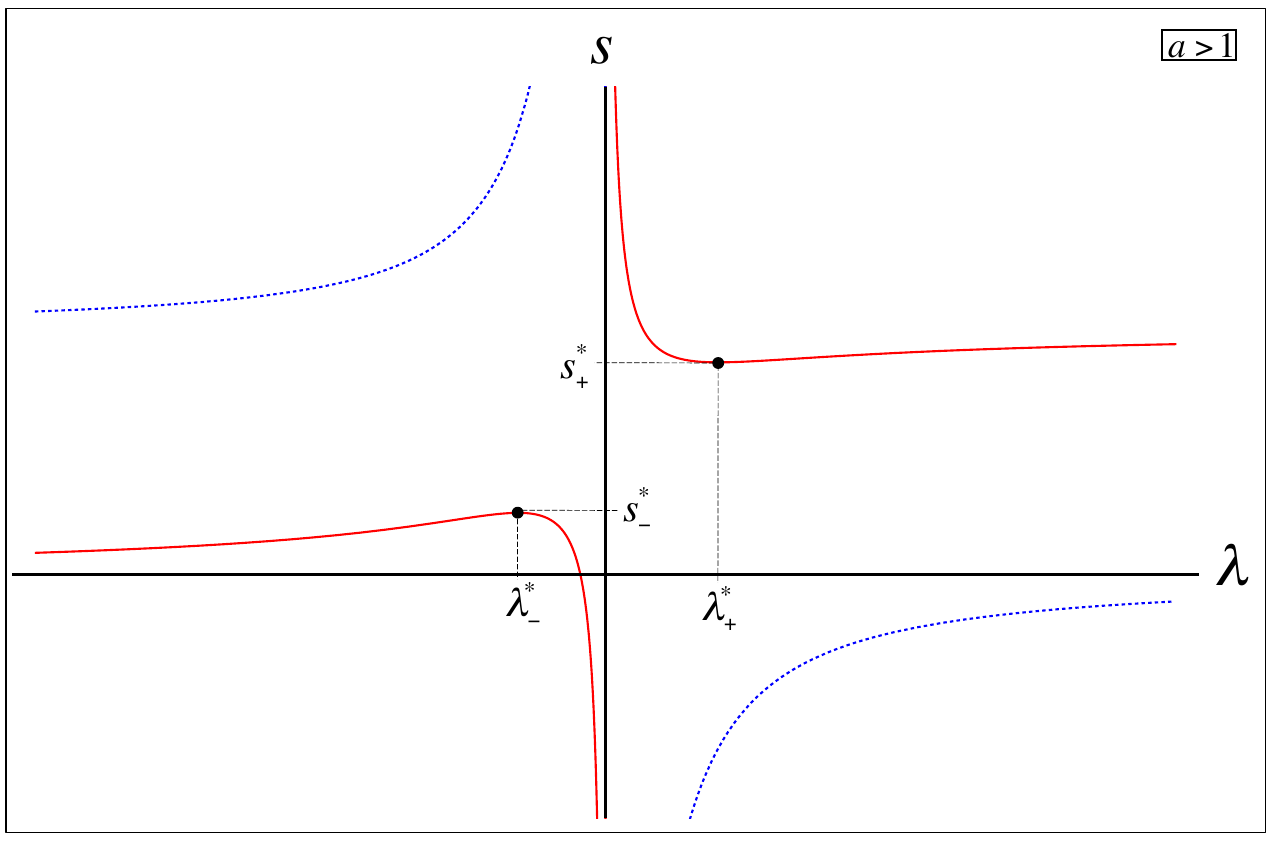}
\caption{(Color online) Typical structure of the branches $s_1(\lambda)$ - red solid curve, and $s_2(\lambda)$ - blue dotted curve.  The points $(\lambda^*_+, s^*_+)$ and $(\lambda^*_-, s^*_-)$ are also noted.  The qualitative picture remains for any $b>0$.  The value of $\lambda^*_-$ goes to $-\infty$ as $a \rightarrow 1$ from above, and stays at $-\infty$ for $a<1$, corresponding to a lack of propagation of the upwind front for $a<1$.  There is a horizontal asymptote always taking place at $s = 1$.}
\label{fig:1SM}
\end{figure}

When $\mathcal{D}=0$, $s_{1}(\lambda)$ and $s_{2}(\lambda)$ are given explicitly by 
\begin{equation}
\label{eq:s_of_lambda_zero_D}
s_{1,2}(\lambda) =  \frac{1-a-b+\lambda \pm \sqrt{(\lambda-1 +a-b)^2 + 4ab}}{2 \lambda},
\end{equation}
with $s_1$ is the $+$ solution and $s_2$ is the $-$ solution.  These relations give the speed as a function of the decay rate of the solution.  This is plotted in Fig.~\ref{fig:1SM}.  Note that positive $\lambda$ describe a downwind front - it decays with increasing $x$, while the negative $\lambda$ describe the upwind front, which grows with increasing $x$.

The question of the selected $(\lambda,s)$ from a given IC is a problem in front selection.  In this work we focused on ICs with a finite support, including a $\delta$-function.  The extremal points of $s(\lambda)$ take place precisely at the locations  predicted by Eqs.~(4)-(5) in the main text, $(\lambda^*_{\pm},s^*_{\pm})$, derived in Section SII without requiring a UTF assumption; that prediction remains true for any IC with a finite support (see Section SIII below). 
The reason that the extrema of $s(\lambda)$ are located at these $(\lambda^*_{\pm},s^*_{\pm})$ is not coincidental, but is consistent with the general theory of fronts \cite{von Saarloos review_SM} in the long-time asymptotic regime.  

The review on front propagation \cite{von Saarloos review_SM} states that sufficiently steep ICs will select the leading front (i.e. decaying with increasing $x$) with a characteristic decay rate being the $\lambda>0$ at which the minimum point of the top-most branch of $s(\lambda)$ occurs, and the speed is given by $s$ at that $\lambda$.  The exception - corresponding to ``pushed'' fronts - occurs when there exists a nonlinear solution that at low density matches exactly the eigen-solution with the non-minimum eigenvalue, but this is a rather special case.  

The discussion in \cite{von Saarloos review_SM} was based on the leading front (i.e. decaying with increasing $x$).  However, the trailing front propagating with a certain speed $s$ becomes a leading front propagating with the speed $-s$, i.e. $\lambda \rightarrow - \lambda$ and $s \rightarrow -s$, upon the spatial mirror-reflection.  Therefore, steep ICs will select the trailing front with a characteristic growth rate being the $\lambda<0$ at which the maximum point of the bottom-most branch of $s(\lambda)$ occurs.  As already mentioned, our study of the upwind front for $\mathcal{D}=0$ confirmed this.  For $\mathcal{D} \neq 0$ this has also been verified against numerical simulations of the model, and supported by saddle-point calculations in Section III.  

The notation $(s^*, \lambda^*)$ will now be used in two ways - denoting the position of the extrema of $s(\lambda)$, as well the properties of the selected state.

Having defined the characteristic width of the front by the eigenvalues $\lambda$, we can say what the ``vicinity'' of the front is:  it is a region of $|x-x_{front}| \lesssim 1/|\lambda|$, where $x_{front}$ can be defined, for example, as the inflection point of $\rho(x,t)$ or $\sigma(x,t)$, although the precise definition is unimportant. 

We can also heuristically argue that a front converges to a UTF.  The speed and decay rate of the leading edge of a solution to Eqs.~(1)-(2) of the main text, is selected by the IC.  However, if the initial evolution leads to a UTF, the front width of the full, nonlinear profile can be estimated from the eigenvalue $\lambda(s)$ of solutions to Eqs.~(\ref{eq: TW_system1})-(\ref{eq: TW_system2}) around the attractor at $(0,0)$.  Although these are obtained from the linearization of Eqs.~(\ref{eq: TW_system1})-(\ref{eq: TW_system2}), they are properties of the solutions of the full, nonlinear profile.  On the other hand, we have obtained the speeds $s^*_{\pm}$ and the widths $\lambda^*_{\pm}$ (Eqs.~(5)-(6) of the main text) for a specific IC without assuming a UTF.  As already mentioned, these $(\lambda^*_{\pm},s^*_{\pm},)$ lie on the $\lambda(s)$ curve produced by the UTF assumption.  Although not a rigorous proof, it is an argument for the solution to approach a UTF.


\subsection{Non-zero diffusion.}
\label{sec:UTFsecD!=0}
\begin{figure}[ht]
\includegraphics[width=3.5in]{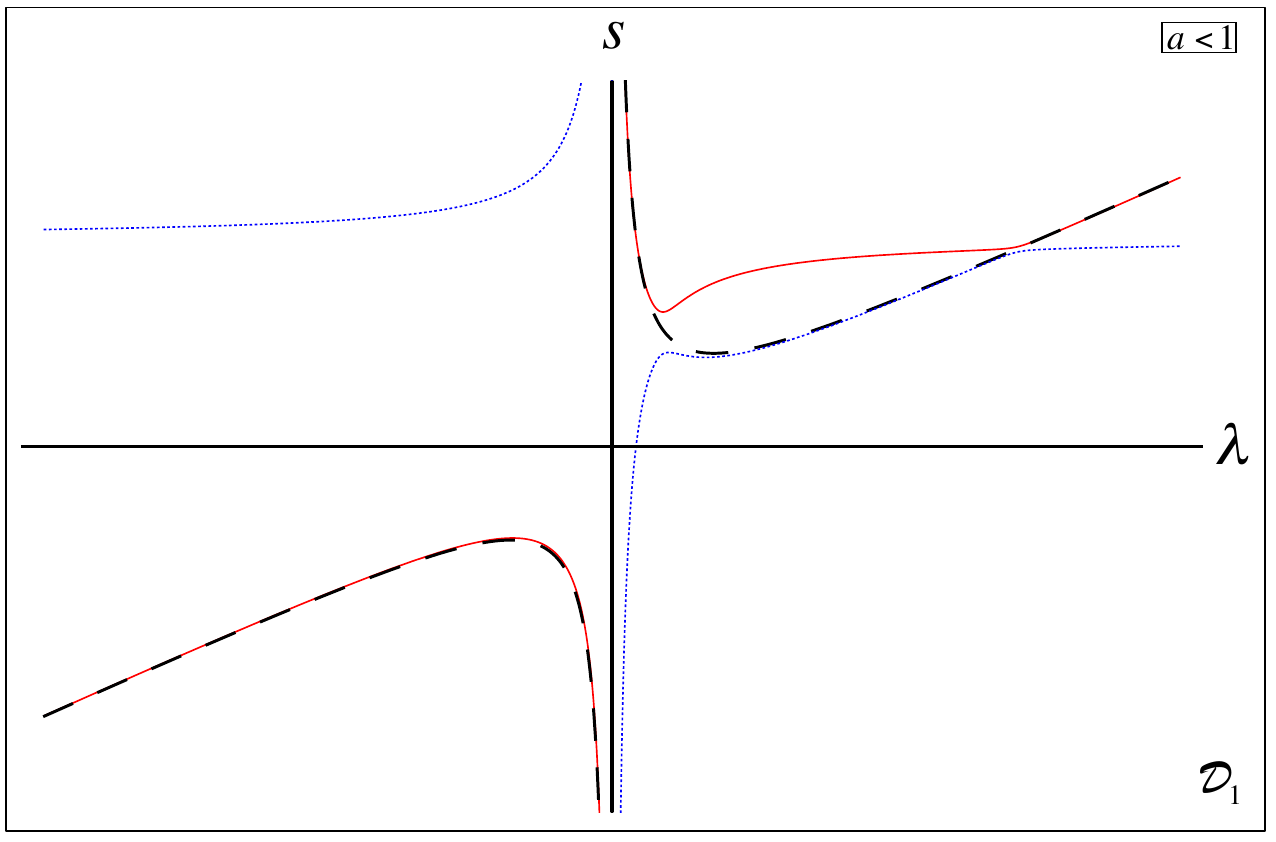}
\includegraphics[width=3.5in]{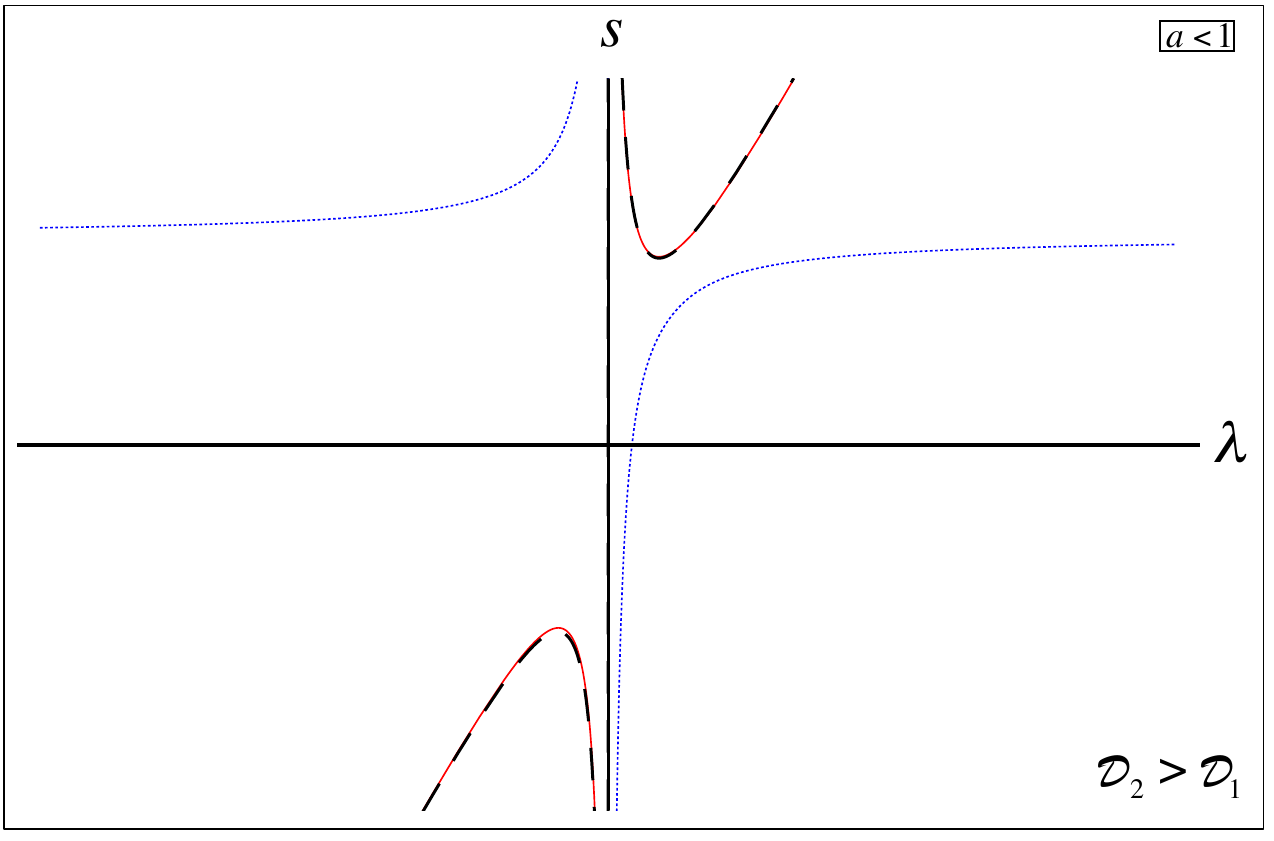}
\caption{(Color online) Typical structure of the branches $s_1(\lambda)$ - red solid curve, and $s_2(\lambda)$ - blue dotted curve with a non-zero $\mathcal{D}$ when $a<1$. }
\label{fig:2SM}
\end{figure}

\begin{figure}[ht]
\includegraphics[width=3.5in]{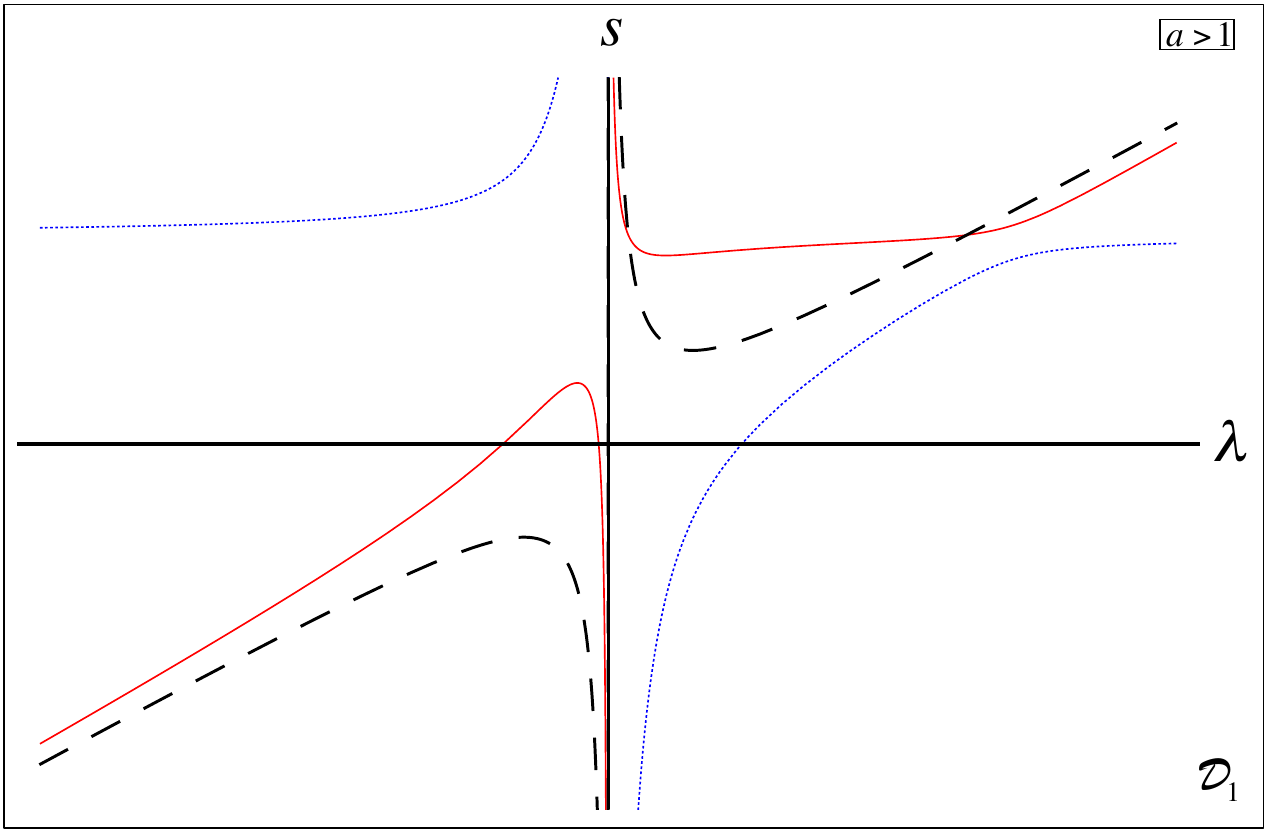}
\includegraphics[width=3.5in]{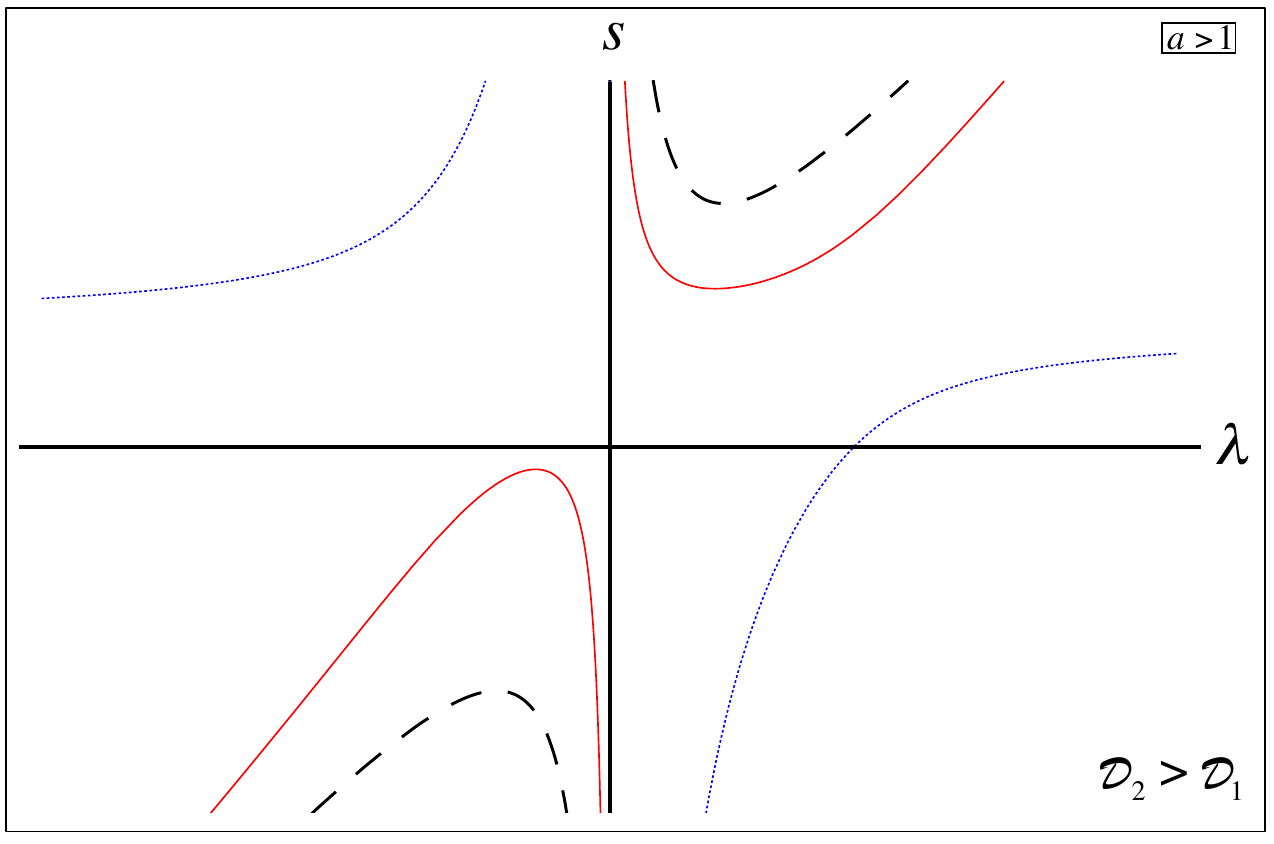}
\caption{(Color online) Typical structure of the branches $s_1(\lambda)$ - red solid curve, and $s_2(\lambda)$ - blue dotted curve with a non-zero $\mathcal{D}$ when $a>1$.}
\label{fig:3SM}
\end{figure}

Below we plot $s(\lambda)$ curves when $\mathcal{D} \neq 0$; Fig.~\ref{fig:2SM} is for $a<1$ and Fig.~\ref{fig:3SM} is for $a>1$.  These figures do not all have the same scale - they are meant to demonstrate qualitative changes in the structure of $s(\lambda)$ as parameters $a$ and $\mathcal{D}$ vary (the parameter $b$ was set to $1$ in these figures).  These plots are meant to demonstrate the evolution of the branches of $s(\lambda)$ from the $\mathcal{D}=0$ case, seen in Fig.~\ref{fig:1SM}.  The black dashed curve represents $\mathcal{D} \lambda + \lambda^{-1} = s_{FK}(\lambda)$ of the single-variable FKPP model $\dot{\phi} = f(\phi) + \mathcal{D} \frac{d^2 \phi}{dx^2}$.
The selected downwind speed that appears in Fig.~4a of the main text is taken from the minimum of the top solid (red) curve for $\lambda>0$, whereas the  selected upwind speed that appears in Fig.~4b of the main text is taken form the maximum of the bottom solid (also red) curve for $\lambda <0$. 

We next study the extrema of $s_1(\lambda)$ analytically in the regime of large and small $\mathcal{D}$. We will continue to use the extremal points of $s(\lambda)$ to predict the selected states $(\lambda^*_{\pm},s^*_{\pm})$ from a compact IC.  When $\mathcal{D}\neq0$, $s_{1}(\lambda)$ and $s_{2}(\lambda)$ are given explicitly by 
\begin{eqnarray}
\label{eq:s_of_lambda_nonzero_D}
& s_{1,2}(\lambda) = \left(1-a-b+\lambda + \mathcal{D}\lambda^2\pm \sqrt{\psi}\right)/\left(2 \lambda \right) \\
& \mbox{where } \psi = a^2+2 a \left(b-\mathcal{D} \lambda ^2+\lambda -1\right)+(1+b+\lambda(\mathcal{D} \lambda -1))^2 \nonumber
\end{eqnarray}
with $s_1$ is the $+$ solution and $s_2$ is the $-$ solution.  The solution $s_1(\lambda)$ describes the aforementioned solid red branch in Figs.~\ref{fig:2SM}-\ref{fig:3SM}, i.e. is the largest solid branch for $\lambda>0$ and the lowest solid branch for $\lambda<0$.  
We are interested in the minimum of this branch for $\lambda>0$ and maximum for $\lambda<0$.  We have
\begin{equation}
\label{eq:deriv}
\frac{ds_1}{d\lambda} = \left(\frac{\lambda  (2 \mathcal{D} \lambda -1) (-a+b+\lambda  (\mathcal{D} \lambda -1)+1)}{\sqrt{\psi}}-\sqrt{\psi}+a+b+\mathcal{D} \lambda^2-1\right)/\left(2 \lambda^2\right).
\end{equation}
Over the next several pages we discuss asymptotic scaling behaviors and their crossovers for $\lambda^*_{\pm}$ and $s^*_{\pm}$, separating the discussion into large-$\mathcal{D}$ and small-$\mathcal{D}$ regimes, as defined below.

\subsubsection{Large $\mathcal{D}$.}
We would like to find $\lambda^*$ that satisfy $\frac{ds_1}{d\lambda} =0$.  For a purely FKPP model, the solutions are $\lambda^*_{\pm} = \pm \frac{1}{\sqrt{\mathcal{D}}}$.  Note that $\lambda^{-1} + \mathcal{D}\lambda$ is the asymptote of the full solution in Eq.~(\ref{eq:s_of_lambda_nonzero_D}) at large $\lambda$, and they both vary like $\sim \lambda^{-1}$ at small $\lambda$ (and finite $a,b$), so the positions of the extrema of $\lambda^{-1} + \mathcal{D}\lambda$ should serve as a first guess for the positions of extrema in the full problem.  In fact, we noticed numerically that the extrema in the full problem at large $\mathcal{D}$ usually do lie very close to the extrema of the FKPP model. This suggests an ansatz 

\begin{equation}
\label{eq:inverse_D_series}
\lambda^* =  \pm \frac{c_1}{\sqrt{\mathcal{D}}} + \frac{c_2}{\mathcal{D}} + ...
\end{equation}

The $\sim \mathcal{D}^{-1/2}$ leading behavior of both $\lambda^*_+$ and $\lambda^*_-$ at large $\mathcal{D}$ has been verified from the numerically computed extrema of $s_1(\lambda)$ from Eq.~(\ref{eq:s_of_lambda_nonzero_D}). The coefficients $c_1$, $c_2$, etc. can be found iteratively - first seek the coefficient of the first term, by substituting $\lambda^* = \pm \frac{c_1}{\sqrt{\mathcal{D}}}$ into Eq.~(\ref{eq:deriv}), expanding in $\frac{1}{\sqrt{\mathcal{D}}}$, and solving for $c_1$ that eliminates the $\mathcal{D}^{0}$ term in this expansion.  By doing this, we find a $c_1$ that makes this ansatz asymptotically-exact as $\mathcal{D} \rightarrow \infty$, i.e. it will give us a large-$\mathcal{D}$-approximation to $\lambda^*$ for arbitrary $a$ and $b$.  We may then repeat this with the second term in Eq.~(\ref{eq:inverse_D_series}) included, and seek $c_2$ that eliminates the $\mathcal{D}^{-1/2}$ term - it will improve the asymptotic approximation to the true $\lambda^*$, and make it more accurate down to smaller $\mathcal{D}$, and so on.

\begin{figure}[ht]
\includegraphics[width=4.2in]{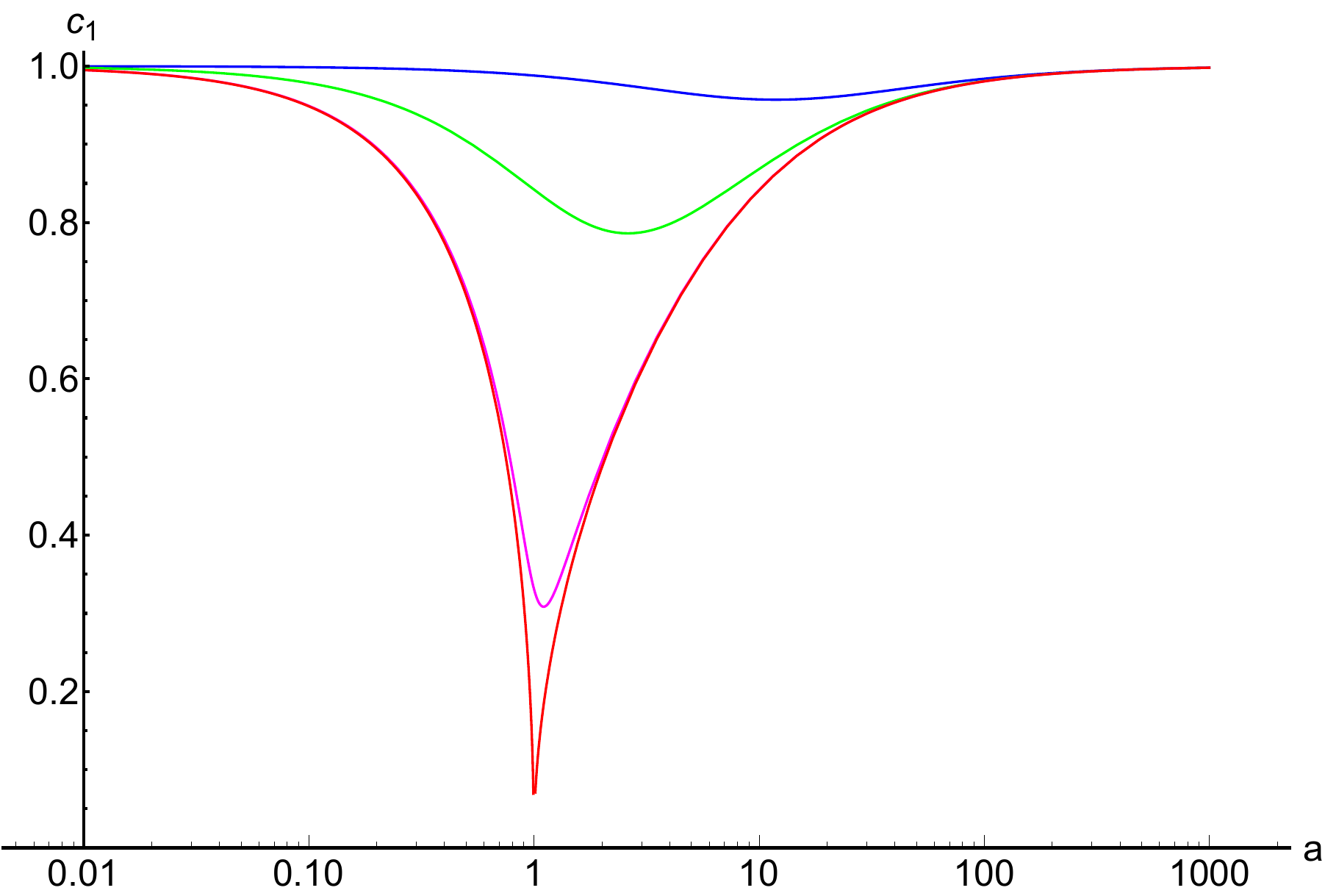}
\caption{(Color online) $c_1$ versus $a$ for several values of $b$.  From top to bottom: $10$ (blue), $1$ (green), $10^{-2}$ (magenta), $10^{-7}$ (red).}
\label{fig:dip}
\end{figure}

When this procedure was implemented, the following solution for $c_1$ was found with the help of Mathematica

\begin{eqnarray}
\label{eq:c1_soln}
c_1(a,b) &=& \sqrt{\frac{(2+a+b)^2-(1+2a+2b)\chi^{1/3}(a,b) + \chi^{2/3}(a,b)}{3\chi^{1/3}(a,b)}},\\
\mbox{where  } \chi(a,b) &=& a^3 +3a^2(2+b)+(2+b)^3+3a(b-1)(b+5)-27a+6i\sqrt{3}\sqrt{a^4 +3a^3(2+b)+a(2+b)^3+3a^2(b-1)(b+5)}. \nonumber
\end{eqnarray}
So at the lowest order in $\frac{1}{\sqrt{\mathcal{D}}}$, the coordinates of the two extrema, $\lambda^*_+$ and $\lambda^*_-$, remain symmetric about $0$.  In contrast, $c_2$ is identical for both fronts, so overall $\lambda^*_+$ and $\lambda^*_-$ are slightly asymmetrical about $0$.  We plot $c_1$ versus $a$ for several values of $b$ in Fig.~
\ref{fig:dip}.

There is a prominent dip at $a=1$ when $b$ becomes small.  The value of $c_1 \rightarrow 0$ as $b \rightarrow 0$ at $a=1$.  At small $b$, $c_1(a=1,b) = \frac{\sqrt{2}}{3^{1/4}} b^{1/4}\left( 1 - \frac{5}{12\sqrt{3}}b^{1/2} + ...\right)$.  This suggests that $\lambda^*_{\pm}$ will go to zero - or equivalently, the width of the front will diverge - at $a=1$ as $b\rightarrow 0$.  However, the corrections to $c_1/\sqrt{\mathcal{D}}$ may protect $\lambda^*_{\pm}$ from reaching zero.  

The correction term $c_2/\mathcal{D}$ does improve the match with the exact solution of $\lambda^*_{\pm}$ as $c_1$ gets smaller, but this improvement is perturbative - at a given $b$, it becomes worse with smaller $\mathcal{D}$, so more and more terms in the series are needed.  To understand the behavior of $\lambda^*_+$ and $\lambda^*_-$ at $(a=1, b\rightarrow 0)$, the following perturbative analysis will be used.  We Taylor expand the right hand side of Eq.~(\ref{eq:deriv}) around $a=1, b=0$ to first order in $b$ (see the comment about the singular limit in \cite{Footnote_SM}).  For  $\lambda<0$, result - at arbitrary $\mathcal{D}$ is
\begin{displaymath}
\frac{ds_1}{d\lambda} =\frac{1}{2\lambda^2}\left[\frac{4 b - 6 b \mathcal{D} \lambda+ 2 \mathcal{D}\lambda^3 - 4 \mathcal{D}^2\lambda^4 + 2 \mathcal{D}^3\lambda^5}{\lambda(1-\lambda \mathcal{D})^2 } + O(b^2)\right]
\end{displaymath}

We seek a set of $\lambda$s that are zeros of the right hand side of this equation.  These zeros are the zeros of the numerator when the latter do not include $0$ and $1/\mathcal{D}$, which is true whenever $b$ is not strictly zero.   The negative roots approach $0$ as $b \rightarrow 0$.  As $b$ deviates from $0$ slightly, the roots that approaches $0$ will take place at very small and negative $\lambda$, and are given asymptotically by the solution to $4b+2\mathcal{D}\lambda^3=0$.  All the other terms become less relevant as $b$ and $\lambda$ approach $0$.  The only real solution is 
\begin{equation}
\lambda^*_- = -\left(\frac{2b}{\mathcal{D}}\right)^{1/3}.
\end{equation}
\begin{figure}[h]
\includegraphics[width=4.2in]{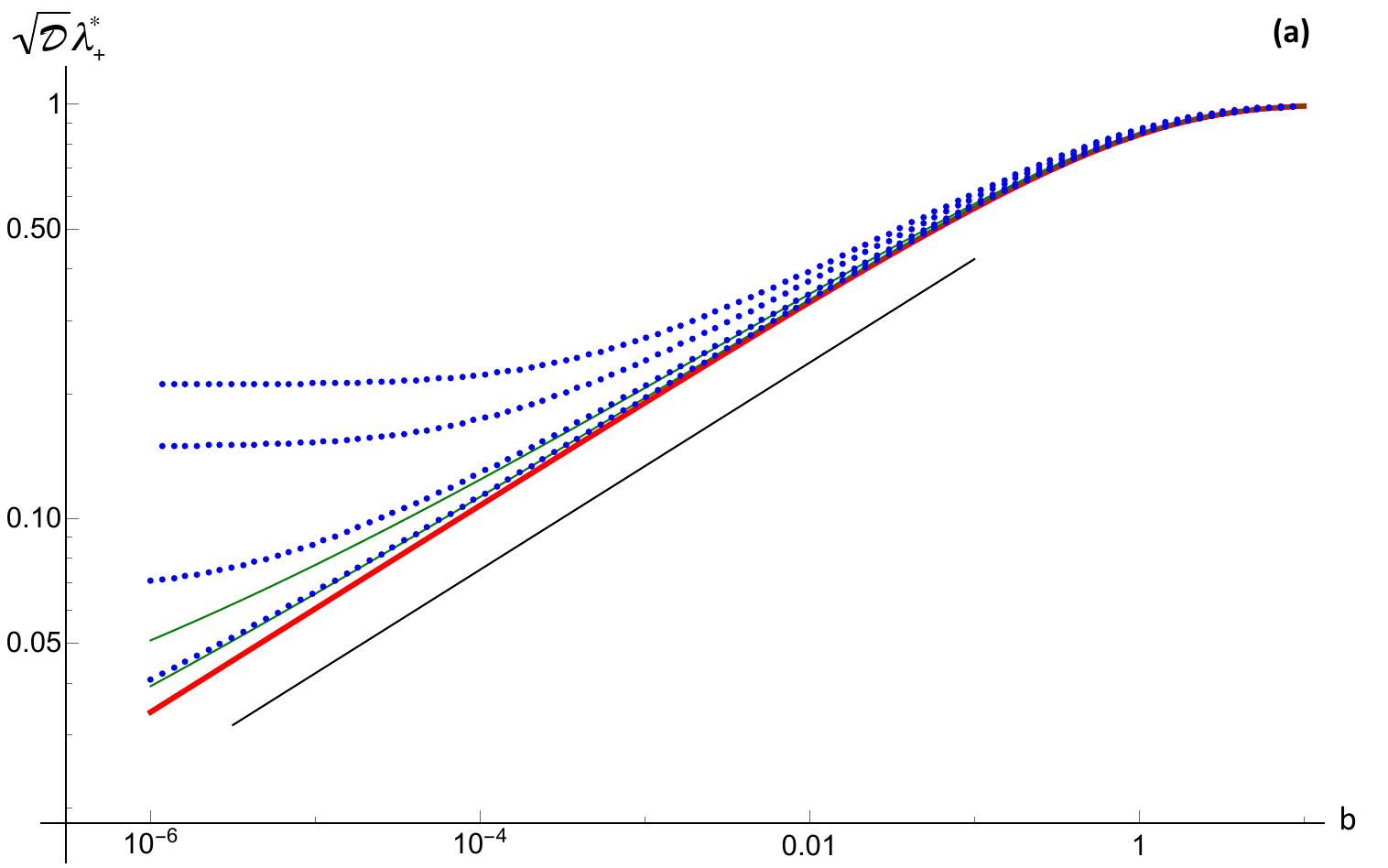}
\includegraphics[width=4.4in]{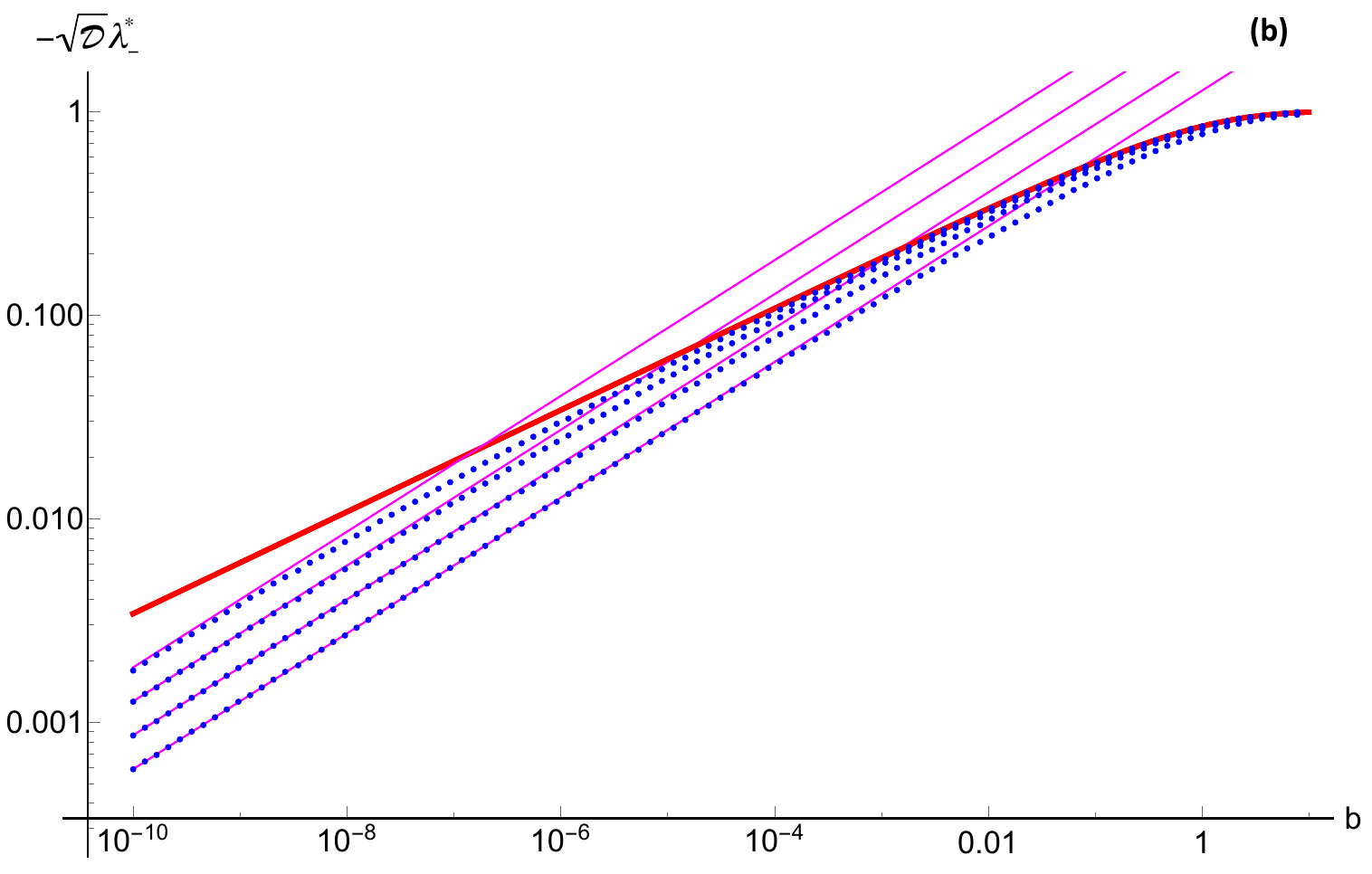}
\caption{(Color online) $\sqrt{\mathcal{D}} \lambda^*_{\pm}$ versus $b$ at $a=1$.  The dots represent an exact numerical solution to $\lambda^*_{\pm}$ that satisfy $ds_1/d\lambda = 0$: in panel (a) for $\mathcal{D} = 10$, $20$, $100$ and $1000$, from top to bottom, and in panel (b) for $\mathcal{D} = 1$, $10$, $100$ and $1000$, from bottom to top.  The thick red curve represents $\lambda^*_{\pm}$ from Eq.~(\ref{eq:inverse_D_series}) with only the first term.  It has an asymptotic slope $1/4$ on the log-log scale, represented by the solid black segment.  The thin green curves in panel (a) represent $\lambda^*_+$ from Eq.~(\ref{eq:inverse_D_series}) with the second term included for $\mathcal{D} = 100$, and $1000$, indicating that corrections improve the results in the right direction.  The magenta lines panel (b) represent the $-\left(\frac{2b}{\mathcal{D}}\right)^{1/3}$ asymptotic limit to $\lambda^*_-$ at small $b$.}
\label{fig:AsymptoticTheory}
\end{figure}
\\
This analysis did not rely on the largeness of $\mathcal{D}$.  As a result, this same solution will appear in the small-$\mathcal{D}$ regime.  For positive $\lambda$, there is a different approximation to the right hand side of Eq.~(\ref{eq:deriv})  at small $b$ \cite{Footnote_SM}.  A similar analysis would lead us to conclude that $\lambda^*_+ \rightarrow \frac{2}{3\mathcal{D}}$ for $\mathcal{D}$ comparable to $1$ or greater.  Both of these conclusions - concerning $\lambda^*_-$ and $\lambda^*_+$ - were confirmed by the comparison with the numerically computed extrema of $s_1(\lambda)$ from Eq.~(\ref{eq:s_of_lambda_nonzero_D}), see Fig.~\ref{fig:AsymptoticTheory}.

Using the ``$\sim$'' notation to denote asymptotic behavior, we can summarize the $a=1$ situation as follows.  As $b$ decreases, $\lambda^*_+ \sim c_1(a=1,b)\mathcal{D}^{-1/2}$ ($\rightarrow \frac{\sqrt{2}b^{1/4}}{3^{1/4}}\mathcal{D}^{-1/2}$ for $b \ll 1$) - it decreases until crossing over to saturate at $\frac{2}{3\mathcal{D}}$ (for $\mathcal{D} \ll1$ the saturation value is not given by this simple value).  This represents an anomalously large front width as $b \rightarrow 0$.  In contrast, $\lambda^*_- \sim -c_1(a=1,b)\mathcal{D}^{-1/2}$ (so $|\lambda^*_-|$ decreases), 
and then experiences a crossover to $\sim -\left(\frac{2b}{\mathcal{D}}\right)^{1/3}$.  This represents a true divergence of the front width as $b \rightarrow 0$ for any $\mathcal{D}$.  

Equivalently, if we hold $b$ fixed and decrease $\mathcal{D}$ we will crossover from the large-$\mathcal{D}$ regime with $\lambda^*_{\pm} \sim \pm c_1(a=1,b)\mathcal{D}^{-1/2}$ to the small-$\mathcal{D}$ regime with $\lambda^*_+ \sim \frac{2}{3\mathcal{D}}$ and the $\lambda^*_- \sim -\left(\frac{2b}{\mathcal{D}}\right)^{1/3}$.  For $\mathcal{D}$ above this crossover, $s^*_{\pm} \sim \mathcal{D}^{1/2}$, as we shall see below.  
For $b \ll 1$, this crossover $\mathcal{D}$ scales as $b^{-1/2}$, while for $b \gg1$, it scales as $b^{-2}$.  
So as $b \rightarrow 0$, this large-$\mathcal{D}$ regime onsets at larger and larger $\mathcal{D}$.  We can now assign a precise definition to the term ``large-$\mathcal{D}$'' - even for $a \neq 1$.  It denotes the parameter regime when $\lambda^*_{\pm} \sim \pm c_1(a,b)\mathcal{D}^{-1/2}$.  ``Small-$\mathcal{D}$'' may have several meanings.  For example, for $\lambda^*_-$ at $a=1$, it means the regime when $\lambda^*_{\pm} \sim -\left(\frac{2b}{\mathcal{D}}\right)^{1/3}$.  We note that when $a \neq 1$, $\lambda^*_-$ and  $\lambda^*_+$ reach a constant value as $b \rightarrow 0$ - front widths do not diverge.


The prediction of an unusually large, or even divergent front width 
is non-intuitive.  There are no diverging length scales in either the FKPP model at finite $\mathcal{D}$ or in the purely advection-only model.  Therefore, this is a non-trivial consequence of the competition between the two transport mechanisms.  
As $a$ increases, more of the produced mass is removed from the GL, effectively lowering $\mathcal{D}$.  Lower $\mathcal{D}$ in FKPP model implies a longer length scale of the front.  However, this argument does not explain the rebound of $\lambda^*$ for $a>1$.  The phenomenology is somewhat reminiscent of resonance - with $|a-1|$ akin to detuning, and $b$ akin to damping.  At this stage, however, this is only a metaphor.
\\

In returning to the discussion of a general $a \neq 1$, we remind the reader that the expansion in Eq.~(\ref{eq:inverse_D_series}) is only meant for large $\mathcal{D}$.  Therefore, Fig.~\ref{fig:dip} represents approximations to $\left|\lambda^*_{\pm}\right|\sqrt{\mathcal{D}}$ only at large $\mathcal{D}$.  As $\mathcal{D}$ is decreased, $\lambda^*_{+}$ and $\lambda^*_{-}$ versus $a$ will change, but in different ways.  At $\mathcal{D} = 1/4$, $\lambda^*_+= 1/\sqrt{\mathcal{D}}=2$ for all $a$ and $b$.   For $\mathcal{D} \ll 1/4$, the plot of $\lambda^*_+$ versus $a$ would resemble the solution to Eq.~(5) in the main text up until a certain crossover value of $a$, when the increase of $\lambda^*_+$ slows down, and eventually saturates at $1/\sqrt{\mathcal{D}}$.  This crossover point grows larger with smaller $\mathcal{D}$.  Therefore, at any finite $a$, there is a continuous change in $\lambda^*_{\pm}$ as $\mathcal{D}$ is tuned up from $0$.  

On the other hand, there is nothing special about $\lambda^*_-$ at $\mathcal{D} = 1/4$.  Also, the $\mathcal{D}=0$ system has a diverging $\lambda^*_-$ as $a$ approaches $1$ from above - see Eq.~(5) in the main text (as a reminder, in the $\mathcal{D} \neq 0$ model, the upwind front propagates against advection when $a<1$, whereas it stands still when $\mathcal{D}=0$).  At non-zero $\mathcal{D}$, $\lambda^*_-$ stays finite, even at $a=1$.  Close to $a=1$, the divergence is replaced by a rapidly growing $|\lambda^*_-|$ as $a$ decreases past $1$, but this growth slows down at smaller $a$, and $\lambda^*_-$ eventually saturates at $-1/\sqrt{\mathcal{D}}$ as $a \rightarrow 0$.  For $a>1$,  $\lambda^*_-$ from Eq.~(5) in the main text becomes closer and closer to the true $\lambda^*_-$ as $\mathcal{D} \rightarrow 0$ up until a certain large crossover value of $a$. At this point the growth in  $\lambda^*_-$ versus $a$ slows down, and it eventually saturates at $-1/\sqrt{\mathcal{D}}$ - similar to what happens with $\lambda^*_+$.  Again, this crossover point grows larger with smaller $\mathcal{D}$.  We shall examine the small-$\mathcal{D}$ behavior of both fronts below.

When $c_1 \neq 0$, i.e. in the exception of $(a=1, b \rightarrow 0)$, we may substitute $\lambda^*_{\pm} = \pm\frac{c_1}{\sqrt{\mathcal{D}}}$ into $s_1$ from Eq.~(\ref{eq:s_of_lambda_nonzero_D}), and expand the numerator in $\frac{1}{\sqrt{\mathcal{D}}}$.  We would obtain:
\begin{eqnarray}
\label{eq:largeDasymptotic} s^*_{\pm} &=& v_{\mathrm{eff}} \pm 2\sqrt{\mathcal{D}_{\mathrm{eff}}} +O\left(\frac{1}{\sqrt{\mathcal{D}}}\right) \\
\mbox{where } \label{eq:Scumbersome1} v_{\mathrm{eff}} &=& \frac{1}{2} + \frac{-1+a-b-c_1^2}{2\sqrt{-2a+a^2+2ab-2ac_1^2+(1+b+c_1^2)^2}}, \\ 
\mbox{and } \label{eq:Scumbersome2} \mathcal{D}_{\mathrm{eff}} &=& \left(\frac{1-a-b+c_1^2+\sqrt{-2a+a^2+2ab-2ac_1^2+(1+b+c_1^2)^2}}{4c_1}\right)^2 \mathcal{D},
\end{eqnarray}
and $c_1$ is given in Eq.~(\ref{eq:c1_soln}).  The expressions in Eqs.~(\ref{eq:Scumbersome1})-(\ref{eq:Scumbersome2}) 
\begin{figure}[ht]
\includegraphics[width=3.1in]{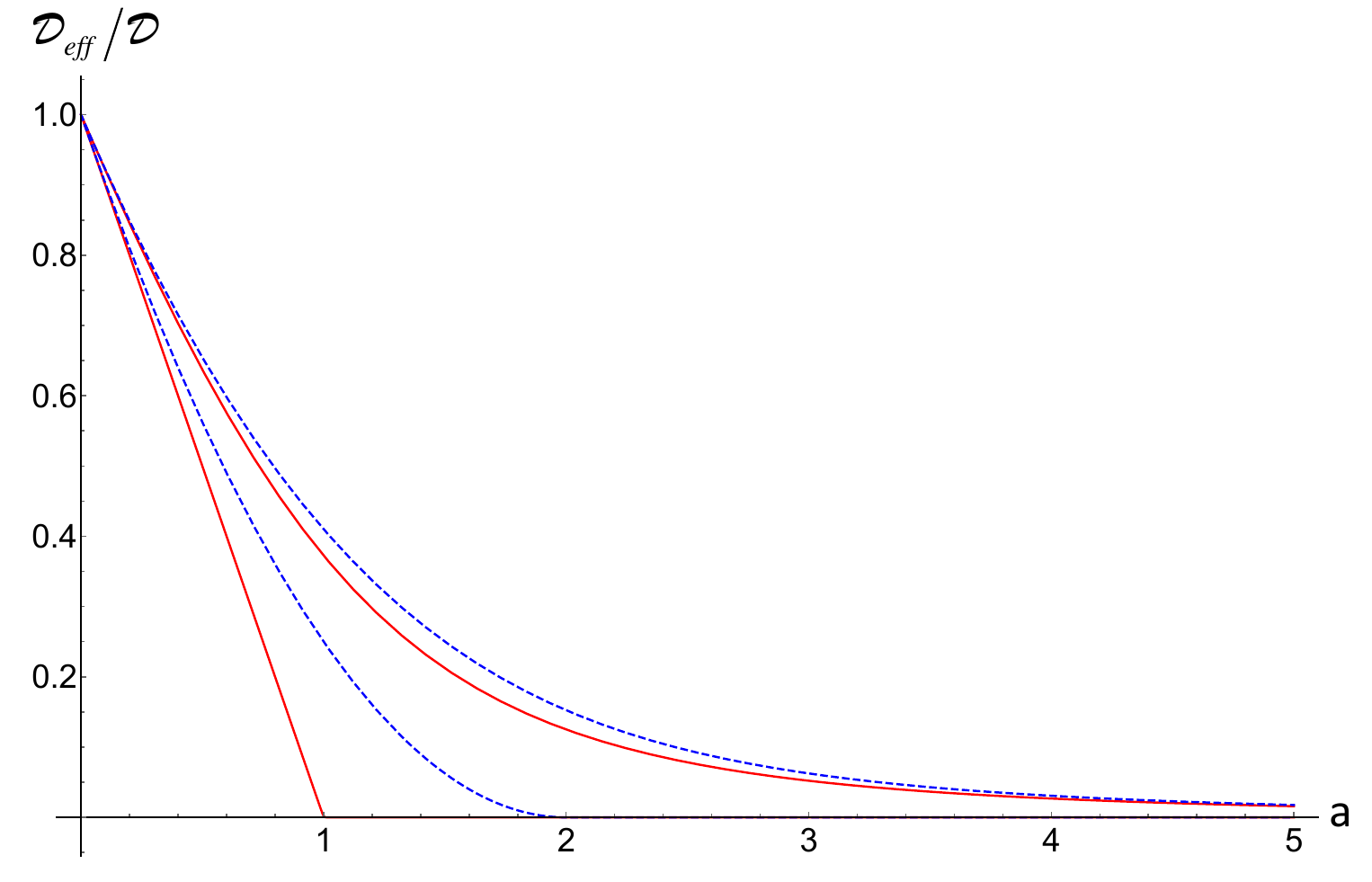}
\includegraphics[width=3.1in]{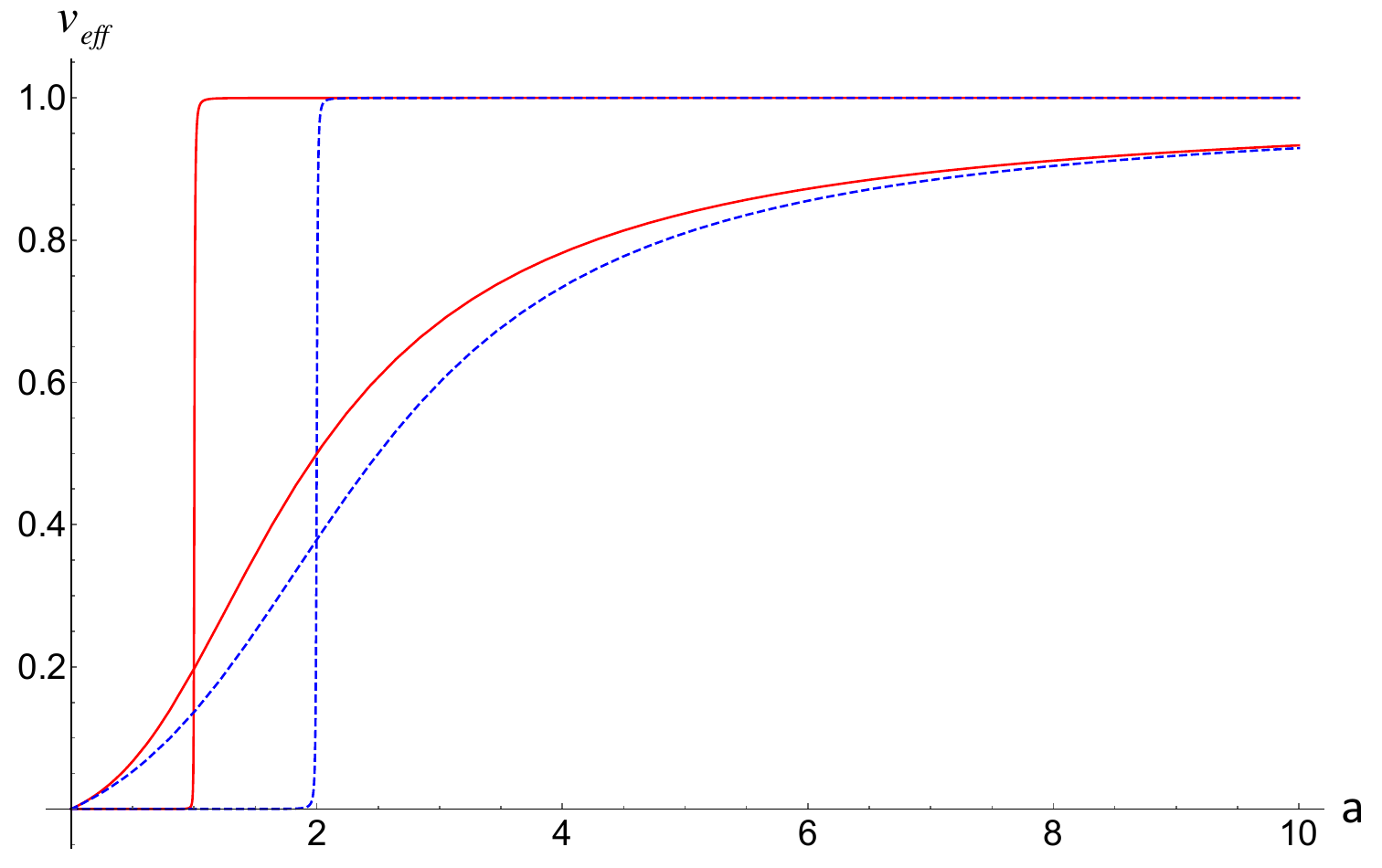}
\caption{(Color online) $\mathcal{D}_{\mathrm{eff}}/\mathcal{D}$ and $v_{\mathrm{eff}}$ versus $a$ using Eqs.~(\ref{eq:Scumbersome1})-(\ref{eq:Scumbersome2}) and using $c_1$ from Eq.~(\ref{eq:c1_soln}) (red, solid curve) or using the simpler approximations obtained by setting $c_1=1$, resulting in Eqs.~(8)-(9) in the main text (blue, dashed curves).  The results are shown for $b=10^{-5}$ (bigger discrepancy), and $b=0.5$ (smaller discrepancy).}
\label{fig:veff_and_Deff}
\end{figure}
are cumbersome, but we notice from Fig.~\ref{fig:dip} that sufficiently far from $(a=1, b=0)$, $c_1$ can be approximated by $1$, corresponding to the FKPP $\lambda^*_{\pm}$.  This leads to much simpler formulas for $s^*_{\pm}$, quoted in the main text as Eqs.~(8)-(9).  We plot $\mathcal{D}_{\mathrm{eff}}/\mathcal{D}$ and $v_{\mathrm{eff}}$ obtained using both of these methods in Fig.~\ref{fig:veff_and_Deff}.

\begin{figure}[ht]
\centering
\begin{tabular}{ccc}
\includegraphics[width=60mm]{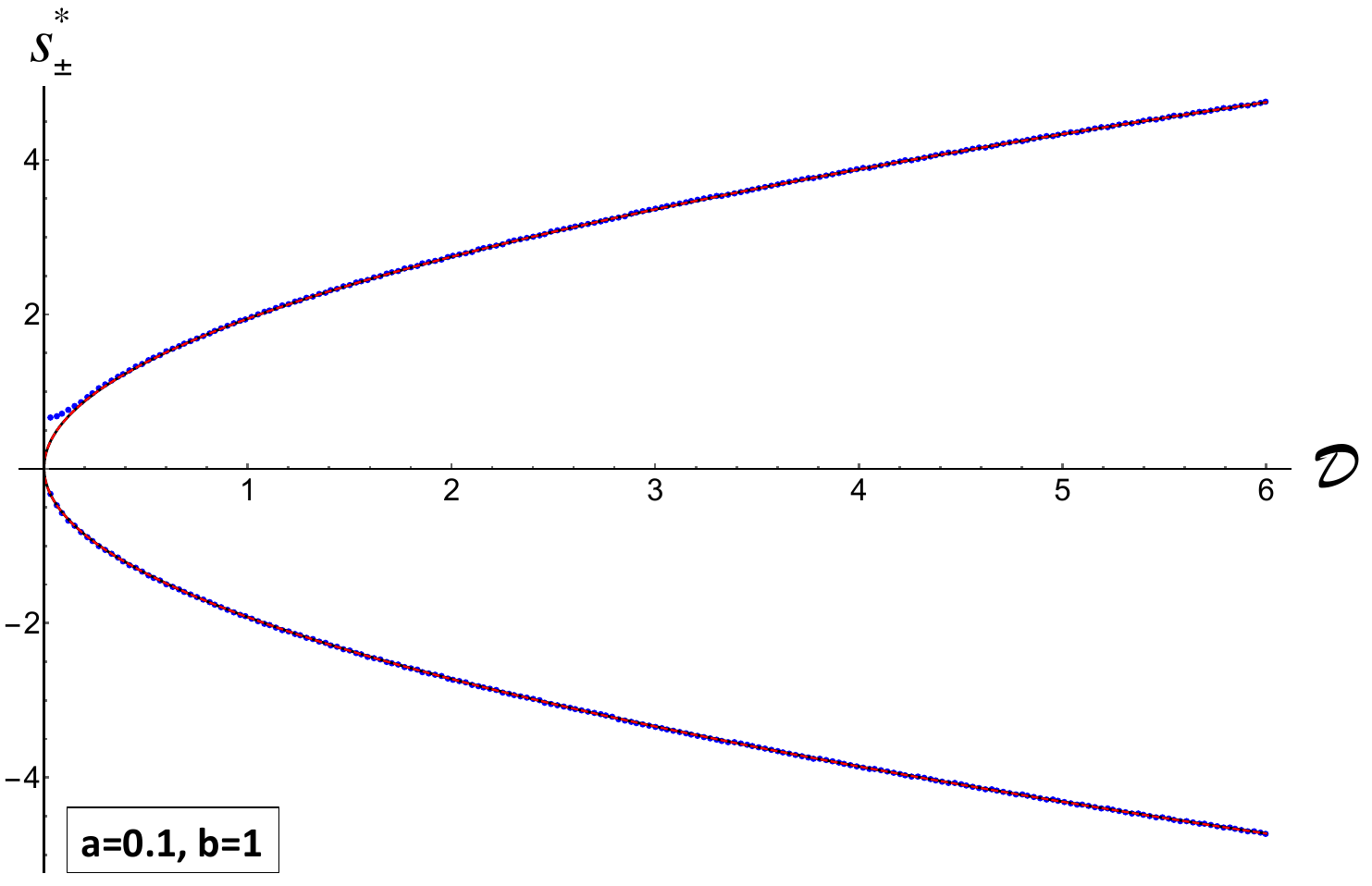}&
\includegraphics[width=60mm]{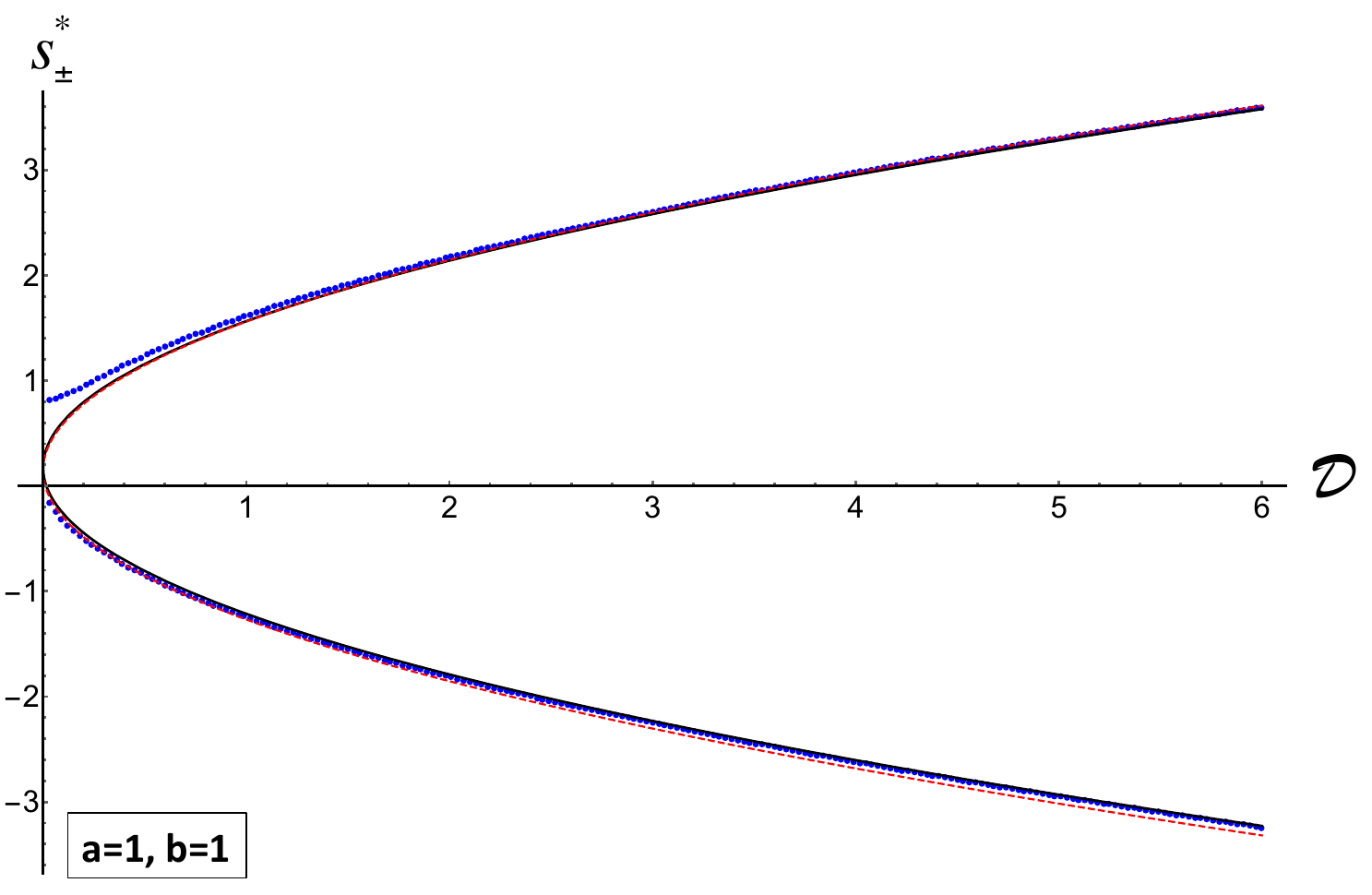}&
\includegraphics[width=60mm]{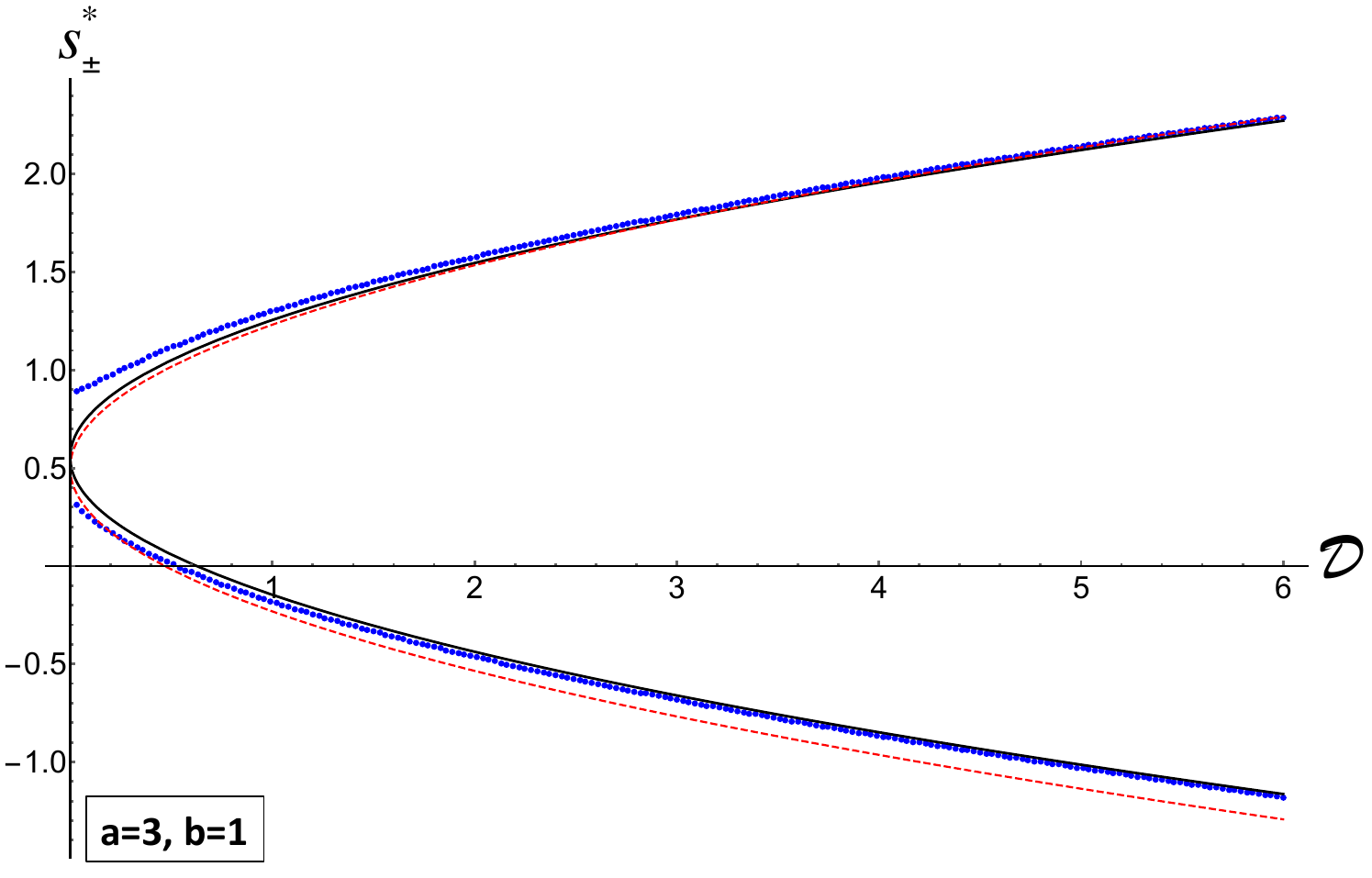}\\
\includegraphics[width=60mm]{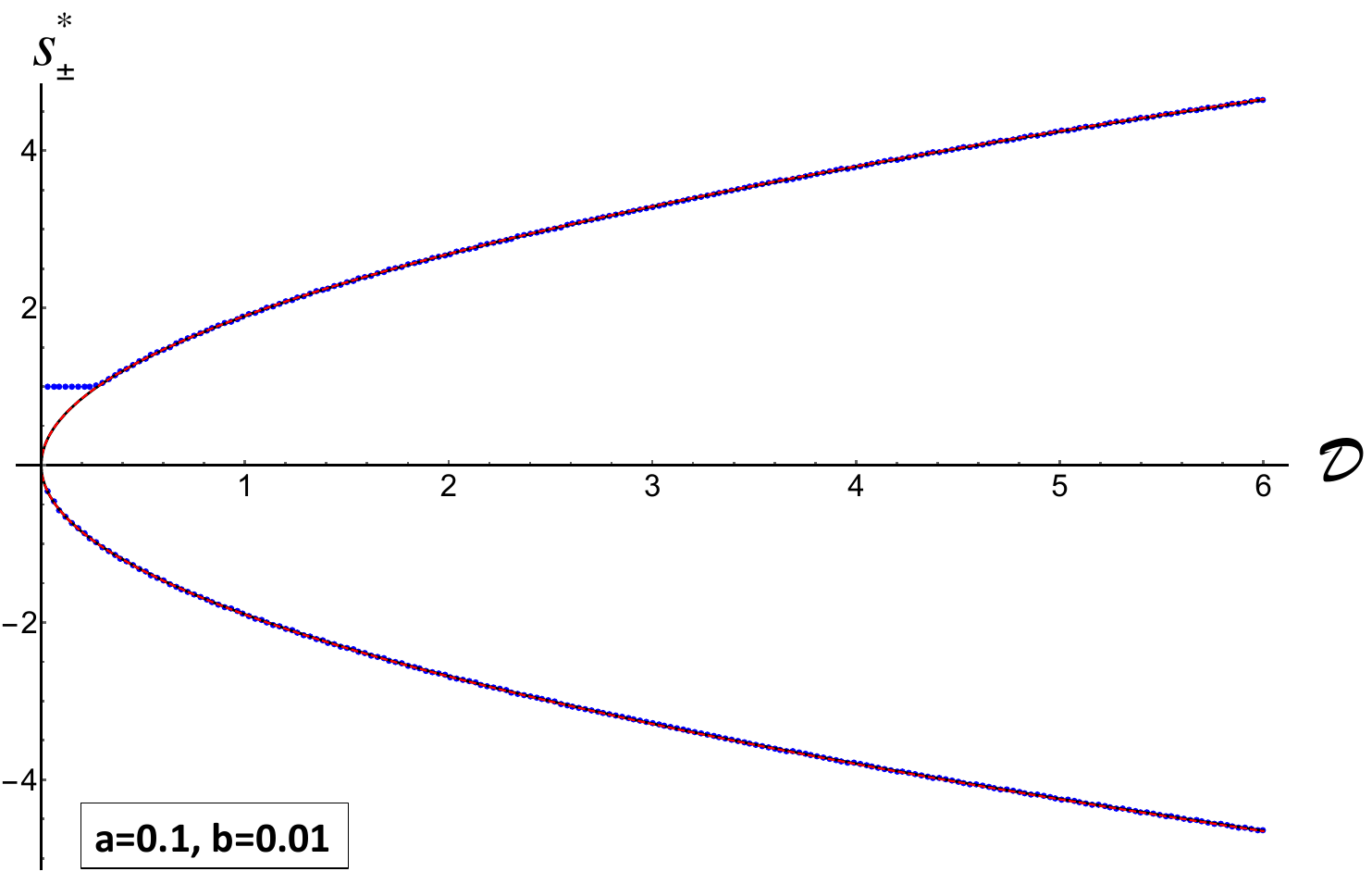}&
\includegraphics[width=60mm]{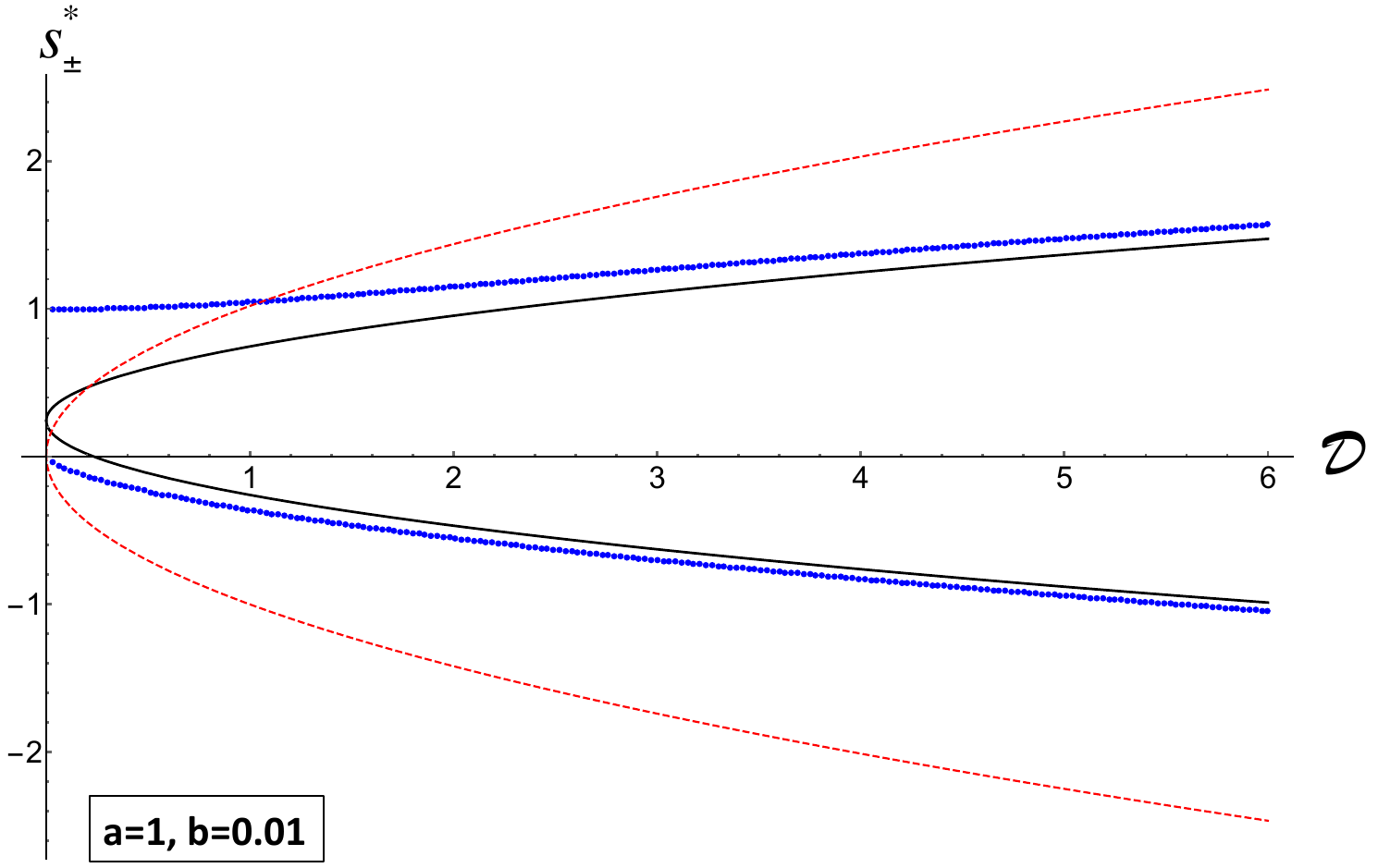}&
\includegraphics[width=60mm]{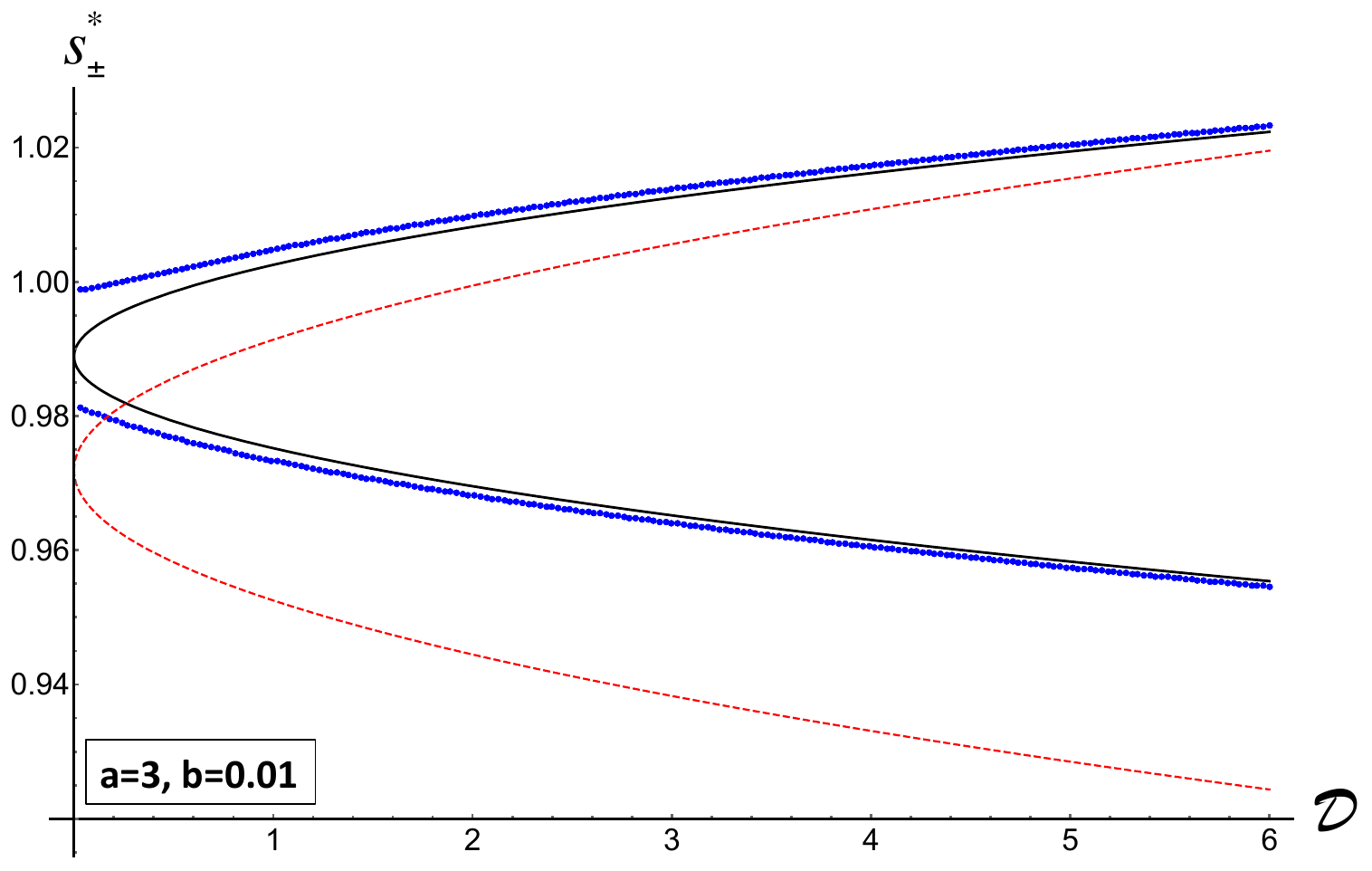}\\
\end{tabular}
\caption{(Color online).  Comparison of $s^*_{\pm}$ obtained from the exact minimum of $s_1(\lambda)$ versus the large-$\mathcal{D}$ asymptotic Eq.~(\ref{eq:largeDasymptotic}) where $v_{\mathrm{eff}}$ and $\mathcal{D}_{\mathrm{eff}}$ are given respectively by Eqs.~(\ref{eq:Scumbersome1})-(\ref{eq:Scumbersome2}) and using $c_1$ from Eq.~(\ref{eq:c1_soln}) (solid black) or with simpler approximations obtained by setting $c_1=1$, resulting in Eqs.~(8)-(9) in the main text (dashed red). Note the scale on the $a=3, b=0.01$ panel.  It is important to point out that $v_{\mathrm{eff}}$ is the offset of the large-$\mathcal{D}$ asymptotic $s^*_{\pm}$, and is not meant to represent $s^*_{\pm}$ at $\mathcal{D}=0$, which will be predicted by the $\mathcal{D}=0$ model (see also next section).  For example, $s^*_{\pm}$ at $\mathcal{D}=0$ will be zero for $a\leq 1$.  
}
\label{fig:speed_test}
\end{figure}

Fig.~\ref{fig:speed_test} compares $s^*_{\pm}$ from Eqs.~(\ref{eq:Scumbersome1})-(\ref{eq:Scumbersome2}) with $s^*_{\pm}$ obtained with the numerically computed extrema of $s_1(\lambda)$ from Eq.~(\ref{eq:s_of_lambda_nonzero_D}).  For mathematically-typical parameters, i.e. away from the $(a=1,b=0)$ point, such asymptotic theory works remarkably well, even for $\mathcal{D}$ comparable to $1/4$.  The simpler theory based on setting $c_1=1$ has a larger exclusion region around $(a=1,b=0)$ where it does not perform well.  

As expected from the discussion above, the comparison with the asymptotic theory based only on the first term in Eq.~(\ref{eq:inverse_D_series}) is worst at $a=1$ as $b$ becomes small.  However, at any finite $b$, the approximation $\lambda^*_{\pm} \approx c_1/\sqrt{\mathcal{D}}$, and the resulting expression for $s^*_{\pm}$, will always become asymptotically-accurate at large-enough $\mathcal{D}$.  This was clearly seen in Fig.~\ref{fig:AsymptoticTheory} - as $\mathcal{D}$ grows, $\sqrt{D}\lambda^*_{\pm} \rightarrow c_1(a=1,b)$ at any finite $b$ (this is especially so for $a \neq 1$, where $c_1$ reaches a finite value as $b \rightarrow 0$).  
Again, ``large-enough'' refers to the value of $\mathcal{D}$ where the crossover into the $\mathcal{D}^{-1/2}$ behavior of $\lambda^*_{\pm}$ lies.  This crossover for $a \neq 1$ is discussed separately below.  As $\mathcal{D}$ is lowered, the $\pm c_1/\sqrt{\mathcal{D}}$ behavior of $\lambda^*_{\pm}$ crosses over to $\lambda^*_+ \sim \frac{2}{3\mathcal{D}}$ and $\lambda^*_- \sim -\left(\frac{2b}{\mathcal{D}}\right)^{1/3}$ (at $a=1$).  In all cases, the speed $s^*_{\pm}$ is obtained by substituting $\lambda^*_{\pm}$  into Eq.~(\ref{eq:s_of_lambda_nonzero_D}).  There are simple approximations in the limit of small $\mathcal{D}$, which we will see in the following subsection.  
%

\newpage
\subsubsection{Small $\mathcal{D}$.}
The main goal of this subsection is to derive the behavior of $s^*_{\pm}$ versus $\mathcal{D}$ as $\mathcal{D} \rightarrow 0$, and to show when this behavior is linear in $\mathcal{D}$.  The linear behavior will indicate that the FKPP-type of behavior does not extend to zero $\mathcal{D}$.  At the end we will synthesize the information from this and the previous subsections to establish the lower limits in $\mathcal{D}$ for the breakdown of the FKPP-type behavior. 

When $\mathcal{D}=0$, Eq.~(\ref{eq:s_of_lambda_nonzero_D})  turns into Eq.~(\ref{eq:s_of_lambda_zero_D}).  The appearance of new $\mathcal{D}$-terms in Eq.~(\ref{eq:s_of_lambda_nonzero_D}) does not change the positions of extrema discontinuously  - if they take place at finite values in the $\mathcal{D}=0$ problem (the smaller is the $\mathcal{D}$, the larger the $\lambda$ should be in order for the effect of new terms to take place).  

This is true for $\lambda^*_+$, so we can study its change due to the appearance of diffusion perturbatively - by tracking the change in the position of $\lambda^*_+$.  We follow the following procedure.  First, substitute the ansatz (see Eq.~(5) in the main text):
\begin{displaymath}
\lambda^*_+ = \frac{1+a+b+2\sqrt{a}}{1+\sqrt{a}} + c\mathcal{D}.
\end{displaymath}
into Eq.~(\ref{eq:deriv}), and Taylor-expand the resulting expression to linear order in $\mathcal{D}$, giving an expression of the form $A\mathcal{D}$ (as expected, there is no $\mathcal{D}^0$ term).  Next, we solve for $c$ that makes this $A$ zero.    The result is
\begin{displaymath}
c=\frac{\left(1+a+b+2 \sqrt{a}\right)^2}{ \left(1+ \sqrt{a}\right)^2}  \frac{\left(2+(3-a-b)\sqrt{a}\right)}{2 \left(1+ \sqrt{a}\right)^2}.
\end{displaymath} 
We then substitute the ansatz for $\lambda^*_+$ with this $c$ into Eq.~(\ref{eq:s_of_lambda_nonzero_D}) and again expand in $\mathcal{D}$ to first order.  We get
\begin{equation}
\label{eq:s_starplussmallD}
s^*_+ = \left(1+\frac{b}{\left(1+\sqrt{a}\right)^2}\right)^{-1} + \left(\frac{b}{\sqrt{a}+1}\right)\mathcal{D} + ...
\end{equation}
The first term unsurprisingly matches $s^*_+$ from the $\mathcal{D}=0$ model (see Eq.~(4) in the main text).  The correction is the quantity we seek - the slope of $s^*_+$ versus $\mathcal{D}$ - see Fig.~4a in the main text.  This demonstrates that unless $b$ goes to infinity - when particles spend essentially all of the time on the GL - $s^*_+$ scales like an integer power of $\mathcal{D}$ instead of the $\sqrt{\mathcal{D}}$ scaling from the FKPP model.
\\

With $\lambda^*_-$, the situation is more tricky, as the position of the maximum at negative $\lambda$ moves to $-\infty$ as $a \rightarrow 1$ from above when $\mathcal{D}=0$.  So the approach by tracking the maximum of $s_1(\lambda)$ (for $\lambda <0$) as $\mathcal{D}$ is turned up from zero while $a$ crosses $1$ (see Figs.~\ref{fig:1SM}-\ref{fig:3SM}) will not work - there is no maximum in the $\mathcal{D}=0$ problem for $a<1$, whereas it exists in the $\mathcal{D} \neq 0$ system. 
Instead, we Taylor-expand the right hand side of Eq.~(\ref{eq:deriv}) in $\mathcal{D}$ around $0$ and find the leading-order asymptotic expression at large negative $\lambda$.  
The magnitude of $\lambda$ has to be large, because that is where the zeros are located at small $\mathcal{D}$ and $a$ close to $1$.  The resulting equation for $\lambda^*_-$ is a simple cubic:
\begin{equation}
\label{eq:cubic}
0 = (a-1)+\frac{2ab}{\lambda} + \lambda^2 \mathcal{D}.
\end{equation}
Negative $\lambda$ that satisfies Eq.~(\ref{eq:cubic}) is the $\lambda^*_-$ we seek.    Note that the first term merely offsets the function in the $y$-direction, so the solution for $\lambda$ is a smooth function of $a$.  When $a=1$, the solution is simple: $\lambda^*_- = -\left(\frac{2b}{\mathcal{D}}\right)^{1/3}$.  We have already seen this very formula in the discussion of the large-$\mathcal{D}$ regime.  We have argued that this asymptotic is valid below a crossover which separates this small-$\mathcal{D}$ regime and the large-$\mathcal{D}$ regime when $\lambda^*_- \sim -c_1(a=1,b)\mathcal{D}^{-1/2}$ (and that this crossover $\sim b^{-1/2}$ for $b \ll 1$).  To get $s^*_-$ in this regime we substitute this into Eq.~(\ref{eq:s_of_lambda_nonzero_D}), which gives a rather cumbersome expression.  An Expansion in $\mathcal{D}^{1/3}$, has the following leading-order term:
\begin{equation}
\label{eq:three-halves}
s^*_-(a=1) = -\frac{3b^{1/3}\mathcal{D}^{2/3}}{2^{2/3}}.
\end{equation}
The result in Eq.~(\ref{eq:three-halves}) was confirmed by the comparison with the numerically-obtained maximum of the exact $s_1(\lambda)$ for $\lambda <0$.  This numerical solution also confirmed $\lambda^*_- = -\left(\frac{2b}{\mathcal{D}}\right)^{1/3}$.  Echoing our findings at large $\mathcal{D}$, the upwind front width diverges as $b \rightarrow 0$ at $a=1$.  This is not so for $\lambda^*_+$ (see above).

We also see from Eq.~(\ref{eq:cubic}) that when $\mathcal{D}=0$, $\lambda^*_- = \frac{-2b}{a-1} + ...$.  This is in fact the leading-order term in the expansion of $\lambda^*_-(\mathcal{D}=0) = \frac{1+a+b-2\sqrt{a}}{1-\sqrt{a}}$ (see Eq.~(5) in the main text) around $a=1$.

Next, we are going to set $\epsilon = a-1$ in the first term of Eq.~(\ref{eq:cubic}), and approximate $a$ by $1$ in the second term (Eq.~(\ref{eq:cubic}) is not exact - it is a consequence of a low-order expansion, so keeping higher order terms in subsequent calculations  is pointless).  Thus, we seek a solution to 
\begin{displaymath}
\epsilon+\frac{2b}{\lambda} + \lambda^2 \mathcal{D}  = 0.
\end{displaymath}
From the structure of the function of $\lambda$ on the left hand side, and the fact that $b$ and $\mathcal{D}$ are always positive, we see that there can be at most one negative root.  It helps to rewrite this in a standard form for a cubic equation, 
\begin{equation}
\label{eq:cubic_equation}
\lambda^3  + \left(\frac{\epsilon}{\mathcal{D}}\right)\lambda + \left(\frac{2b}{\mathcal{D}}\right) = 0.
\end{equation}
The solution for positive $\epsilon$ is  
\begin{equation}
\label{eq:solution_positive_epsilon}
\left(\frac{b}{\mathcal{D}}\right)^{1/3}\left[\left(-1 + \sqrt{1+ \left(\frac{\epsilon}{3b^{2/3} \mathcal{D}^{1/3}}\right)^3}\right)^{1/3} - \frac{\left(\frac{\epsilon}{3b^{2/3} \mathcal{D}^{1/3}}\right)}{\left(-1 + \sqrt{1+ \left(\frac{\epsilon}{3b^{2/3} \mathcal{D}^{1/3}}\right)^3}\right)^{1/3}}\right],
\end{equation}
which is real and negative over that domain, so it is indeed $\lambda^*_-$ when $\epsilon > 0$.  We are working in the neighborhood of small $\epsilon$ and $\mathcal{D}$.  Evidently, $\frac{\epsilon}{3b^{2/3} \mathcal{D}^{1/3}}$ is a natural measure of prevalence of each of these parameters.  One can easily show that when $\mathcal{D} \ll \frac{\epsilon^3}{27b^2}$,  $\lambda^*_-$ saturates to $-\frac{2b}{\epsilon}$, and when $\mathcal{D} \gg \frac{\epsilon ^3}{27b^2}$, $\lambda^*_- \sim  -\left(\frac{2 b}{\mathcal{D}}\right)^{1/3}$.  
These two limits match the expressions that we have just discussed separately.  As $\mathcal{D}$ is increased further, the $-\left(\frac{2 b}{\mathcal{D}}\right)^{1/3}$ behavior of $\lambda^*_-$ eventually meets a second crossover and gives way to the $\sim c_1(1+\epsilon,b)\mathcal{D}^{-1/2}$ asymptotic behavior.  We have shown that the function $c_1(a,b) $ goes to a finite value as $b\rightarrow 0$ when $a \neq 1$, so this suggests that both crossovers continues to increase, in contradiction to the expectation that $\lambda^*_{\pm} \sim \mathcal{D}^{-1/2}$ at large-enough $\mathcal{D}$ when $a \neq 1$ (see the previous subsection on Large $\mathcal{D}$).  
However, the calculations here are based on an expansion around $\mathcal{D}=0$, and will break down at large-enough $\mathcal{D}$.  Thus, in practice - as verified by the numerical calculations of Eq.~(\ref{eq:deriv}) - the crossover to the $\sqrt{\mathcal{D}}$ scaling behavior does not continue to increase as $b$ is lowered, if $\epsilon \neq 0$.

It is also worth mentioning that empirically, the expression in Eq.~(\ref{eq:solution_positive_epsilon}) is well approximated by $-\left(\left(\frac{\epsilon}{2b}\right)^2 + \left(\frac{\mathcal{D}}{2b}\right)^{2/3}\right)^{-1/2}$.  The expression $[...]$ appearing in Eq.~(\ref{eq:solution_positive_epsilon}) is not real and negative for $\epsilon <0$, so will need to choose another solution there; the two solutions must join in a continuous fashion, since as already mentioned, $\lambda^*_-$ changes smoothly as $\epsilon$ crosses zero.  We first study the positive $\epsilon$ case.

As was the case with $\lambda^*_-$, we are interested in the functional behavior of $s^*_-$ versus very small $\mathcal{D}$, primarily to demonstrate that a correction to  $s^*_-$ at $\mathcal{D}=0$ is proportional to an integer power of $\mathcal{D}$, and therefore, strongly departs from any FKPP-like predictions.  To do this, we expand the $[...]$ quantity in Eq.~(\ref{eq:solution_positive_epsilon}) to the first two terms in small $\frac{\epsilon}{3b^{2/3} \mathcal{D}^{1/3}}$ , and get
\begin{equation}
\label{eq:lambda_minus_expansion}
\lambda^*_- = -\frac{2b}{\epsilon}\left(1-\frac{4b^2 \mathcal{D}}{\epsilon^3}\right).
\end{equation}
The leading-order term is $ -\frac{2b}{\epsilon}$ as can be seen.  When this expression is substituted into Eq.~(\ref{eq:s_of_lambda_nonzero_D}), and the result is expanded in $\epsilon$ and $\mathcal{D}$, we get $s^*_- = \left(\frac{\epsilon^2}{4b} + O(\epsilon^3) \right) - \left(\frac{2b}{\epsilon} + O(\epsilon^0) \right)\mathcal{D}$.  The calculation presented in this subsection was based on an expansion in $\epsilon = a-1$ and $\mathcal{D}$.  Therefore, we can not hope to capture the $\mathcal{D}^{0}$ term fully.  Now, $s_1$ in Eq.~(\ref{eq:s_of_lambda_nonzero_D}) upon which this perturbative analysis is based, reduces when $\mathcal{D}=0$, to $s_1$ in Eq.~(\ref{eq:s_of_lambda_zero_D}) for which $s^*_- = \left(1+ \frac{b}{(1 - \sqrt{1+\epsilon})^2}\right)^{-1}$ (Eq.~(4) in the main text).  Moreover the leading term in the $\epsilon$-expansion of this quantity is indeed $\frac{\epsilon^2}{4b}$ (the cubic terms also agree).  

The slope $-\frac{2b}{a-1}$ in the expression for $s^*_-$ versus $\mathcal{D}$ was derived with the assumption that $a-1>0$ is small.  Had we used the procedure used for $\lambda^*_+$ (Eq.~(\ref{eq:s_starplussmallD})), we would find that the slope is $-\frac{b}{\sqrt{a}-1}$, which does not rely on the smallness $a-1$.  However, the two results become equivalent when $a-1 \ll 1$.  Therefore, we can conclude that
\begin{equation}
\label{eq:s_minus_expansion2}
s^*_- =  \left(1+ \frac{b}{(1 - \sqrt{a})^2}\right)^{-1} - \left(\frac{b}{\sqrt{a}-1} \right)\mathcal{D} + ...
\end{equation}
This result agrees with $s^*_-$ computed numerically from the exact maximum of $s_1(\lambda)$ from Eq.~(\ref{eq:s_of_lambda_nonzero_D}) for $\lambda <0$.  The $\mathcal{D}$-correction is the quantity we seek - the slope of $s^*_-$ versus $\mathcal{D}$ at zero $\mathcal{D}$.  This slope goes to infinity as $\epsilon \rightarrow 0$, since at $\epsilon =0$, $s^*_- \sim -\mathcal{D}^{1/3}$, which has an infinite slope at zero $\mathcal{D}$.  

We now address the negative $\epsilon$ case (i.e.~$a<1$).  In this domain of $\epsilon$, the following is the root of Eq.~(\ref{eq:cubic_equation}) that is real and negative:
\begin{equation}
\label{eq:solution_negative_epsilon}
\left(\frac{b}{\mathcal{D}}\right)^{1/3}\left[(-1)^{2/3}\left(-1 + \sqrt{1+ \left(\frac{\epsilon}{3b^{2/3} \mathcal{D}^{1/3}}\right)^3}\right)^{1/3} + \frac{(-1)^{1/3}\left(\frac{\epsilon}{3b^{2/3} \mathcal{D}^{1/3}}\right)}{\left(-1 + \sqrt{1+ \left(\frac{\epsilon}{3b^{2/3} \mathcal{D}^{1/3}}\right)^3}\right)^{1/3}}\right],
\end{equation}
One can show that when $\mathcal{D} \ll \frac{\epsilon^3}{27b^2}$,  $\lambda^*_- \sim \frac{\sqrt{-\epsilon}}{\mathcal{D}^{1/2}}$, and when $\mathcal{D} \gg \frac{\epsilon ^3}{27b^2}$, $\lambda^*_- \sim  -\left(\frac{2 b}{\mathcal{D}}\right)^{1/3}$ as before.  Again, at larger $\mathcal{D}$ there is a second crossover to the $\sim c_1(1+\epsilon,b)\mathcal{D}^{-1/2}$ asymptotic behavior.  The position of the first crossover scales like $\epsilon^3$, and at $\epsilon=0$, the $\left(\frac{2 b}{\mathcal{D}}\right)^{1/3}$ asymptotic extends all the way to $\mathcal{D}=0$.  
Interestingly, there is a $\mathcal{D}^{-1/2}$ scaling on both sides of the $\mathcal{D}^{-1/3}$ scaling, which disappears altogether with large-enough $\epsilon$, when the two crossovers meet.  In $\mathcal{D} \ll \frac{\epsilon^3}{27b^2}$ regime - the very first scaling behavior around $\mathcal{D}=0$ before the first crossover - the speed will behave as $s^*_- \sim -2\sqrt{1-a}\sqrt{\mathcal{D}}$.  This result also agrees with the low$-\mathcal{D}$ tail of $s^*_-$ computed numerically from the exact maximum of $s_1(\lambda)$ from Eq.~(\ref{eq:s_of_lambda_nonzero_D}) for $a < 1$.

If, instead of $b$ being fixed, it was some given fraction of $a$, i.e. $b = ra = r(1+\epsilon)$ (for example, $r=1$ in Fig.~4 in the main text), all of the above results for $\lambda^*_-$ and $s^*_-$ that are given as series in $\epsilon$ (or $a-1$) would hold at the leading-order in $\epsilon$ by replacing $b \rightarrow r$. 


\subsubsection{Stalling condition for the upwind front.}
Equation~(\ref{eq:s_minus_expansion2}) lets us derive $\mathcal{D}_{stall}$. We can see that for $a-1>0$, 
\begin{equation}
\mathcal{D}_{\mathrm{stall}} \approx \frac{(a-1)^3}{8b^2}.
\end{equation}
As $a$ increases, the position of $\mathcal{D}_{\mathrm{stall}}$ also increases.  Eventually,  it enters the regime where large-$\mathcal{D}$ theory should apply, so $\mathcal{D}_{\mathrm{stall}}$ will come from the condition $v_{\mathrm{eff}} = 2\sqrt{\mathcal{D}_{\mathrm{eff}}}$ (recall that $\mathcal{D}_{\mathrm{eff}}$ is proportional to $\mathcal{D}$).  Using the simpler formulas, Eq.~(8)-(9) from the main text, we predict
\begin{equation}
\mathcal{D}_{\mathrm{stall}} = \frac{\left(1+\frac{a-b-2}{\sqrt{a^2+2 a (b-2)+(b+2)^2}}\right)^2}{\left(2-a-b+\sqrt{a^2+2 a (b-2)+(b+2)^2}\right)^2}.
\end{equation}
This agrees well with $\mathcal{D}_{\mathrm{stall}}$ obtained from the more exact Eqs.~(\ref{eq:Scumbersome1})-(\ref{eq:Scumbersome2}), except for $b$ very close to $0$ and $a$ close to $1$ (although, the latter is also an approximation that works worst in these conditions).  At large $a$ and $b$, both results have the following leading-order behavior:
\begin{equation}
\mathcal{D}_{\mathrm{stall}} = \left(\frac{a}{2b}\right)^2.
\end{equation}
This shows that when $a$ and $b$ are both varied in such a way so as to keep their ratio constant, $\mathcal{D}_{\mathrm{stall}}$ will have this limiting value.  For example, when $a=b$, $\mathcal{D}_{\mathrm{stall}}$ will approach $1/4$ that we see in Fig.~4b in the main text.


\subsubsection{Common intersection point at $\mathcal{D} = 1/4$.}
When $\mathcal{D}=1/4$, $\lambda^*_+ = 2$ solves Eq.~(\ref{eq:deriv}) for any $a$ and $b$.  Substituting these values into Eq.~(\ref{eq:s_of_lambda_nonzero_D}) gives the speed of $1$, for any $a$ and $b$.  More details can be found in our ``Large-$\mathcal{D}$'' discussion.


\subsubsection{Crossover into the $\sqrt{\mathcal{D}}$ behavior}
We have discussed this crossover at length for $a=1$, but not for general parameters.  This crossover is important because it addresses the question under what conditions the behavior of the model becomes FKPP-like.  
This is a challenging topic, because a crossover can be defined in multiple ways, each predicting a somewhat critical  $\mathcal{D}$.  Moreover, in addition to the characteristic position of the crossover itself, the width of the crossover region is an additional quantity that characterizes crossover physics.  The following discussion will consider crossovers in different regions of $(a,b)$ space, from which an overall picture should emerge.  We will show that the crossover surface, as a function of $a$ and $b$, is complicated - reflecting the richness of the phenomenology that arises when advection and diffusion are in competition - something that does not happen in a reaction-diffusion model with an advective term.
\\

We first discuss the downwind front.  When both $a \ll 1$ and $b \ll 1$ there is a sharp crossover at $\mathcal{D} = 1/4$, such that $s^*_+ \rightarrow 1$ for $\mathcal{D} < 1/4$ and $s^* \rightarrow 2\sqrt{\mathcal{D}}$ for $\mathcal{D} > 1/4$ as both $a \rightarrow 0$ and $b \rightarrow 0$.   To see this note that when $a=0$ and $b=0$, Eq.~(\ref{eq:s_of_lambda_equation}) predicts two branches: $s=1$ and $s = 1/\lambda + \mathcal{D} \lambda$.  The minimum value of the latter is $2\sqrt{D}$.  These two branches will intersect for $\mathcal{D} < 1/4$, and the intersection points take place at $\lambda = (1\pm\sqrt{1-4\mathcal{D}})/2\mathcal{D}$.  On the other hand, as $a \rightarrow 0$, $b \rightarrow 0$, the choice of branches is such that in between these two intersection points, $s_1$ is given by $1$, and otherwise, it is given by $1/\lambda + \mathcal{D}\lambda$ (see remark about singular limits in \cite{Footnote_SM}).  Making $a$ and $b$ nonzero will create a gap between the branches, but as $a,b \rightarrow 0$, the gap is very small, and thus, for $\mathcal{D} < 1/4$, $s^*_+ \approx 1$.  Therefore, when $a\ll 1$ and $b \ll 1$, $s^*_+ \approx 1$ for $\mathcal{D} < 1/4$, and $s^*_+ \approx 2\sqrt{\mathcal{D}}$ for $\mathcal{D} > 1/4$, i.e.  the crossover to the $\sqrt{\mathcal{D}}$ behavior is very sharp, and takes place at $\mathcal{D} = 1/4$.  This is the regime when the mechanism that gives the biggest speed dominates.  By a similar argument, when only $a$ is small, the crossover will take place at $\mathcal{D} = (4(1+b))^{-1}$.  For $\mathcal{D}$ somewhat below this value, $s^*_+ \approx 1- \frac{2b\mathcal{D}}{1-\sqrt{1-4\mathcal{D}(1+b)}}$, while for $\mathcal{D}$ above this value, $s^*_+ \approx 2\sqrt{\mathcal{D}}$, and the transition between these behaviors takes place in a very narrow region of $\mathcal{D}$, i.e. it is also a sharp crossover.  As $b$ increases, there is a greater tendency for the GL to dominate - particles spend less and less time in the AL, and we expect the crossover to shift to lower and lower $\mathcal{D}$.  We will examine the regime of large $b$ below.  

\begin{figure}[ht]
\includegraphics[width=4in]{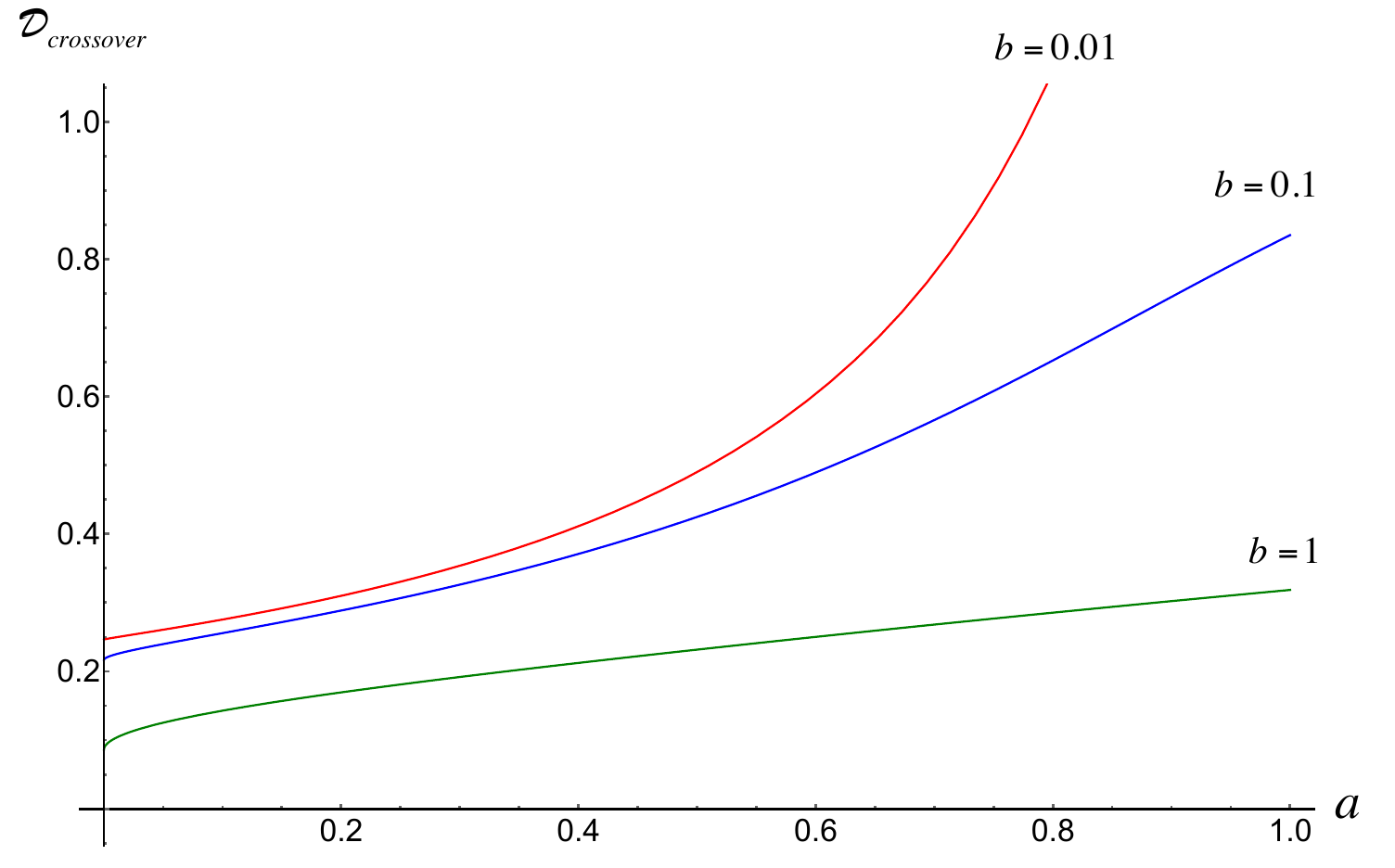}
\caption{Crossover defined from the intersection of $\frac{1}{1+\frac{b}{\left(1 \pm \sqrt{a}\right)^2}} \pm \frac{b \mathcal{D}}{\sqrt{a} \pm 1}$ and $v_{\mathrm{eff}} \pm 2\sqrt{\mathcal{D}_{\mathrm{eff}}}$.}
\label{fig:Crossover_Fig}
\end{figure}

Another approach to estimating the crossover is by examining the intersection of the small-$\mathcal{D}$ limit given by $s^*_{\pm} = \frac{1}{1+\frac{b}{\left(1 \pm \sqrt{a}\right)^2}} \pm \frac{b \mathcal{D}}{\sqrt{a} \pm 1}$ (see previous sub-section, where this result is derived;  note that at $a=0$, these are the first two terms in the Taylor expansion of $s^*_+$ from the previous paragraph) with the large-$\mathcal{D}$ limit $v_{\mathrm{eff}} \pm 2\sqrt{\mathcal{D}_{\mathrm{eff}}}$, where the parameter dependence for $\mathcal{D}_{\mathrm{eff}}$ and $v_{\mathrm{eff}}$ given by Eqs.~(\ref{eq:Scumbersome1})-(\ref{eq:Scumbersome2}).  For example, recall that in the large-$\mathcal{D}$ asymptotic, $\mathcal{D}_{\mathrm{eff}} \rightarrow \mathcal{D}$ and $v_{\mathrm{eff}} \rightarrow 0$ as $a \rightarrow 0$ for any $b$ (see also Fig.~\ref{fig:veff_and_Deff}).  Such asymptotic matching argument also suggests a sharp crossover for small $a$ and $b$, as well as the lowering of the crossover point as $b$ grows.  This method of estimating the crossover will fail for $b$ somewhat above $1$, as these two asymptotics will not actually intersect.  However, it can be used to gain a qualitative understanding of how the crossover varies with parameters in the neighborhood of the $(a,  b) = (0,0)$ corner.  This is shown in Fig.~ \ref{fig:Crossover_Fig}.  In the previous paragraph, we used a different definition of a crossover, and applied it to $a \ll 1$.  Taking this alternative definition, and setting $a=0$, will not produce the identical result - it will lie somewhat below $(4(1+b))^{-1}$.  This is a consequence of the fact that crossovers are often ambiguous and can be defined in various ways.  Whereas the previous definition will work very well for $a \ll 1$, the current definition can be extended to values larger than $1$.  

When $a \gg 1$, the slope of $s^*_+(\mathcal{D})$ at $\mathcal{D}=0$ is small when $b \ll \sqrt{a}$.  In this regime, $\mathcal{D}_{\mathrm{eff}} \rightarrow 0$ and $v_{\mathrm{eff}} \rightarrow 1$. A log-log plot of the numerically-obtained $s^*_+$ reveals that the crossover moves to lower $\mathcal{D}$ as $a$ increases.  However, this crossover is not very meaningful, because $s^*_+$ grows very slowly, and remains $\approx 1$ until a very large $\mathcal{D}$ (see inset in Fig.~4a in the main text).  As $b$ increases, eventually becoming $\gg \sqrt{a}$, the crossover moves to an even smaller value of $\mathcal{D}$, as expected physically. 

In summary, the crossover $\mathcal{D}$ into the FKPP regime for the downwind front $\rightarrow 1/4$ when $a \ll 1$ and $b \ll 1$.   Moving away from this corner in parameter space, the crossover will decrease with increasing $b$ and increases with increasing $a$, but when $a$ is large, increasing $a$ at fixed $b$ leads to $s^*_+$ that is essentially $1$ until a very large $\mathcal{D}$.  When $b$ increases at fixed $a$ or at co-varying $a$ while their ratio $a/b$ is fixed, the crossover decreases as $b^{-2}$ (see below).  As discussed in the previous section, the crossover diverges at $a=1$ as $b \rightarrow 0$ - indicating an extreme departure from the FKPP behavior.  When both $a$ and $b$ grow much beyond $1$, the crossover decreases.  Evidently, the crossover surface is quite complicated.  The breaking of the Gallilean invariance due to addition of a separate competing transport channel breaks the applicability of a reaction-diffusion picture of front propagation in a highly non-trivial way.  
\\

We briefly discuss the upwind front.  As discussed above, when $a<1$, the diffusive mechanism is essential for propagating the front.  Therefore, with the exception of a small region of $(a,b)$ parameter space, $s^*_-$ will scale as $-\sqrt{\mathcal{D}}$ when $a<1$.  For $a \gg 1$, all of the downwind front conclusions hold - we know this from the numerical study of $s^*_-$, the asymptotic matching argument and the large-$b$ analysis that we will now describe.
\\

As $b \rightarrow \infty$, both $\lambda^*_{\pm}$ and $s^*_{\pm}$ approach FKPP values.  Letting $b = 1/\epsilon$, $\lambda^*_{\pm} = \frac{\pm 1}{\sqrt{\mathcal{D}}} + c' \epsilon^2$, substituting into Eq.~(\ref{eq:deriv}) for $s_1$ and expanding in $\epsilon$, we obtain
\begin{displaymath}
\left(8a + \frac{2a}{\mathcal{D}} - \frac{8a}{\sqrt{\mathcal{D}}} \pm 4c' \sqrt{\mathcal{D}} \right)\epsilon^2 = 0
\end{displaymath}
Solving for $c'$ we have
Thus,
\begin{equation}
\label{eq:lambda_large_b}
\lambda^*_{\pm} = \pm \frac{1}{\sqrt{\mathcal{D}}} \mp \frac{a\left(1-4\sqrt{\mathcal{D}} \mp 4\mathcal{D}\right)}{2b^2\mathcal{D}^{3/2}}
\end{equation}
(if instead we guessed that a correction term is $O(1/b)$, the resulting $c'$ will have to be $0$).   As $b \rightarrow \infty$ at fixed $\mathcal{D}$, the FKPP result dominates.  As $b$ is decreased, this result begins to break down.  Equating the two terms suggests that at a given $\mathcal{D}$, the correction becomes important when $b$ is 
\begin{displaymath}
b = \sqrt{2a \left(1 \mp \frac{1}{\sqrt{4\mathcal{D}}}\right)^2}
\end{displaymath}
Thus, sitting at a given $\mathcal{D} \ll 1/4$, $b$ must scale as $\mathcal{D}^{-1/2}$ in order for the FKPP scaling of $\lambda^*_{\pm}$ to begin to break down.  Equivalently, the crossover value of $\mathcal{D}$ scales like $\sim b^{-2}$ at large $b$ and small $\mathcal{D}$  (i.e. $ \mathcal{D} \ll 1/4$).

Substituting the result in Eq.~(\ref{eq:lambda_large_b}) into Eq.~(\ref{eq:s_of_lambda_nonzero_D}), and Taylor-expanding in $\epsilon$ we find.
\begin{equation}
s^*_{\pm} = \pm \sqrt{\mathcal{D}} + \frac{a \mp 2a \sqrt{\mathcal{D}}}{b}.
\end{equation}
Again, for the FKPP result to start to break down at a given $a$ and $\mathcal{D} \ll 1/4$, $b$ has to scale as $\mathcal{D}^{-1/2}$.

A similar procedure with $a = rb$, $b \gg 1$ will give 
\begin{equation}
s^*_{\pm} = \frac{r}{1+r} \pm \frac{2\sqrt{\mathcal{D}}}{1+r}.
\end{equation}
This will hold for all $\mathcal{D}$ as $b \rightarrow 0$.


\newpage
\section{SII. Full solution of the linearized model with zero diffusion}
\label{sec:FullLin}
In this section we study the linearized problem.  An exact solution to a $\delta$-function IC will be given.  The linearization of the non-dimensionalized Eqs.~(1)-(2) of the main text around the unstable state $\rho = \sigma=0$, gives
\begin{eqnarray}
\label{eq:Appendix1} \frac{\partial \rho}{\partial t} &=& -\frac{\partial \rho}{\partial x} + a \sigma - b \rho \\
\label{eq:Appendix2} \frac{\partial \sigma}{\partial t}  &=& (1- a) \sigma + b \rho.
\end{eqnarray}
We solve the problem by a Fourier Transform method.  Let 
\begin{eqnarray}
\label{eq:rho-eq} \rho &=& A_{\rho}(k) e^{i\omega(k) t} e^{-ikx} \\
\label{eq:theta-eq} \sigma &=& A_{\sigma}(k) e^{i \omega(k) t}e^{-ikx}
\end{eqnarray}
The $\omega(k)$ and $\vec{A}(k)$ satisfy the following eigen-problem:
\begin{equation}
\omega \left(\begin{array}{c} A_{\rho} \\ A_{\sigma}\end{array} \right) = \left(\begin{array}{cc}k + ib & -ia \\ -ib & i(a-1) \end{array}\right) \left(\begin{array}{c} A_{\rho} \\ A_{\sigma}\end{array} \right)
\end{equation}
The eigenvalues are given by
\begin{equation}
\label{eq:omega-eq} \omega = \frac{k-i(1-a-b)}{2} + \frac{1}{2}\sqrt{\left(k + i(1-a+b)\right)^2 - 4ab}.
\end{equation}
We now define the two branches.  The square root term can be expressed as 
\begin{eqnarray}
&\frac{1}{2}\sqrt{(k-k_1)(k-k_2)}, \\
\mbox{where   } &k_1 = -i(1-a+b) - 2\sqrt{ab}, \nonumber \\
&k_2 = -i(1-a+b) + 2\sqrt{ab}, \nonumber
\end{eqnarray}
These $k=k_1$ and $k=k_2$ are branch points.  We have the freedom in how we place the branch cut - a construction that ensures single-valuedness.   Let $k-k_1 = \rho_1 e^{i\phi_1}$ and $k-k_2 = \rho_2 e^{i\phi_2}$, Fig.~\ref{fig:kplane}.   We define each $\phi$ to be in $[-\pi,\pi]$.  With this definition of angles, a path along a loop that encloses \emph{both} branch points will not encounter multi-valuedness, but a path around each single branch point will encounter a discontinuity of the exponential factor along a segment between $k_1$ and $k_2$.  Therefore, with this definition of $\phi$s, the branch cut is a straight segment located between $k_1$ and $k_2$.  Then
\begin{equation}
\label{eq:Discontinuities}
\begin{array}{ccccccc}
 & ``+" \mbox{branch of the} \sqrt{\mbox{  }} &  & ``-" \mbox{branch of the} \sqrt{\mbox{  }}  &  & &  \Delta_- - \Delta_+ \\
\mbox{Immediately above the cut} & \frac{i}{2}\sqrt{\rho_1 \rho_2}  &  & -\frac{i}{2}\sqrt{\rho_1 \rho_2} & & &  -i \sqrt{\rho_1 \rho_2}\\
\mbox{Immediately below the cut} & -\frac{i}{2}\sqrt{\rho_1 \rho_2}  &  & \frac{i}{2}\sqrt{\rho_1 \rho_2} & & & i \sqrt{\rho_1 \rho_2}
\end{array}
\end{equation}
Had we chosen a different definition of $\phi$s, the definition of a cut (and of branches) would also change.  We will denote the two branches by $\pm$.

\begin{figure}[ht]
\includegraphics[width=2.6in]{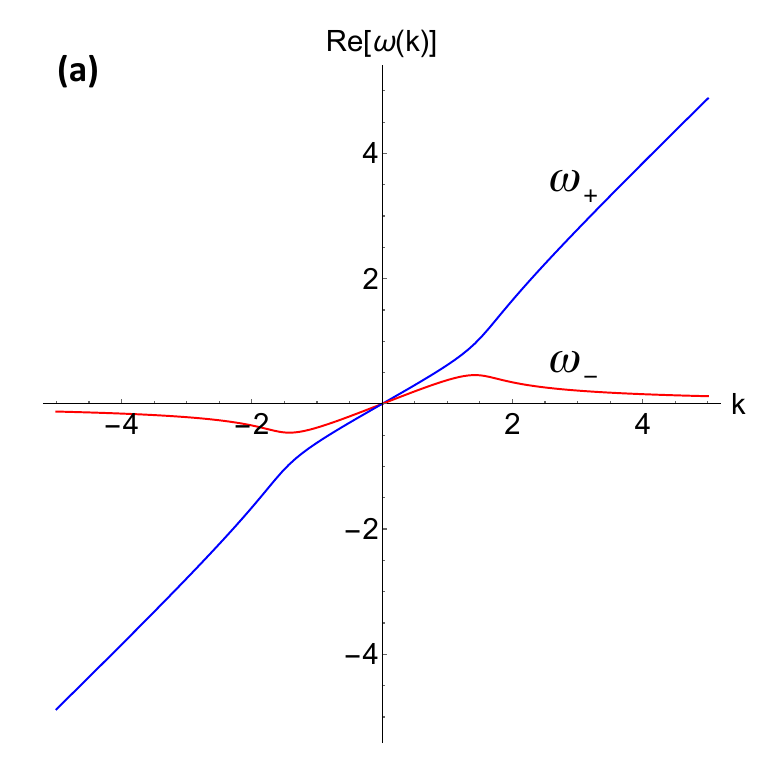}
\includegraphics[width=2.6in]{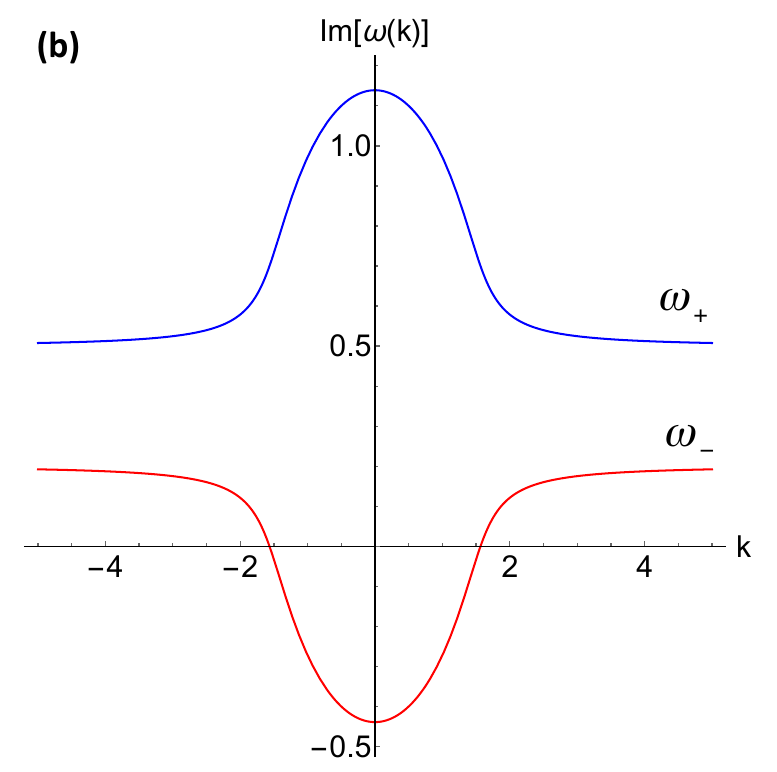}
\caption{(a): $Re(\omega)$ vs. $k$.  (b):  $Im(\omega)$ vs. $k$.   Here the parameters were $a=1.2$, $b=0.5$.  }
\label{fig:DispRel}
\end{figure}

\begin{figure}[ht]
\includegraphics[width=3in]{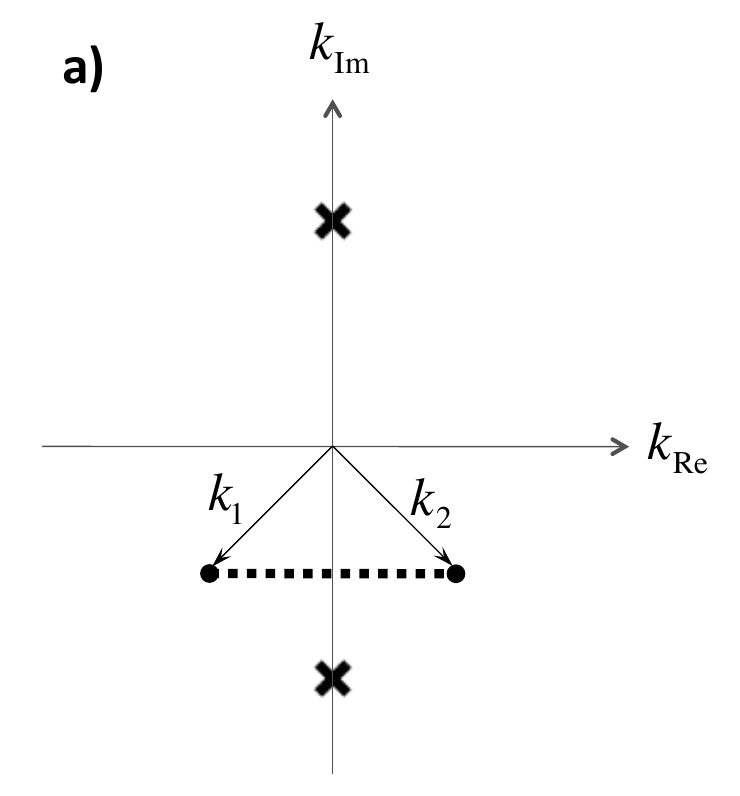}
\includegraphics[width=3in]{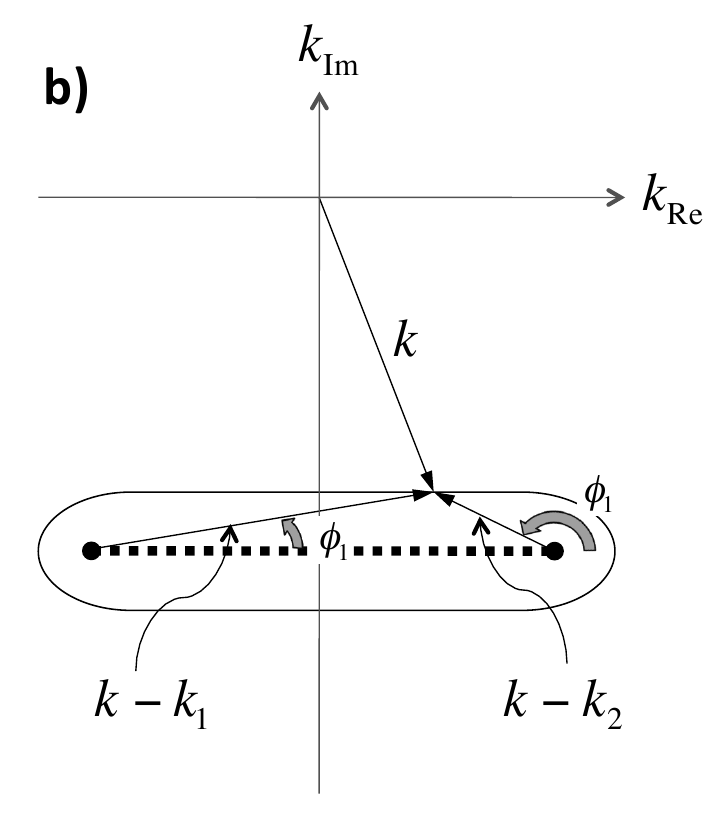}
\caption{a) Features in the k-space: dashed - branch cut, crosses - poles.  b) Cartoon of a contour around the branch cut.  The contour lies infinitesimally close to the cut.}
\label{fig:kplane}
\end{figure}

Unless $a=1+b$, a branch cut is not located on the real axis.  When $a=1+b$, the portion of the real axis from $k=-\sqrt{ab}$ to $k = \sqrt{ab}$ still belongs to either one or the other branch.  Thus, in plotting a dispersion relation versus the real $k$, no branch is crossed.  A typical plot of a dispersion relation - $\omega_{\pm}$ versus (a real) $k$, is shown in Fig.~\ref{fig:DispRel}.
The ``$+$'' branch of the $Im[\omega(k)]$ curve is always positive.  For $a<1$, the ``$-$'' branch is entirely negative, and for $a>1$, it is negative only over a range of $|k|$ below a certain value.  Since the growth rate of a $k$-mode is given by $e^{-Im[\omega(k)]t}$, this indicates that all modes are unstable for $a<1$, but large $k$ modes become stable when $a>1$.  Physically, $a>1$ means the rate of biotic mass production is less than the rate of leaving into the advective layer (AL).  Thus, as $a$ becomes larger and larger, biotic particles (such a spores) spend less and less time on the growth layer (GL).  In the limit of infinite $a$  they do not spend any time on the GL, and do not contribute to the growth, so $\sigma=\rho=0$ becomes a stable state.  In fact, the whole ``$-$'' branch becomes marginally-stable (zero).  Similarly, if $a>1$ and $b$ goes to zero, the state $\sigma=\rho=0$ also approaches marginality.  The lowest value of the ``$-$'' branch is 
\begin{equation}
-Im[\omega_{-}(k=0)] = \frac{1-a-b}{2} + \frac{1}{2}\sqrt{\left(1-a+b\right)^2 + 4ab}.
\end{equation}
It is the inverse of the characteristic time scale for the growth of the most unstable $(k=0)$ mode.  Notice that this equals $\lambda s_1(\lambda)$ at $\lambda=0$, where $s_1$ is the $+$ solution in Eq.~(\ref{eq:s_of_lambda_zero_D}).  

The corresponding eigenvectors are given by
\begin{eqnarray}
\left(\begin{array}{c} A^{\pm}_{\rho} \\ A^{\pm}_{\sigma}\end{array} \right) = \left(\begin{array}{c} \frac{C_{\pm}}{\sqrt{a^2 - \Delta^2_{\pm}}} \\  \frac{iC_{\pm}\Delta_{\pm}}{\sqrt{a^2 - \Delta^2_{\pm}}}  \end{array}\right), \\
\nonumber \\
\mbox{where     } \Delta_{\pm} \equiv \omega_{\pm}(k) - k - ib \nonumber
\end{eqnarray}
The $C_{\pm}$ are sign factors, and they will cancel out with sign factors in Fourier coefficients below.  The general solution is an an integral over all $k$ of a linear combination of these two solutions:
\begin{equation}
\label{eq:sol-general}
\left(\begin{array}{c} \rho(x,t) \\ \sigma(x,t) \end{array}\right) = \frac{1}{2\pi}\int_{-\infty}^{\infty} \left(\tilde{\alpha}(k) \vec{A}^+(k) e^{i\omega_+(k) t} + \tilde{\beta}(k)\vec{A}^-(k) e^{i\omega_-(k) t}  \right) e^{-ikx}\,dk.
\end{equation}
The coefficients $\tilde{\alpha}(k)$ and $\tilde{\beta}(k)$ are determined from the ICs.  Let the Fourier Transform of the IC be $\tilde{\rho}_0(k)$ and $\tilde{\sigma}_0(k)$.  Then
\begin{eqnarray*}
\tilde{\rho}_0(k) &=& \tilde{\alpha}(k) A^+_{\rho}(k) + \tilde{\beta}(k) A^-_{\rho}(k), \\
\tilde{\sigma}_0(k) &=& \tilde{\alpha}(k) A^+_{\sigma}(k) + \tilde{\beta}(k) A^-_{\sigma}(k). 
\end{eqnarray*}
Solving for $\tilde{\alpha}(k)$ and $\tilde{\beta}(k)$, and substituting into Eq.~(\ref{eq:sol-general}) we end up with 
\begin{equation}
\rho(x,t) = \frac{1}{2\pi} \int_{-\infty}^{\infty} \frac{(\tilde{\rho}_0\Delta_- + i a \tilde{\sigma}_0)e^{i\omega_+t} - (\tilde{\rho}_0\Delta_+ + i a \tilde{\sigma}_0)e^{i \omega_-t}}{\Delta_- -  \Delta_+}e^{-ikx}\,dk
\end{equation}
and $\tilde{\rho}_0$, $\tilde{\sigma}_0$, $\Delta_{\pm}$ and $\omega_{\pm}$ are functions of $k$, as defined above.  There is also (a more complicated) expression for $\sigma(x,t)$, but it is easier to extract $\sigma$ using Eq.~(\ref{eq:Appendix1}) if we know $\rho$.  The integral for $\rho$ can be re-written as
\begin{eqnarray}
\rho(x,t) &=& \rho_{AL}(x,t) + \rho_{GL}(x,t) \\
\label{eq:rhoadv} \rho_{AL}(x,t) &=& \frac{1}{2\pi} \int_{-\infty}^{\infty}  \tilde{\rho}_0(k) \left(\frac{\Delta_- e^{i\omega_+ t} - \Delta_+ e^{i\omega_- t}}{\Delta_- -  \Delta_+}\right) e^{-ikx} \,dk  = \rho^+_{AL} - \rho^-_{AL},\\
\label{eq:rhognd} \rho_{GL}(x,t) &=& \frac{ia}{2\pi} \int_{-\infty}^{\infty}  \tilde{\sigma}_0(k) \left( \frac{e^{i\omega_+ t} - e^{i\omega_- t}}{\Delta_- -  \Delta_+}\right) e^{-ikx} \,dk  = \rho^+_{GL} - \rho^-_{GL}.
\end{eqnarray}
Here $\rho_{AL}$ is a contribution to $\rho(x,t)$ from the IC in the AL, and $\rho_{GL}$ is a contribution to $\rho(x,t)$ from the IC on the GL.  In this paper we will only be concerned with ICs on the GL.  Therefore, to lighten the notation, the subscript ``GL'' in $\rho_{GL}$ will be dropped, unless stated explicitly. 

We will consider a special point-source initial distribution,
\begin{equation}
\sigma_0(x) = M\delta(x),
\end{equation}
that has a fourier transform given by $M$ in all of $k$-space.  An exact solution will be given for this type of IC.
We will also consider an exponentially-localized ICs
\begin{equation}
\label{eq:localized_IC}
\sigma_0(x) = \frac{M\mu}{2} e^{-\mu|x-x_0|}.
\end{equation}
$x_0$ can be set to $0$ without loss of generality, since in this problem the coefficients $a$ and $b$ are homogeneous.  The fourier transform of such an IC is
\begin{equation}
\tilde{\sigma}_0(k) =  \frac{M}{1+\left(k/\mu\right)^2}.
\end{equation} 
The solution with this IC in the limit $\mu \rightarrow \infty$ should be identical to the solution with $\delta$-function IC.  The behavior of other IC that have a finite, but non-point support should approach the behavior of solutions with a $\delta$-function IC at distances much greater than the extent of this support.  Since the main interest of this paper concerns with long-range transport, we will not make explicit calculations for other compact IC.  We remark that ICs with power law tails gives rise to accelerating wave-fronts, while gaussian ICs behave as a point-sources.   
\\

The integrals in Eqs.~(\ref{eq:rhoadv}) and (\ref{eq:rhognd}) are taken along the real line in $k$-space, but close the contour (which is possible, since the branch cuts are finite segments with our definition of branches of the square root) we have to discuss the behavior of $\omega_{\pm}$ as $|k| \rightarrow \infty$.  There are two branches of $\omega$ and they differ by a sign.  Thus, as $|k|$ gets large, $\omega_{\pm} \sim \frac{k}{2} \pm \frac{k}{2} + O(\frac{1}{k})$.  So $\omega_+ \sim k$, and 
\begin{displaymath}
e^{i\omega_+ t}e^{-ikt} \sim e^{ik(t-x)}
\end{displaymath}
at large $|k|$.  Evidently, the contour of the $\rho^+$-integral will have to be closed in the lower half-plane for $x>t$ and in the upper half-plane for $x<t$.  The $\omega_-$ branch does not have an important $k$-dependence at large $|k|$, so 
\begin{displaymath}
e^{i\omega_- t}e^{-ikx} \sim e^{-ikx}
\end{displaymath}
at large $|k|$.  The contour of the $\rho^-$-integral will have to be closed in the lower half-plane for $x>0$ and in the upper half-plane for $x<0$.   Table \ref{Contours} summarizes the contours.  
\begin{table}[ht]
\begin{ruledtabular}
\begin{tabular}{lccc}
  & $x<0$ & $0<x<t$ &  $x>t$ \\
  \hline
  $\rho^+$-integral & Above & Above & Below \\
  $\rho^-$-integral & Above & Below & Below  \\
\end{tabular}
\end{ruledtabular}
\caption{Summary of integration contours}\label{Contours}
\end{table}

The two types of features of the integrand that these contours may enclose are: \emph{poles} at $k = \pm i\lambda$ that are present only for exponential ICs, but not compact ICs, and a \emph{branch cut segment} that is present for any IC, Fig.~\ref{fig:kplane}.   Its center is located at position $-i(1-a+b)$, so it will be located in the upper half-plane for $a>1+b$ and in the lower half-plane for $a<1+b$.   A semi-circular contour may be shrunk to enclose only these features.  Thus, if a contour encloses a pole and a branch cut, there will be a pole contribution and a branch cut contribution.  

With these considerations in mind, the total integral (this is in fact true for either $\rho_{GL}$ or $\rho_{AL}$) is summarized in Table \ref{Total rho}:
\begin{table}[ht]
\begin{ruledtabular}
\begin{tabular}{l|c|c|c}
  & $x<0$ & $0<x<t$ &  $x>t$ \\
  \hline
  $a<1+b$ & $\rho^+$-pole$_+ - \rho^-$-pole$_+$ &$\rho^+$-pole$_+ - (\rho^-$-pole$_- + \rho^-$-BC$_-)$ & $\begin{array}{c} (\rho^+\mbox{-BC}_- + \rho^+\mbox{-pole}_-) \\
   - (\rho^-\mbox{-BC}_- + \rho^-\mbox{-pole}_-)\end{array}$ \\
   \hline
  $a>1+b$ & $\begin{array}{c} (\rho^+\mbox{-BC}_+ + \rho^+\mbox{-pole}_+) \\
   - (\rho^-\mbox{-BC}_+ + \rho^-\mbox{-pole}_+)\end{array}$ & $(\rho^+$-BC$_+ + \rho^+$-pole$_+) - \rho^-$-pole$_-$ & $\rho^+$-pole$_- - \rho^-$-pole$_-$ 
\end{tabular}
\end{ruledtabular}
\caption{Total $\rho$ for different regions of space.}\label{Total rho}
\end{table}
\\
Here the pole$_+$ and pole$_-$ refers to the position of the pole - the one in the upper half-plane or in the lower half-plane respectively;  same for a branch cut.  It will turn out that the branch cut contributions for either $x<0$ or $x>t$ will cancel and only poles contribute to the solution in these regions.  Thus, if we start with a localized IC (when there are no poles), $\rho$ (and $\sigma$) will be zero in these regions, as expected, since the wind cannot carry material backwards, and material also cannot arrive to a point $x$ faster than the wind (which has speed $1$ in these units).  Note: when $a=1+b$ the branch cut segment lies right on the real axis.  However, we may move the original contour off the real-axis by an infinitesimal amount, close the contour as specified above, and shrink it to enclose the branch cut segment and the pole.  Alternatively, we may treat the case $a=1+b$ as a limit, because it is unique - as we will see, the resulting limit for $\rho$ and $\theta$ is the same, whether the limit approaches $1+b$ from above or from below.

The contour around the cut consists of a straight line segment right above the cut, the straight line segment right below the cut, and two infinitesimal semi-circles around each end of the cut.  It is easy to show that their contributions goes to zero in the limit as the radius of these semi-circles go to zero.  The directions of integration above and below the cut are opposing each other, but these pieces do not cancel because because the value of both branches of $\omega$ differs right above and right below the cut, as specified in Eq.~(\ref{eq:Discontinuities}).  Then
\footnotesize
\begin{eqnarray}
\rho^+\mbox{-BC}_- = \frac{ia}{2\pi} \int_{k_1}^{k_2} \frac{M}{1+\left(k/\mu\right)^2} \frac{e^{-ikx} e^{i\frac{\left[k-i(1-a-b) \right]t}{2}} e^{-\frac{t}{2} \sqrt{\rho_1 \rho_2}}}{-i \sqrt{\rho_1 \rho_2}} \,dk 
+ \frac{ia}{2\pi} \int_{k_1}^{k_2} \frac{M}{1+\left(k/\mu\right)^2} \frac{e^{-ikx} e^{i\frac{\left[k-i(1-a-b) \right]t}{2}} e^{\frac{t}{2} \sqrt{\rho_1 \rho_2}}}{-i \sqrt{\rho_1 \rho_2}} \,dk, \nonumber \\  \\ \nonumber \\
\rho^-\mbox{-BC}_- = \frac{ia}{2\pi} \int_{k_1}^{k_2} \frac{M}{1+\left(k/\mu\right)^2} \frac{e^{-ikx} e^{i\frac{\left[k-i(1-a-b) \right]t}{2}} e^{\frac{t}{2} \sqrt{\rho_1 \rho_2}}}{-i \sqrt{\rho_1 \rho_2}} \,dk 
+\frac{ia}{2\pi} \int_{k_1}^{k_2} \frac{M}{1+\left(k/\mu\right)^2} \frac{e^{-ikx} e^{i\frac{\left[k-i(1-a-b) \right]t}{2}} e^{-\frac{t}{2} \sqrt{\rho_1 \rho_2}}}{-i \sqrt{\rho_1 \rho_2}} \,dk. \nonumber \\
\end{eqnarray}
\normalsize 
If the branch cut is above the real axis, the integrals gain a minus sign, since the contour is oriented in the opposite direction, i.e. $\rho^+$-BC$_+ = -\rho^+$-BC$_-$ and $\rho^-$-BC$_+ = -\rho^-$-BC$_-$.  We see immediately that $\rho^+$-BC$_\pm= \rho^-$-BC$_\pm$.    All these considerations allow us to simplify Table~\ref{Total rho} as follows:
\begin{table}[ht]
\begin{ruledtabular}
\begin{tabular}{l|c|c|c}
  & $\mbox{Region-I}: x<0$ & $\mbox{Region-II}: 0<x<t$ &  $\mbox{Region-III}: x>t$ \\
  \hline
  \emph{Any} $a$ or $b$ & $\rho^+$-pole$_+ - \rho^-$-pole$_+$ &$(\rho^+$-pole$_+ - \rho^-$-pole$_-) - \rho^-$-BC$_-$ & $\rho^+$-pole$_- - \rho^-$-pole$_-$  
 \end{tabular}
\end{ruledtabular}
\caption{Total $\rho$ for different regions of space.}\label{Total rho2}
\end{table}
\\
Only poles contribute outside of $0<x<t$, as expected.  Letting $k = k_c + l \xi$, where $k_c = -i(1-a+b)$ and $l = 2\sqrt{ab}$ - half of the width of the branch cut, we have
\small
\begin{equation}
\label{eq:partial_soln}
-\rho^- \mbox{-BC}_- = \frac{aMe^{-\kappa (x-w t)}}{2\pi}\left[\int_{-1}^{1} \frac{e^{-iA\xi}e^{-B\sqrt{1-\xi^2}}}{\sqrt{1-\xi^2}} \frac{d\xi}{1+\left(\frac{k_c + l \xi}{\mu}\right)^2} + \int_{-1}^{1} \frac{e^{-iA\xi}e^{B\sqrt{1-\xi^2}}}{\sqrt{1-\xi^2}} \frac{d\xi}{1+\left(\frac{k_c + l \xi}{\mu}\right)^2}\right] 
\end{equation}
\normalsize
\begin{eqnarray}
\kappa &=& 1-a+b \nonumber \\
w &=& \frac{1-a}{1-a+b} \nonumber \\
A &=& 2\sqrt{ab}\left(x-\frac{t}{2}\right) \nonumber \\
B &=& t\sqrt{ab} 
\end{eqnarray}

We now confront the integrals in Eq.~(\ref{eq:partial_soln}).  The parameter $B$ becomes greater than $1$ for $t > 1/\sqrt{ab}$, after which time the second integral becomes rapidly dominant.  Now, let $\xi = \sin{y}$.  Then, the remaining integral is
\begin{equation}
\label{eq:SecondIntegral}
\mathcal{I} = \int_{-\pi/2}^{\pi/2} \frac{e^{-iA \sin{y}}e^{B \cos{y}} }{1+\left(\frac{k_c + l \sin{y}}{\mu}\right)^2} \,dy.
\end{equation}
We may extend the limits of integration to $[-\pi,\pi]$ with very little error, because $\cos{y}$ is negative in this extra region, and the exponent contains a large positive $B$.  Then, using the trigonometric identity we have
\begin{equation}
\label{eq:SecondIntegral2}
\mathcal{I} \approx \int_{-\pi}^{\pi} \frac{e^{\sqrt{B^2 - A^2}\cos{(y - y_0)}}}{1+\left(\frac{k_c + l \sin{y}}{\mu}\right)^2} \,dy 
\end{equation}
\begin{displaymath}
\mbox{where } \sin{y_0} = \frac{-iA}{\sqrt{B^2 - A^2}}, \mbox{    and } \cos{y_0} = \frac{B}{\sqrt{B^2 - A^2}} \nonumber
\end{displaymath}
The factor in the exponent is 
\begin{equation}
\sqrt{B^2 - A^2} = 2\sqrt{ab}\sqrt{x(t-x)}.
\end{equation}


\subsubsection{Point IC in the GL}
For a $\delta$-function IC, the denominator in the integrand in Eq.~(\ref{eq:SecondIntegral}) or (\ref{eq:SecondIntegral2}) will be simply $1$.  In this special case the answer turns out to be 
\begin{equation}
\mathcal{I} = 2\pi I_0\left(2\sqrt{ab}\sqrt{x(t-x)}\right),
\end{equation}
where $I_0$ is the modified Bessel function of the first kind.  And thus, 
\begin{equation}
\label{eq:rho_soln_linear_full}
\rho(x,t) = -\rho^- \mbox{-BC}_- \approx  aMe^{-\kappa (x-w t)} I_0\left(2\sqrt{ab}\sqrt{x(t-x)}\right).
\end{equation}
(see Eq.~(\ref{eq:partial_soln})).  Although this is technically an approximation, it works very well for all but the very early times ($\ll 1/\sqrt{ab}$).  We can substitute this result into Eqs.~(\ref{eq:Appendix1})-(\ref{eq:Appendix2}) and obtain 
\begin{equation}
\label{eq:sigma_soln_linear_full}
\sigma(x,t) \approx  \sqrt{ab}Me^{-\kappa (x-w t)} I_1\left(2\sqrt{ab}\sqrt{x(t-x)}\right)\frac{t-x}{\sqrt{x(t-x)}}.
\end{equation}

\subsection{Propagation speed and decay rates of tails}
We can extract the speed of the propagation as well as the decay rate of solution tails.  First, we use the asymptotic approximation for both $I_0(z)$ and $I_1(z) \sim \frac{e^z}{\sqrt{2\pi z}}$, so both $\rho(x,t)$ and $\sigma(x,t)$ have the following exponential behavior in $x$ and $t$: 
\begin{equation}
\rho(x,t), \sigma(x,t) \sim e^{-\kappa(x-w t) + 2\sqrt{ab}\sqrt{x(t-x)}}
\end{equation} 
We will now solve for $x_c(t)$ - the movement of the intersection of $\rho(x,t)$ with a contour of constant value $c$.  First ignoring the non-exponential factors, we have
\begin{displaymath}
\frac{2 a b t+\kappa  (c+\kappa  t w) \pm 2 \sqrt{a b \left(a b t^2-(c+\kappa  t (w-1)) (c+\kappa  t w)\right)}}{4 a b+\kappa ^2}
\end{displaymath} 
In the long time limit we get
\begin{equation}
x_c(t) = \frac{t}{1+\frac{b}{\left(1 \pm \sqrt{a}\right)^2}}
\end{equation}
The value of $c$ enters into the corrections that grow slower than $c$; it affects the time required to develop this asymptotic behavior linear in time.   Had we included the non-exponential prefactor, there would be a logarithmic in time correction to $x_c(t)$.  Thus, the sped will still relax to the above value.
\\

We can now switch to the co-moving variables $x = \frac{t}{1+\frac{b}{\left(1 \pm \sqrt{a}\right)^2}} + z$.  Ignoring the non-exponential prefactor, the result is 
\begin{equation}
\rho(z,t), \sigma(z,t) \sim e^{-\lambda^*_{\pm}z \mp \frac{\sqrt{a}\left(\lambda_{\pm}\right)^3}{4bt}z^2 + ...},
\end{equation}
with $\lambda^*_{\pm}$ being given by Eq.~(5) from the main text.  Thus, in the long-time limit, the leading-order term is $e^{-\lambda^*_{\pm}z}$ - a function of $x-s^*_{\pm}t$.  We also see the power-law relaxation to both speed and the stationary tail shape, in agreement with the general theory of fronts \cite{von Saarloos review_SM}.   Thus, we reproduced the formula for the front speed $s^*_{\pm}$ (Eq.~(4) in the main text) and the exponential contribution to the shape of the density tails (Eq.~(5) in the main text) - both of which we obtained by other methods (see the previous and the following sections).


\subsection{Zero growth case}
We now address an important special case of zero growth.  In terms of physical parameters, this is described by
\begin{eqnarray}
\label{eq:Appendix1_dimensional} \frac{\partial \rho}{\partial t} &=& -v_0\frac{\partial \rho}{\partial x} + \alpha \sigma - \beta \rho \\
\label{eq:Appendix2_dimensional} \frac{\partial \sigma}{\partial t}  &=& (1- \alpha) \sigma + \beta \rho.
\end{eqnarray}
We can retrace the entire derivation involving the contour integration.  For example, 
\begin{equation}
\rho \approx \frac{M \alpha}{v_0} e^{\left(\frac{\alpha-\beta}{v_0}\right)\left(x-\frac{\alpha v_0}{\alpha - \beta} t\right)}I_0\left(\frac{2\sqrt{\alpha \beta}}{v_0}\sqrt{x\left(v_0 t - x\right)}\right).
\end{equation}
We will find the identical result if we switch Eq.~(\ref{eq:rho_soln_linear_full}) to physical variables.  This solution describes a pulse, the width of which grows as $\sqrt{t}$  - so it has no defined stationary limit.  Note that if we take the general formula for profile width and switch to physical units, we get a profile width that grows as $1/\sqrt{\delta}$ as $\delta \rightarrow 0$.  So this is a permanently transient solution.  The speed of the profile peak is $\frac{v_0}{1+\beta/\alpha}$ - exactly as what the general formula predicts by taking $\delta \rightarrow 0$.   The magnitude of the peak decays as $1/t$.  In the vicinity of the peak, the solution is
\begin{equation}
\rho \approx \frac{\alpha M}{2v_0 \sqrt{\pi} t}\sqrt{\frac{\alpha + \beta}{\alpha \beta}} \exp{\left(-\frac{(\alpha + \beta)^3 \xi^2}{4\alpha \beta v_0^2 t}\right)}.
\end{equation}
The total mass under this profile is conserved - it is given by $\frac{M}{1+\beta\alpha}$.  Similarly to the speed, this is simply related to the fraction of the time spent in the advective layer.  We can also obtain this number by integrating Eqs.~(\ref{eq:Appendix1_dimensional})-(\ref{eq:Appendix2_dimensional}) over all space, and imposing the constraint that the sum of the mass in both layers is a constant.


\newpage
\section{SIII. Alternative derivation of speeds and decay lengths using the saddle-point method}
\label{sec:SaddlePtMethod}

We here show how to find the speed of pulled density fronts using a saddle point approximation, as an alternate technique. The motivation for doing this is to corroborate the results derived in the main paper. This technique was taken from a comprehensive review on front propagation by W. von Saarloos  \cite{von Saarloos review_SM}.  We first recite the derivation of the technique, and then apply it to our problem.

\subsubsection{Method Summary}

Consider a solution of the full, nonlinear equation propagating into the zero-density linearly-unstable state.  The smallness of the density in the leading part of the propagating profile suggests that dynamics of those regions, along with their properties (speed, decay rate, etc.) may be extracted from the linearized equations of motion.   This is not always true, because these leading tails are matched to the part of the density profile where the nonlinearities do become important.  However, in many cases this idea is correct.  A front of the nonlinear partial differential equation (PDE) is said to be \emph{pulled}, if its speed - defined by the speed measured at a constant density -  is identical to the speed under the linearized dynamics.

In light of this, we consider a scalar field $\phi(x,t)$, whose dynamics is determined by a translationally invariant linear PDE, obtained by linearizing the full equation of motion around the state $\phi=0$, and express this solution as a Fourier Transform:
\begin{equation}
\phi(x,t) = \frac{1}{ 2\pi} \int_{-\infty}^{\infty} dk \bar{\phi}_0(k) e^{-i \big( k x - \omega(k) t \big)}. 
\label{fourier_transform} 
\end{equation}
Here, $\omega(k)$ is the dispersion relation, which can be found, for instance, by substituting the Fourier ansatz $e^{i (kx-\omega t)}$ into the governing linear equations.   We assume $\phi=0$ is a linearly unstable solution, i.e. the amplitude of some of the Fourier modes grow in time under the linearized equations.  From  Eq.~\eqref{fourier_transform}, these are the modes with wavenumber $k$ for which $\text{Im} \; \omega(k) > 0$.   Because these mode are unstable, a typical localized IC will give rise to a disturbance that grows and spreads out in time under the linearized dynamics. We define the speed of the profile to be the asymptotic speed of the point of constant contour:
\begin{equation}
 s_0 = \lim_{t \rightarrow \infty} \frac{d x_{c_0}}{dt},
 \end{equation}
where $\phi(x_{c_0}, t) = c_0$.  The resulting speed is independent of the value of $c_0$ due to linearity of the governing PDE.  In general, disturbances could propagate to the left and to the right.  The method outlined here is general, and we would need to distinguish between multiple solutions for the speed.  

The speed $s_0$ can be determined self-consistently by making the following key observation: it is the speed of such a reference frame, from which the density profile looks stationary after the transients decay.  Let $z$ denote the coordinate in the co-moving frame: $z = x - s_0t$.  Then
\begin{align}
\phi(z,t) &= \frac{1}{ 2\pi} \int_{-\infty}^{\infty} dk \bar{\phi}_0(k) e^{-i \big[ (k x - k s_0 t) - \big(\omega(k) t - s_0 k t \big)  \big]} \nonumber \\
 &=   \frac{1}{ 2\pi} \int_{-\infty}^{\infty} dk \bar{\phi}_0(k) e^{-i k z + i  t \big[\omega(k) - s_0 k    \big]}. 
 \label{ft_z}
\end{align}

\noindent
If $\bar{\phi}_0$ is analytic everywhere in the complex plane, we can compute this integral when $t$ is large using a saddle point approximation - finding the $k^*$ at which the term $\omega(k) - s_0 k $ has a saddle point, and expanding that term to quadratic order (the function $\omega(k)$ is also assumed to be analytic in the vicinity of its extrema, so this extrema can only be saddles).  This $k^*$ is given by
\begin{equation}
\label{saddle_eq}
\frac{d}{dk} \Big[\omega(k) - s_0 k  \Big]_{k^*} = 0. 
\end{equation}
This leads to our first expression for the speed $s_0$,
\begin{equation}
s_0 = \frac{d \omega}{ dk} \Bigr|_{k^*}.
\label{v1}
\end{equation}

\noindent
The integrand in Eq.~\eqref{ft_z} will be proportional to: $e^{-i k^* z} e^{i t \big( \omega(k^*) - s_0 k^* \big)}$.  From our earlier observation, we require that $\phi(z)$ neither grows nor decays. The means that,
\begin{equation}
\text{Im} \Big[\omega(k^*) - s_0 k^* \Big] = 0,
\label{v1}
\end{equation}
which leads to our second expression for $s_0$,
\begin{equation}
 s_0 = \omega_i^* / k_i^*.
\label{v2}
\end{equation}
Here the $r$ and $i$ subscripts denote real and imaginary part of complex quantities.  We can find $k^*$ by equating Eqs.~\eqref{v1} and \eqref{v2}, and then substitute this back into Eq.~\eqref{v2} to obtain our desired expression for $s$. \begin{align}
& \frac{d \omega}{ dk} \Bigr|_{k^*} = \omega_i^* / k_i^* \hspace{0.2 in} \Rightarrow \text{Find} \;  k^* \label{f1} \\
& s = \omega_i^* / k_i^* \label{f2}     \hspace{0.5 in} \Rightarrow \text{Find} \;  s
\end{align}
We can also compute an approximation to the wave profile $\phi$.  Expanding the term $\omega(k) - s_0 k$ in Eq.~\eqref{ft_z} to second order around $k^*$ and taking into account Eqs.~(\ref{saddle_eq}) and (\ref{v1}), results in the following saddle-point approximation:
\begin{align}
\phi(z,t) & \approx \frac{1}{ 2\pi} \bar{\phi}_0(k^*) e^{-ik^*z} \int_{-\infty}^{\infty} dk  e^{\big[it\left( \omega_r^*  - s_0k_r^*\right)  - Dt  (\Delta k)^2 \big]} \nonumber \\
& =   \frac{1}{\sqrt{4 \pi D t}} \bar{\phi}_0(k^*) e^{ -i(k_r^* z- \omega_r^* t + s_0k_r^*t) } e^{k_i^* z} e^{\frac{- z^2} {4 D t}},
\label{phi_approx}
\end{align}
where $\Delta k = k - k^*$ and  $D = -(i/2) \omega''(k^*)$.  
We must also prohibit oscillatory solutions, since the density cannot be negative.  Therefore, we require that $(k_r, \omega_r, D_i) = (0,0,0)$.  This result will help us to eliminate certain solutions when $s_0$ is multivalued.  
The resulting non-oscillatory expression can be written as 
\begin{align}
\phi(z,t) \approx \frac{1}{\sqrt{4\pi D}}\bar{\phi}_0(k^*)e^{k_i^* z - \frac{1}{2}\ln{t} - \frac{z^2}{4D_rt}}.
\end{align}
Aside from the logarithmic error, which is a consequence of a Gaussian approximation of the integrand, this function becomes time-independent at large times, as planned.  One must also check that $D_r >0$, for physically-meaningful solutions.  The sign of $k^*_i$ helps us to distinguish between the downwind and upwind flanks of the solution of the linear equation (these are respectively the analogues of the downwind and upwind fronts, which are properties of solutions of the parent nonlinear equation).  For the downwind flank, we require $k^*_i < 0$, so that the profile is exponentially decaying for increasing $x$. For the upwind flank, we require $k^*_i > 0$, so that the profile is exponentially increasing for increasing $x$.  

Eqs.~\eqref{f1}, \eqref{f2}, \eqref{phi_approx} summarize our results. They allow us to determine the properties of the tail of a pulled front, including its front speed $s$, and the growth or decay rate $k^*_i$. The only requirement for their use is the dispersion relation $\omega(k)$ and analyticity of $\bar{\phi}_0(k)$.
\\

In summary:
\begin{enumerate}
\item Find dispersion relation $\omega(k)$: Substitute $\phi(x,t) = e^{-i \big( kx - \omega(k) t \big)}$ into linearized PDE. 
\item Find $k^*$: $\frac{d \omega}{ dk} \Bigr|_{k^*} = \omega_i^* / k_i^*$ \\
\item Find speed $s_0 = \omega_i^* / k_i^*$, and decay (for the downwind flank) or growth (for the upwind flank) rate $k^*_i$.
\item Enforce $(\omega_r^*, k_r^*,D_r) = (0,0,0)$ for non-oscillating solution.
\end{enumerate}
Multiple solutions for $k^*$ are possible, stemming from the fact that the method is not restricted to a specific IC, as long as the Fourier Transform of the IC is an entire function in $k$-space.


\subsubsection{Application to our problem}
We now apply these results to our problem. The front speed will depend on the three parameters: $s_0(a,b,\mathcal{D})$. We first solve the $\mathcal{D}=0$ case, and then consider the more realistic situation when $\mathcal{D} \neq 0$.  The $^*$ will be dropped.

\textbf{\textit{Without Diffusion}}:
The dispersion relation $\omega(k)$ has already been found for this case - see Eqs.~(\ref{eq:rho-eq})-(\ref{eq:omega-eq}).  It satisfies 
\begin{equation}
\label{eq:main_eqns1} \omega ^2 - i \omega  (a+b-i k-1)  + b-i k(1-a) =  0.
\end{equation}
We can follow the procedure advertised above:  find $(k_r,k_i)$ of the saddle point, and compute $\omega$ at this $k$.  However, $\omega$ is a multi-valued function, and this would require keeping track of the branches.  On the other hand, note that Eqs.~(\ref{eq:main_eqns1}) and \eqref{f1} constitute a set of four algebraic equations in the variables $(\omega_i, \omega_r, k_r, k_i)$. From Eq.~\eqref{f2}, we can substitute $\omega_i = s_0 k_i$, to obtain a set of equations in the variables $(s_0, \omega_r,  k_r, k_i,)$.  There are six solutions, two of which have $(k_r = 0 = \omega_r)$.  These are
\begin{equation}
\begin{array}{cccc}
\mbox{Solution} & \mathbf{s_0} & \mathbf{k_i} & \mathbf{D} \\
\hline\hline 
\\
1 & \frac{1}{1+\frac{b}{\left(1-\sqrt{a}\right)^2}} & \frac{b}{\sqrt{a}-1}+\sqrt{a}-1 & \frac{\left(\sqrt{a}-1\right)^3 b}{\sqrt{a} \left(a-2 \sqrt{a}+b+1\right)^3} \\
2 & \frac{1}{1+\frac{b}{\left(1+\sqrt{a}\right)^2}} & -\frac{b}{\sqrt{a}+1}-\sqrt{a}-1 & \frac{\left(\sqrt{a}+1\right)^3 b}{\sqrt{a} \left(a+2 \sqrt{a}+b+1\right)^3} 
\end{array}
\end{equation}
We identify the two speeds as those of the upwind and downwind fronts, $s_{\pm}$, cf. Eq.~(4) in the main text.   The leading term in the spatial profile is given by $\phi \sim e^{k_i z} \equiv e^{-\lambda z}$.  The corresponding $\lambda$ also match with Eq.~(5) of the main text.  

To be consistent, we require that $k_i^{(2)} < 0$ for $s_0^{(2)}$ to correspond to the speed of the downwind flank. Inspecting the table above, we see that this is indeed the case. We similarly require that $k_i^{(1)} > 0$ for $s_0^{(1)}$ to be the speed of the upwind flank. Here the situation is not as clear-cut.  One can show that $k_i^{(1)} > 0$ only when $a > 1$. This is consistent with our observation in the main text that the upwind flank moves only when $a>1$.  But for $a < 1$, $k_i^{(2)} < 0$, indicating that $s_0^{(2)}$ is \textit{another} possible wave speed for the downwind flank.  Here the sign of $D$ helps to select the branch:  $D^{(1)}$ is negative for $a<1$, so this is an unphysical solution, and we must select $s^{(2)}$ as the speed of the downwind flank for any $a$.

\textbf{\textit{With Diffusion}}:
We repeat the same procedure as before. With the inclusion of diffusion, the dispersion relation is now satisfied by
\begin{equation}
\omega ^2 - i \omega  \left(a+b+\mathcal{D} k^2-i k-1\right) -i k \left(1-a-\mathcal{D} k^2\right)-b \mathcal{D} k^2+b = 0.
\end{equation}
As before, this determines a set of four algebraic equations in the variables $(s_0, \omega_r, k_r, k_i)$. By imposing $(\omega_r, k_r) = (0,0)$, we get four solutions for the speed $(s^{(1)},s^{(2)},s^{(3)},s^{(4)})$, which are all functions of the parameters $(a,b,\mathcal{D})$. The expressions are complicated, involving roots of 6th degree polynomials. 

The first of these solutions has $k_i^{(1)} > 0$, for all parameter ranges, indicating $s^{(1)}$ corresponds to the speed of the upwind flank. From plotting this solution, we find it corresponds exactly to the solution plotted in Fig.~4b of the main text.   The remaining three solutions never have $k_i > 0$, so they refer to the downwind flank.  Unlike in the $\mathcal{D}=0$ case, the constraint that $D$ is positive and real does not narrow down the candidates to a single solution.  We point out, however, that the values of $-k_i$ match the values of $\lambda$ at which the extrema of the $s(\lambda)$ take place for $\lambda >0$ (see Fig.~\ref{fig:2SM} and \ref{fig:3SM}).  According to \cite{von Saarloos review_SM} - for pulled fronts, ICs with a compact support will select the $\lambda$ at which the branch of $s(\lambda)$ with the largest speed has a minimum.


\section{SIV. Numerical Method}
\label{sec:NumMeth}
\subsubsection{Preliminary calculations}
Before discussing the method, it will be useful to prove that the model without diffusion does not admit shocks.  This is important, because it makes the use of special numerical methods, that otherwise must be employed to keep track of the movement of shock waves, unnecessary.  The dimensionless version of Eqs.~(1)-(2) from the main text can be converted to a single second-order equation for either $\sigma$ or $\rho$.  For example, 
\begin{equation}
\frac{\partial^2 \sigma}{\partial t^2} + \frac{\partial^2 \sigma}{\partial t \partial x} + \frac{\partial \sigma}{\partial t}\Big(a +b - 1 + 2\sigma\Big) + \frac{\partial \sigma}{\partial x} \Big(a-1+2\sigma\Big) = 2f(\sigma)
\end{equation}
A similar equation can be derived for $\rho$.  In both cases, they have the form
\begin{equation}
\frac{\partial^2 \phi}{\partial t^2} + \frac{\partial^2 \phi}{\partial x \partial t} = F\left(\phi, \frac{\partial \phi}{\partial t}, \frac{\partial \phi}{\partial x} \right)
\end{equation}
From this, it can be easily shown \cite{PDESource_SM} that characteristics $x(t)$ obey 
\begin{equation}
\left(\frac{dx}{dt}\right)^2 - \left(\frac{dx}{dt}\right) = 0.
\end{equation}
The two pairs of families of characteristics are thus $x=c_1$ and $x=t + c_2$, where $c_1$ and $c_2$ are arbitrary real constants.  Evidently, characteristics in each family do not intersect each other, proving the absence of shocks.


\subsubsection{Outline of the numerical methods and parameters used}
We used the first-order time-differencing scheme
\begin{equation}
\left.\frac{\partial \phi}{\partial t}\right|_{x_m, t_n} \rightarrow \frac{\phi(x_m, t_n) - \phi(x_m, t_{n-1})}{\Delta t},
\end{equation}
and an upwind spatial differencing scheme:
\begin{equation}
\left.\frac{\partial \phi}{\partial x}\right|_{x_m, t_n} \rightarrow \frac{\phi(x_m, t_n) - \phi(x_{m-1}, t_n)}{\Delta x},
\end{equation}
With this discretization, our equations become
\begin{align}
\rho(x_m, t_n) &= \rho(x_m, t_{n-1}) \Big( 1 - b \Delta t - \Delta t / \Delta x \Big)  + \sigma(x_m, t_{n-1}) a \Delta t  + \rho(x_{m-1}, t_{n-1})  \Delta t / \Delta x.
\end{align}

\begin{align}
\sigma(x_m, t_n) &= \sigma(x_m, t_{n-1}) \Big( 1 - a \Delta t \Big) + f \Big( \sigma(x_m, t_{n-1}) \Big) \Delta t  + \rho(x_m, t_{n-1}) b \Delta t  \nonumber \\
&+ \mathcal{D} \frac{\Delta t}{(\Delta x)^2} \Big[ \sigma(x_{m+1}, t_{n-1}) - 2 \sigma(x_m, t_{n-1}) + \sigma(x_{m-1}, t_{n-1})    \Big],
\end{align}
\noindent
where $x_m = x_0 + m \Delta x$, and $t_n = t_0 + n \Delta t$. Thus, this is an explicit method, i.e. it uses known data at time-step $t_{n-1}$ to march a solution forward in time to $t_n$.

In order for this scheme to be stable, we require $\Delta t / \Delta x < 1$ for the case with $\mathcal{D}=0$ or  $\mathcal{D} \Delta t / ( \Delta x)^2 < 1$ when $\mathcal{D}$ is finite (Courant condition, \cite{NumericalSource_SM}).  To ensure these criteria were met, we chose $\Delta t = \min{ \{ 0.25 \; \Delta x, 0.25 \; (\Delta x)^2 / \mathcal{D} \}}$.  Instabilities were never observed.  Front speeds (and in some select cases, profile shapes) exhibited convergence when $\Delta x$ was made progressively smaller.  The results stated in the main text were obtained with $\Delta x = 0.01$, which is $\ll$ physical characteristic length represented by $1/|\lambda_{\pm}|$ from Eq.~(5) of the main text.

To find the speed of the wavefronts described by $\rho(x,t)$ and $\sigma(x,t)$, we tracked the position of the constant contour $x_C$, which is defined through $\rho(x_C, t) = C$ (Note, since $\rho$ and $\sigma$ have the same speed, we need only consider $\rho$ or $\sigma$ in our calculations).  For example, to extract the speed of the downwind front, we used the following routine:
\begin{itemize}
\item At every time step $t_n$, find position of maximum $\rho(x,t)$, which we call $x_{max}$.
\item Extract all $\rho(x_m,t_n)$ to the right of this maximum: $\tilde{\rho}(x_m, t_n) \equiv \rho(x_m > x_{max}, t_n)$. This function $\tilde{\rho}$ is now unimodal.
\item Then $x_C(t_n) = \min{ |  \tilde{\rho}(x,t_n)  - C |}$ 
\end{itemize}
This results in a list $x_C(t_n)$, which after initial transients, describes a straight line. The wave speed $s$ is the slope of this line.  We ran the calculations until $t=200$, and discarded the first $50 \%$ of the data to remove transient behavior.

\subsubsection{The role of numerical diffusion in long-time asymptotic solutions}
It is known that our differencing scheme is also a lowest-order approximation to an equation with a small diffusion term, even if $\mathcal{D} = 0$ \cite{NumericalSource_SM}.  
To underline the smallness of the role of any effective higher derivative terms on long-time asymptotic results, we show here two plots of the profile shape for $\mathcal{D}=0$ and compare them with the analytical front shape obtained from the UTF ansatz.
\begin{figure}[ht]
\includegraphics[width=4in]{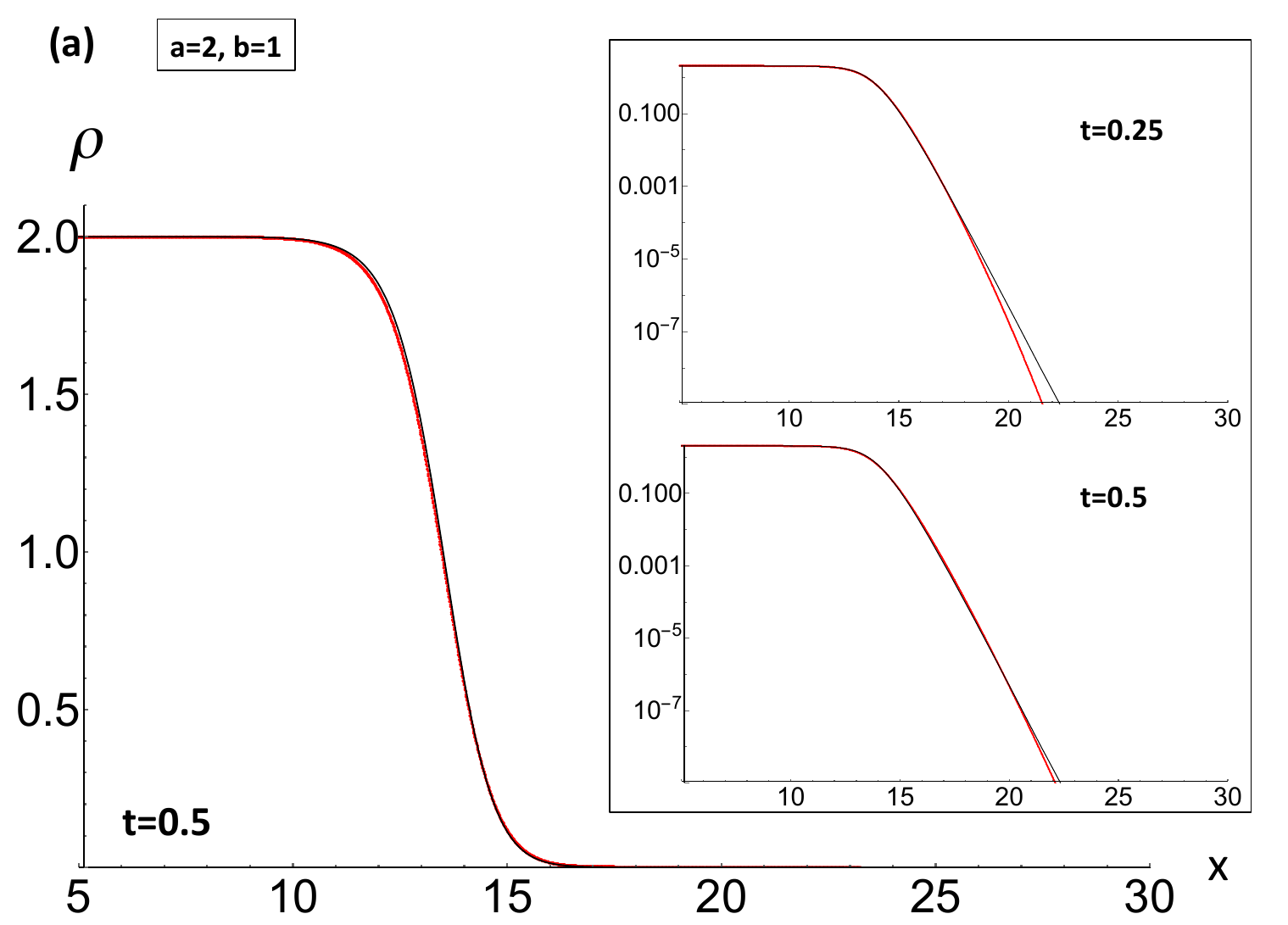}
\includegraphics[width=4in]{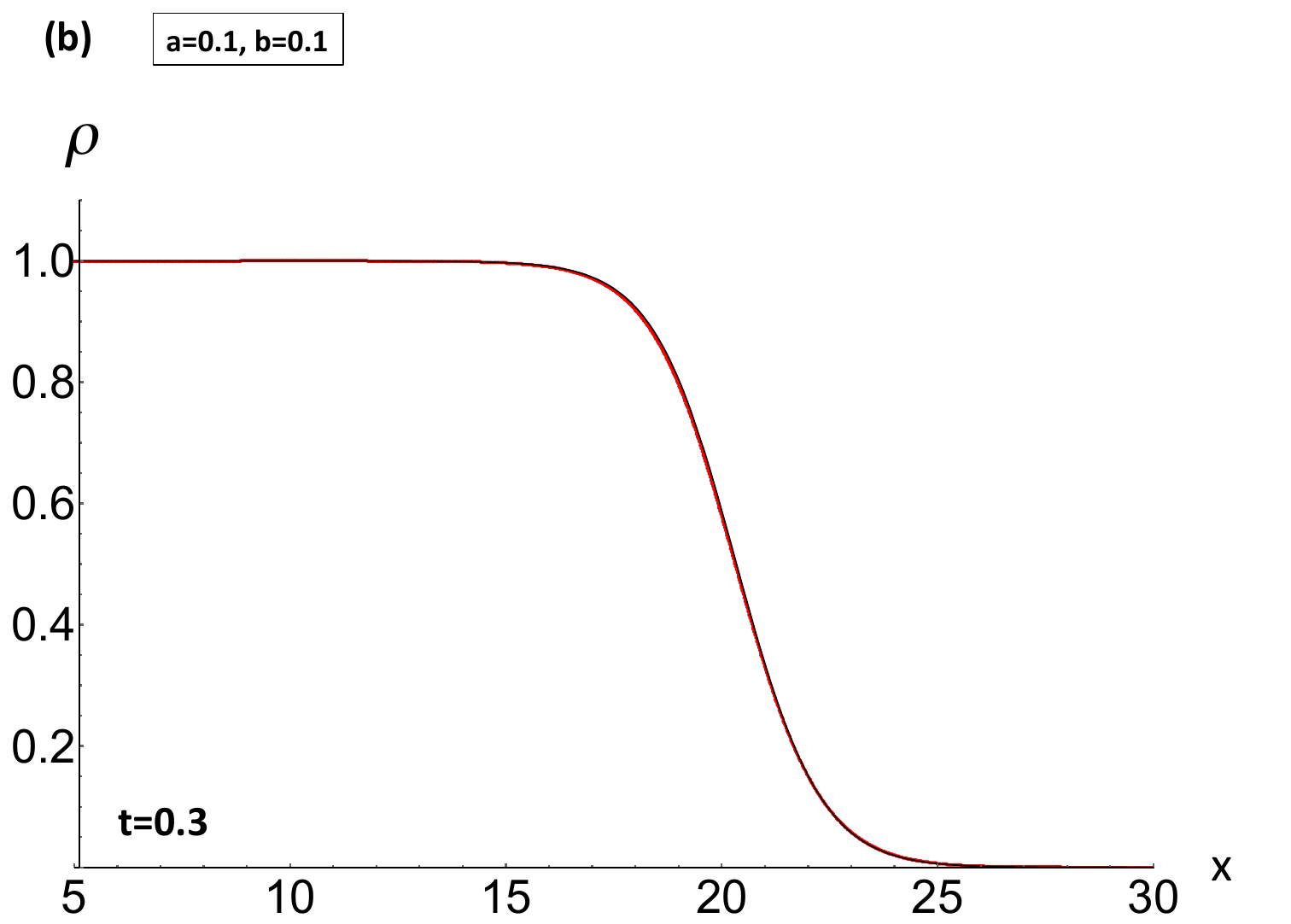}
\caption{(Color online) Comparison of the heteroclinic solution of Eqs.~(\ref{eq: TW_system1})-(\ref{eq: TW_system2}) - UTF profile (thin black curve), with the numerical solution of the dimensionless version of Eqs.~(1)-(2) of the main text (thick, red curve) at a given value of time, shifted to closely match the UTF profile.  (a) $a=2$, $b=1$.  (b)  $a=0.1$, $b=0.1$. The insets: time evolution, logarithmic scale.}
\label{fig:Profilea2b1}
\end{figure}


\section{SV. Application to fungal pathogen transport}
This work was motivated by the problem of wind-driven transport of fungal pathogens on continental scales.  
Wind mediated fungal pathogen spread is a process involving production of spores, lifting, horizontal transport, and deposition \cite{Aerobiology_SM, Production_Review_SM, Plot-scale_SM, LR_transport_SM, Fungal disease overview_SM}.  

In recent years, ``reaction-dispersal'' models \cite{DFisher_SM} have been considered in the context of dispersal of biota, such as seeds and insects \cite{Kernel models_SM}.  Such treatment describes random walks on multiple scales, and may be applicable over distances where the highly turbulent atmospheric boundary layer \cite{ABL_SM} (ABL) - the lowest level of the atmosphere - is the dominant mechanism of dispersal.  However, ABL tends to return the passive scalar back to the ground over the scale of its largest eddies, i.e. its own thickness of $O(1)$ $km$, so it is inefficient at much longer range transport.  The Free Atmosphere (FA) located above the ABL emerges as a competing transport mechanism over longer length scales.  The FA is less random than the ABL, contains persistent advective currents, and can carry passive scalar - including micro-organisms -  across continents \cite{Dust transport_SM} with characteristic speeds of $O(10)$ $km/hr$ \cite{Pedlosky_SM}.   Reaction-dispersal models do not capture the role of the FA.  We were motivated to (i) investigate the validity of ignoring advective transport channels that lie above the ABL and (ii) develop a theory of spatio-temporal dynamics of long-range biotic transport. 

The main insight about the pathogen transport gained from our work is that the advective layer - such as the free atmosphere - can not be ignored, even for very small rates of flow of mass into this advective layer.  To make further statements, it helps to estimate our parameters $a$, $b$ and $\mathcal{D}$.  We know that $a \ll 1$ because for a given amount of spores produced on the ground in a given time interval, most will return to the ground within $100$ meters \cite{Roelfs_SM}, so only a very small fraction will leave the ABL, which has the width of the order of several kilometers \cite{ABL_SM}.  The rate $b$ is generally higher due to gravitational settling, but in the strong turbulence limit, both rates will be comparable (as in a pot on the boiler).  
To estimate the dimensionless diffusion $\mathcal{D}$, recall that it is given by $\frac{\delta D}{v_0^2}$.  Here $v_0$ is the speed of the free atmosphere, and it is of the order of $10$ $km/hr$.  We interpret $D$ to be the eddy diffusion coefficient \cite{Tennekes_SM} of the small-scale turbulence that returns most particles to the ground within the aforementioned $100$ meter radius from the source.  That is, this random transport is accommodated by much smaller eddies than those that contribute to the interlayer transport.   The eddy diffusion coefficient $D$ is expected to scale as $\sim u L$, where $u$ is the characteristic instantaneous velocity in the turbulence, and $L$ is the scale of the eddy, i.e. $0.1$km in this case.  The speed scale is typically used as $1/100$ of the driving speed \cite{Tennekes_SM}, i.e. $u \sim v_0/100 \sim O(0.1)$ $km/hr$.   So $D \sim O(10^{-2})$ $km^2/hr$. 

Finally, we need a growth rate, $\delta$.  
According to \cite{Roelfs1985_SM}, a $5\%$ disease severity amounts to $50$ spore-producing postules per plant tiller.  At this $5\%$ disease severity, a plot will yield approximately two trillion spore/hectare/24-h \cite{Roelfs1985_SM}, i.e. $2 \times 10^8$ spores per $m^2$ per day.  If there are $~O(100)$ plants per $m^2$, the production rate is $4\times 10^5$ spores per postule per day.  If each new spore were to lead to a fungus that produced only one postule, this would imply a multiplication rate of $4\times 10^5$ spores per day produced from a single spore.  In this paper we assumed a logistic growth model, which reduces to exponential growth at low densities.  Thus, an exponential growth model, would thus give $4\times 10^5 = e^{\delta\times \mbox{24 hr}}$, giving $\delta \sim O(0.1)$ $hr^{-1}$.  All together, this gives $\mathcal{D} \sim O\left(10^{-6}\right)$.

Evidently, the problem is completely dominated by the advection, and moreover, the parameters $a$ and $b$ are both $\ll 1$.  For such a parameter regime, our theory predicts the speed is $\sim O($advective speed$)$, or propagation of a front by hundreds of $km$s per day.  This suggests that this model is insufficient for the purpose of the application to fungal pathogen transport, because invasion fronts for observed pathogens, such as Wheat Stem Rust are expected to propagate tens of kms per day \cite{Aylor UV_SM}.  Moreover, these numbers are based on the observations of incidences of disease symptoms on plants, not of spore densities.  We are currently augmenting a model to include both the fungal density (immobile), and spore density (mobile), in addition to the effects of latency.  Our current calculations show that spore death, for instance - a very well-known and important effect \cite{Aylor UV_SM} can decrease the front speeds dramatically.   However, this discussion, in addition to the discussion of other biological specializations of the present theory will appear in a separate publication.
\\

In the main text we argued that the mean-field description will hold when
\begin{equation}
\frac{a}{b} \gg \frac{\delta \lambda(a,b)}{\sigma_{\mathrm{max}} v_0},
\end{equation}
where $\sigma_{\mathrm{max}}$ is the carrying capacity on the GL per unit length.  In application of the fungal pathogen problem, right hand side of this equation is exceptionally small because of the largeness of the carrying capacity.  We already know that $\lambda \rightarrow 1$ when $a,b \ll 1$.  From the discussion above, we see that the order of magnitude of the carrying capacity will be billions of spores per $m^2$.  Using $10^9$ $m^{-2}$ for $\sigma_{\mathrm{max}}$ we arrive at our estimate of $10^{14}$ for the ratio of $a/b$ above which the mean-field model should hold.


{}


\begin{thebibliography}{}
\bibitem{Footnote5prime} We also tested Gaussian ICs with width $\gg$ spatial grid spacing, for select parameters.  The profile speed and the front shape approached the point IC results as time progresses.
\bibitem{Footnote6} We slightly overloaded the notation: now $(\lambda^*_{\pm}, s^*_{\pm})$ represents an extremum point of $s(\lambda)$ and selected downwind/upwind front from a $\delta$-function IC.
\end{thebibliography}
\begin{thebibliography}{}
\bibitem{Footnote_SM} Since $(a=1,b=0)$ is a singular limit, one has to exercise caution.  The limit $b \rightarrow 0$ of the right hand side of Eq.~(\ref{eq:deriv}) reads
\begin{displaymath}
\frac{1}{2\lambda^2} \left[\frac{\mathcal{D} \lambda  \left(\sqrt{\lambda ^2  (\mathcal{D} \lambda -1)^2}+\lambda  (\mathcal{D} \lambda -1)\right)}{\mathcal{D} \lambda -1} \right].
\end{displaymath}
The quantity $\sqrt{\lambda ^2 (\mathcal{D} \lambda -1)^2} = \lambda (\mathcal{D} \lambda -1)$ for $\lambda<0$ and for $\lambda >1/\mathcal{D}$.  Otherwise, $\sqrt{\lambda ^2 (\mathcal{D} \lambda -1)^2} = -\lambda (\mathcal{D} \lambda -1)$.  This choice implies that  $ds_1/d\lambda = \mathcal{D}$ for $\lambda<0$ and for $\lambda >1/\mathcal{D}$, while it is $0$ otherwise.   We could follow a similar argument for $s_1(\lambda)$ and $s_2(\lambda)$ directly, and reach the same conclusion about their piece-wise nature.  Note that setting $b=0$ a priori in Eq.~(\ref{eq:s_of_lambda_equation}) (it would have to be multiplied out by $b+(s-1)\lambda$ first) would give $1$ and $\mathcal{D}\lambda$ as the two branches of $s(\lambda)$ for $\lambda \neq 0$ (and at $a=1$).  This discrepancy is the manifestation of $b=0$ being a singular limit.  Similar remarks apply to  the $a \rightarrow 0$ limit at fixed $b$.  Finally, when Taylor expanding the right hand side of Eq.~(\ref{eq:deriv}) in $b$, the quantity $\sqrt{\lambda ^2 (\mathcal{D} \lambda -1)^2}$ will also make an appearance, which we must treat similarly in order to describe the correct branches at finite $b$.
\end{thebibliography}
\end{document}